# Decentralized Infrastructure for (Neuro)science

*Or, Kill the Cloud in your Mind*

**Jonny L. Saunders**@

University of Oregon
Institute of Neuroscience, Department of Psychology
Eugene, OR 97403, United States

This document was written for the web. The PDF is primarily for
archival purposes, and may be out of date — note the compile
date and version number. Please see the original at:
https://jon-e.net/infrastructure

*Thank you to the many people who made this work possible.*

*I would like to thank my parents for tolerating me for so many years, I've finally found a way to take all the angst and DMCA shutoffs and bricked computers and go pro.*

*Thank you to my lab for supporting me through this project, my advisor Mike Wehr, Lucas Ott who keeps everything running, Nick Sattler, Sam Mehan, Molly Shallows, Alexa Wright, Tillie Morris, Rocky Penick, and Matt Nardoci. Thank you to my dissertation committee, Melissa Baese-Berk, Santiago Jaramillo, and Matt Smear for making sure I stay within the bounds of reason. I would be writing this having been summarily kicked out of academia without the help of Lori Olsen, who shielded and guided me through the viscera of the institution. She is a true testament to how science would simply grind to a halt without those who organize the work.*

*My view on the world has been profoundly shaped by the Janet Smith House and all of the people that I have had the blessing of living with here. There is nothing that has given me hope for the future more than what I have learned in cooperation with you all.*

*Thank you to the many people who were willing to talk with me at length, contributing their wisdom and perspective on the many fields that I would have otherwise stumbled through. In alphabetical order to avoid the appearance of hierarchy: Joon An, Andrey Andreev, Björn Brembs, Nire Bryce, Joel Chan, Kris Chauvin, Jeremy Delahanty, Avery Everhart, Dan Goodman, Olivia Guest, Eartha Mae Guthman, Leslie Harka, Gabriele Hayden, Ceci Herbert, Andrew Hoffman, Os Keyes, Irene Knapp, Mark Laubach & Open Behavior Team, Christine Lemmer-Webber, Gonçalo Lopes, Mackenzie Mathis, Danny Mclanahan, James Meickle, the NWB & DANDI teams, Phil Parker, Ralph Emilio Peterson, Tomasz Pluskiewicz, Chris Rodgers, Manuel Schottdorf, Arnold Schrijver, Sanjay Srivastava & Metascience Class, Petar Todorov, Aad Versteden, Lauren E. Wool, and The Emerging ONICE team. If we talked and I have left you off this list, please let me know, as credit is at the heart of this work.*

*Thank you to my dear Rumbly Tumbly Lawnmower for being the light of my life.*

# Contents

# 1

# Introduction

> If we can make something decentralised, out of control, and of great simplicity, we must be prepared to be astonished at whatever might grow out of that new medium.
>
> Tim Berners-Lee (1998): Realising the Full Potential of the Web

> A good analogy for the development of the Internet is that of constantly renewing the individual streets and buildings of a city, rather than razing the city and rebuilding it. The architectural principles therefore aim to provide a framework for creating cooperation and standards, as a small "spanning set" of rules that generates a large, varied and evolving space of technology.
>
> RFC 1958: Architectural Principles of the Internet

> In building cyberinfrastructure, the key question is not whether a problem is a "social" problem or a "technical" one. That is putting it the wrong way around. The question is whether we choose, for any given problem, a primarily social or a technical solution
>
> Bowker, Baker, Millerand, and Ribes (2010): Toward Information Infrastructure Studies [1]

> Billionaires have squatted on the Magna Cum Lauded / [...] Methodically they plotted against those who fought it / [...] / Now the scientific process got hijacked for profits / It flows in the direction that a silver spoon prodded / We'll get science for the people when we run the economics.
>
> The Coup (2012) The Gods of Science

---

Scientists work in isolation at every scale, reinventing the wheel in parallel. Our knowledge dissemination systems are as nimble as static PDFs[1] served by an extractive publishing turned surveillance industry we can't seem to quit. Experimental instrumentation except for that at the polar extremes of technological complexity or simplicity is designed and built custom, locally, and on-demand[2]. Software for performing experiments is a patchwork of libraries that satisfy some of the requirements of the experiment, sewn together by some uncommented script written years ago by a grad student who left the lab long-since. The technical knowledge to build both instrumentation and software is fragmented and unavailable as it sifts through the funnels of word-limited methods sections and never-finished documentation. Our data is born into this world without coherent form to speak of, indexable only by passively-encrypted notes in a paper lab notebook, and analyzed once before being mothballed in ignominy on some unlabeled external drive.

These problems are typically treated in isolation, but all are symptomatic of a broader deficit in **digital infrastructure** for science. Every routine need that requires heavy technical development, an appeal to a hostile publishing system, or yet another platform subscription is an indicator that infrastructural deficits *define the daily reality of science.* We *should* be able to easily store, share, and search for data; be able to organize and communicate with each other; be able to write and review our work, but we are hemmed in on all sides by looming tech profiteers and chasms of underdevelopment.

[1] Save some complicated half-in flirtation with social media.

[2] In many disciplines, appropriate caveats below.

If the term infrastructure conjures images of highways and plumbing, then surely digital infrastructure would be flattered at the association. Roughly following Star and Ruhleder's (1996) dimensions [2], by analogy they illustrate many of its promises and challenges: when designed to, it can make practically impossible things trivial, allowing the development of cities by catching water where it lives and snaking it through tubes and tunnels sometimes directly into your kitchen. Its absence or failure is visible and impactful, as in the case of power outages. There is no guarantee that it "optimally" satisfies some set of needs for the benefit of the greatest number of people, as in the case of the commercial broadband duopolies. It exists not only as its technical reality, but also as an embodied and shared set of social practices, and so even when it does exist its form is not inevitable or final; as in the case of bottled water producers competing with municipal tap water on a behavioral basis despite being dramatically less efficient and more costly. Finally it is not socially or ethically neutral, and the impact of failure to build or maintain it is not equally shared, as in the expression of institutional racism that was the Flint, Michigan water crisis [3].

Infrastructural deficits are not our inevitable and eternal fate, but the course of infrastructuring is far from certain. It is not the case that "scientific digital infrastructure" will rise from the sea monolithically as a natural result of more development time and funding, but instead has many possible futures [4], each with their own advocates and beneficiaries. Without concerted and strategic development based on a shared and liberatory ethical framework, science will continue to follow the same path as other domains of digital technology down the dark road of platform capitalism. The prize of owning the infrastructure that the practice of science is built on is too great, and it is not hard to imagine tech behemoths buying out the emerging landscape of small scientific-software-as-a-service startups and selling subscriptions to Science Prime.

The possibility of future capture of nascent infrastructure is still too naïve a framing: operating as obligate brokers of (usually surveillance) data [5, 6, 7], prestige, and computational resources naturally relies on *displacing* the possibility of alternative infrastructure. Our predicament is doubly difficult: we both have digital infrastructural deficits, but are also being actively *deinfrastructured.* The harms of deinfrastructuring are bidirectional, comprising both the missed opportunities from decades of free knowledge exchange, and the impacts of the informational regime that exists in its place. One can only imagine what the state of science and medicine might be if NIH's 1999 push to displace for-profit journals [8, 9, 10, 11] had succeeded and we had more than 20 years of infrastructural development built atop a fundamentally free system of scientific knowledge. Instead, our failure to seize the digital infrastructure of science has led to a system where what should be our shared intellectual heritage is yoked to the profit engine of surveillance conglomerates (formerly known as publishers) [5, 12] that repackage it along with a deluge of mined personal data in a circular economy of control [13, 14, 15] that makes us directly complicit in the worst abuses of informational capitalism [16, 17, 18, 19, 20].

We need to move beyond conceptualizing the problems of scientific infrastructure as being unique to science, a sighing hope for some future that "might be nice" to have (built by an always-anonymous "*someone else*"), but one to be pursued gradually after staid and cautious scholars are convinced no risk will come to our precious systems of prestige and peer review. We need to start seeing ours as one of many stories in the digital enclosure movement where adversarial economic entities take ownership of basic digital infrastructure and wipe out a domain of knowledge work, reducing it to a captive market and content farm [7, 21]. We need to see taking control of our digital infrastructure as *essential* to the continued existence of science as we know it.

This paper is an argument that **decentralized** digital infrastructure[3] is the best means of alleviating the harms of infrastructural deficits and building a digital landscape that supports, rather than extracts from science. I will draw from several disciplines and knowledge communities, across and outside academia to articulate a vision of an infrastructure in three parts: **shared data, shared tools, and shared knowledge.** These domains reflect three of the dominant modes of digital enclosure prerequisite for platform capture: **storage, computation, and communication.** The systems we will describe are in conversation with and a continuation of a long history of reimagining the relationship between these domains for a healthier web (see eg. [22, 23]). We depart from it to describe a system of fluid, peer-to-peer social affiliation and folksonomic linked data with lessons primarily from early wikis and Wikipedia, the fissures of the semantic web and linked data communities, the social structure of private bittorrent trackers, and the federation system of ActivityPub and the Fediverse. Approaching this problem from science has its constraints — like the structuring need to rebuild systems of credit assignment — as well as the powerful opportunity of one of the last systems of labor largely not driven by profit developing technology and seeding communities that could begin to directly address the dire, societywide need for digital freedom.

The problems we face are different than they were at the dawn of the internet, but we can learn from its history: we shouldn't be waiting for a new journal-like **platform,** software package, or subscription to save us. We need to build **protocols** for communication, interoperability, and self-governance (see, recently [24]).

I will start with a brief description of what I understand to be the state of our digital infrastructure and the structural barriers and incentives that constrain its development. I will then propose a set of design principles for decentralized infrastructure and possible means of implementing it informed by prior successes and failures at building mass digital infrastructure. I will close with contrasting visions of what science could be like depending on the course of our infrastructuring, and my thoughts on how different actors in the scientific system can contribute to and benefit from decentralization.

I insist that what I will describe is *not utopian* but is eminently practical — the truly impractical choice is to do nothing and continue to rest the practice of science on a pyramid scheme [25] of underpaid labor. With a bit of development to integrate and improve the tools, **every class of technology I propose here already exists and is widely used.** A central principle of decentralized systems is embracing heterogeneity: harnessing the power of the diverse ways we do science instead of constraining them. Rather than a patronizing argument that everyone needs to fundamentally alter the way they do science, the systems that I describe are specifically designed to be easily incorporated into existing practices and adapted to variable needs. In this way I argue decentralized systems are *more practical* than the dream that any one system will be capable of expanding to the scale of all science — and as will hopefully become clear, inarguably *more powerful* than a disconnected sea of centralized platforms and services.

[3] Recently the notion of decentralized digital infrastructure has been co-opted by a variety of swindlers and other motivated parties to refer to blockchain-based technologies like cryptocurrencies, decentralized autonomous organizations, and the like. This work will not discuss them, as their model of artificial scarcity is antithetical to its ethical premises, and they have not been demonstrated to do anything that peer-to-peer technology with adjoining social systems can't do — except use a colossal quantity of fossil fuels and drain a lot of credulous people's bank accounts.

An easy and common misstep is to categorize this as solely a *technical* challenge. Instead the challenge of infrastructure is also *social* and *cultural* — it involves embedding any technology in a set of social practices, a shared belief that such technology should exist, that its form is not neutral, and a sense of communal valuation and purpose that sustains it [26].

The social and technical perspectives are both essential, but make some conflicting demands on the construction of the piece: Infrastructuring requires considering the interrelatedness and mutual reinforcement of the problems to be addressed, rather than treating them as isolated problems that can be addressed piecemeal with a new package or by founding a new journal alternative. Such a broad scope trades off with a detailed description of the relevant technology and systems, but a myopic techno-zealotry that does not examine the social and ethical nature of scientific practice risks reproducing or creating new sources of harm. That, and techno-solutionism never *works* anyway. As a balance I will not be proposing a complete technical specification or protocol, but describing the general form of the tools and some existing examples that satisfy them; I will not attempt a full history or treatment of the problem of infrastructuring, but provide enough to motivate the form of the proposed implementations.

My understanding of this problem is, of course, uncorrectably structured by my training largely centered in systems neuroscience and my position as an early career researcher (ECR). While the core of my argument is intended to be a sketch compatible with sciences and knowledge systems generally, my examples will sample from, and my focus will skew to my experience. In many cases, my use of "science" or "scientist" could be "neuroscience" or "neuroscientist," but I will mostly use the former to avoid the constant context switches. This document is also an experiment in public collaboration on a living scientific document: to try and ease our way out of disciplinary tunnelvision, we invite annotation and contribution with no lower bound — if you'd like to add or correct a sentence or two (or a page or ten), you're welcome as coauthor. I ask the reader for a measure of patience for the many ways this argument requires elaboration and modification for distant fields.

# 2

# The State of Things

## 2.1    The Costs of Infrastructure Deficits

A diagnosis of digital infrastructure deficits gives a common framework to consider many technical and social harms in scientific work that are typically treated separately, and allows us to problematize other symptoms have become embedded as norms.

I will list some of the present costs to give a sense of the scale of need, as well as scope for the problems we intend to address here. These lists are grouped into rough and overlapping categories, but make no pretense at completeness.

Impacts on the **daily experience** of researchers include:

- A prodigious duplication and dead-weight loss of labor as each lab, and sometimes each person within each lab, will reinvent basic code, tools, and practices from scratch. Literally it is the inefficiency of the Harberger's triangle in the supply and demand system for scientific infrastructure caused by inadequate supply. Labs with enough resources are forced to pay from other parts of their grants to hire professional programmers and engineers to build the infrastructure for their lab[1], but most just operate on a purely amateur basis. Many PhD students will spend the first several years of their degree re-solving already-solved problems, chasing the tails of the wrong half-readable engineering whitepapers, in their 6th year finally discovering the technique that they actually needed all along. That's not an educational or training model, it's the effect of displacing the undone labor of unbuilt infrastructure on vulnerable graduate workers almost always paid poverty wages.

> [1] (and usually their lab or institute only)

- At least the partial cause of the phenomenon where "every scientist needs to be a programmer now" as people who aren't particularly interested in being programmers — which is *fine* and *normal* — need to either suffer through code written by some other unlucky amateur or learn several additional disciplines in order to do the work of the one they chose.

- A great deal of pain and alienation for early- career researchers not previously trained in programming before being thrown in the deep end. Learning data hygiene practices like backup, annotation, etc. "the hard way" through some catastrophic loss is accepted myth in much of science. At some scale all the very real and widespread pain, guilt, and shame felt by people who had little choice but to reinvent their own data management system must be recognized as an infrastructural, rather than a personal problem.

- The high cost of "openness" and the dearth of mass-scale collaboration. It is still rare to publish full, raw data and analysis code, often because the labor of cleaning it is too great. We can't expect openness from everyone while it is still so *hard*. The "Open science" movement, roughly construed, has reached a few hard limits from present infrastructure that have forced its energy to leak from the sides as bullying leaderboards or sets of symbols that are mere signifiers of cultural affiliation to openness. "Openness" is not a uniform or universal goal for all science, and even the framing of openness as inspection of results collected and analyzed in private isolation illustrates how infrastructural deficits bound our imagination. Our dreams can be bigger than being able to police each other's data, towards a more continuously collaborative process that renders the need for post-hoc openness irrelevant with mutually beneficial information sharing baked into every stage.

Impacts on the **system of scientific inquiry** include:

- A profoundly leaky knowledge acquisition system where entire PhDs worth of data can be lost and rendered useless when a student leaves a lab and no one remembers how to access the data or how it's formatted.

- The inevitability of continual replication crises because it is often literally impossible to replicate an experiment that is done on a rig that was built one time, used entirely in-lab code, and was never documented

- Reliance on communication platforms and knowledge systems that aren't designed to, and don't come close to satisfying the needs of scientific communication. In the absence of some generalized means of knowledge organization, scientists ask the void[2] for advice or guidance from anyone that algorithmically stumbles by. Often our best recourse is to make a Slack about it, which is incapable of producing a public, durable, and cumulative resource: and so the same questions will be asked again... and again...    [2] (Twitter)

- A perhaps doomed intellectual endeavor as we attempt to understand the staggering complexity of the brain by peering at it through the camera obscura of just the most recent data you or your lab have collected rather than being able to index across the many measurements of the same phenomena. The unnecessary reduplication of experiments becomes not just a methodological limitation, but an ethical catastrophe as researchers have little choice but to abandon the elemental principle of sacrificing as few animals as possible.

- A near-absence of semantic or topical organization of research that makes cumulative progress in science probabilistic at best, and subject to the malformed incentives of publication and prestige gathering at worst. Since engaging with prior literature is a matter of manually reconstructing a caricature of a field of work in every introduction, continuing lines of inquiry or responding to conflicting results is *strictly optional*.

- A hierarchy of prestige that devalues the labor of many groups of technicians, animal care workers, and so on. Authorship is the coin of the realm, but many workers that are fundamental to the operation of science only receive the credit of an acknowledgement. We need a system to value and assign credit for the immense amount of technical and practical knowledge and labor they contribute.

Impacts on the relationship between **science and society**:

- An insular system where the inaccessibility of all the "contextual" knowledge [27, 28] that doesn't have a venue for sharing but is necessary to perform experiments, like "how to build this apparatus," "what kind of motor would work here," etc. is a force that favors established and well-funded labs who can rely on
local knowledge and hiring engineers/etc. and excludes new, lesser-funded labs at
non-ivy institutions. The concentration of technical knowledge magnifies the in-
equity of strongly skewed funding distributions such that the most well-funded
labs can do a completely different kind of science than the rest of us, turning the
positive-feedback loop of funding begetting funding ever faster.

- An absconscion with the public resources we are privileged enough to receive,
where rather than returning the fruits of the many technical challenges we are
tasked with solving to the public in the form of data, tools, collected practical
knowledge, etc. we largely return papers. Since those papers are often impene-
trable outside of their discipline or paywalled outside of academia, we multiply
the above impacts of labor duplication and knowledge inaccessibility by the scale
of society.

- The complicity of scientists in rendering our collective intellectual heritage noth-
ing more than another regiment in the ever-advancing armies of platform capital-
ism. If our highest aspirations are to shunt all our experiments, data, and analysis
tools onto Amazon Web Services, our failure of imagination will be responsible
for yet another obligate funnel of wealth into the system of extractive platforms
that dominate the flow of global information. For ourselves, we stand to have the
practice of science filleted at the seams into a series of mutually incompatible
subscription services. For society, we squander the chance for one of the very
few domains of non-economic labor to build systems to recollectivize the basic
infrastructure of the internet: rather than providing an alternative to the infor-
mation overlords and their digital enclosure movement, we will be run right into
their arms.

Considered separately, these are serious problems, but together they are a damning
indictment of our role as stewards of our corner of the human knowledge project.

We arrive at this situation not because scientists are lazy and incompetent, but be-
cause we are embedded in a system of mutually reinforcing disincentives to cumu-
lative infrastructure development. Our incentive systems are, in turn, coproductive
with a raft of economically powerful entities that would really prefer owning it all
themselves, thanks. Put bluntly, "we are dealing with a massively entrenched set of
institutions, built around the last information age and fighting for its life" [1]

There is, of course, an enormous amount of work being done by researchers and
engineers on all of these problems, and a huge amount of progress has been made
on them. My intention is not to shame or devalue anyone's work, but to try and
describe a path towards integrating it and making it mutually reinforcing.

Before proposing a potential solution to some of the above problems, it is important
to motivate why they haven't already been solved, or why their solution is not nec-
essarily imminent. To do that, we need a sense of the social and technical challenges
that structure the development of our tools.

## 2.2    (Mis)incentives in Scientific Software

The incentive systems in science are complex, subject to infinite variation every-
where, so these are intended as general tendencies rather than statements of irrevo-
cable and uniform truth.

## 2.2.1   Incentivized Fragmentation

Scientific software development favors the production of many isolated, single-purpose software packages rather than cumulative work on shared infrastructure. The primary means of evaluation for a scientist is academic reputation, primarily operationalized by publications, but a software project will yield a single paper (if any). Traditional publications are static units of work that are "finished" and frozen in time, but software is never finished: the thousands of commits needed to maintain and extend the software are formally not a part of the system of academic reputation.

Howison & Herbsleb described this dynamic in the context of BLAST[3]

> In essence we found that BLAST innovations from those motivated to improve BLAST by academic reputation are motivated to develop and to reveal, but not to integrate their contributions. Either integration is actively avoided to maintain a separate academic reputation or it is highly conditioned on whether or not publications on which they are authors will receive visibility and citation. [29]

[3] "Basic Local Alignment Search Tool" - a tool to compare genetic or protein sequences to find potential matches or analogues.

For an example in Neuroscience, one can browse the papers that cite the DeepLab-Cut paper [30] to find hundreds of downstream projects that make various extensions and improvements that are not integrated into the main library. While the alternative extreme of a single monolithic ur-library is also undesirable, working in fragmented islands makes infrastructure a random walk instead of a cumulative effort.

After publication, scientists have little incentive to **maintain** software outside of the domains in which the primary contributors use it, so outside of the most-used libraries most scientific software is brittle and difficult to use [31, 32, 33].

Since the reputational value of a publication depends on its placement within a journal and number of citations (among other metrics), citation practices for scientific software are far from uniform and universal, and relatively few "prestige" journals publish software papers at all, the incentive to write scientific software in the first place is low compared to its near-universal use [34].

## 2.2.2   Domain-Specific Silos

When funding exists for scientific infrastructure development, it typically comes in the form of side effects from, or administrative supplements to research grants. The NIH describes as much in their Strategic Plan for Data Science [35] :

> from 2007 to 2016, NIH ICs used dozens of different funding strategies to support data resources, most of them linked to research-grant mechanisms that prioritized innovation and hypothesis testing over user service, utility, access, or efficiency. In addition, although the need for open and efficient data sharing is clear, where to store and access datasets generated by individual laboratories— and how to make them compliant with FAIR principles—is not yet straightforward. Overall, it is critical that the data-resource ecosystem become seamlessly integrated such that different data types and information about different organisms or diseases can be used easily together rather than existing in separate data "silos" with only local utility.

The National Library of Medicine within the NIH currently lists 122 separate databases in its search tool, each serving a specific type of data for a specific research community. Though their current funding priorities signal a shift away from domain-specific tools, the rest of the scientific software system consists primarily of tools and data formats purpose-built for a relatively circumscribed group of scientists. Every field has its own challenges and needs for software tools, but there is little incentive to build tools that serve as generalized frameworks to integrate them.

### 2.2.3   "The Long Now" of Immediacy vs. Idealism

Digital infrastructure development takes place at multiple timescales simultaneously — from the momentary work of implementing it; through longer timescales of planning, organization, and documenting; to the imagined indefinite future of its use — what Ribes and Finholt call "The Long Now. [36] " Infrastructural projects constitutively need to contend with the need for immediately useful results vs. general and robust systems; the need to involve the effort of skilled workers vs. the uncertainty of future support; the balance between stability and mutability; and so on. The tension between hacking something together vs. building something sustainable for future use is well-trod territory in the hot-glue and exposed wiring of systems neuroscience rigs.

Deinfrastructuring divides the incentives and interests of junior and senior researchers. ECRs might be interested in developing tools they'll use throughout their careers, but given the pressure to establish their reputation with publications rarely have the time to develop something fully. The time pressure never ends, and established researchers also need to push enough publications through the door to be able to secure the next round of funding. The time preference of scientific software development is thus very short: hack it together, get the paper out, we'll fix it later.

The constant need to produce software that *does something* in the context of scientific programming which largely lacks the institutional systems and expert mentorship needed for well-architected software means that most programmers *never* have a chance to learn best practices commonly accepted in software engineering. As a consequence, a lot of software tools are developed by near-amateurs with no formal software training, contributing to their brittleness [37].

The problem of time horizon in development is not purely a product of inexperience, and a longer time horizon is not uniformly better. We can look to the history of the semantic web, a project that was intended to bridge human and computer-readable content on the web, for cautionary tales. In the semantic web era, thousands of some of the most gifted programmers and some of the original architects of the internet worked with an eye to the indefinite future, but the raw idealism and neglect of the pragmatic reality of the need for software to *do something* drove many to abandon the effort (bold is mine, italics in original):

> **But there was no *use* of it.** I wasn't using any of the technologies for anything, except for things related to the technology itself. The Semantic Web is utterly inbred in that respect. The problem is in the model, that we create this metaformat, RDF, and *then* the use cases will come. But they haven't, and they won't. Even the genealogy use case turned out to be based on a fallacy. The very few use cases that there are, such as Dan Connolly's hAudio export process, don't justify hundreds of eminent computer scientists cranking out specification after specification and API after API.

> When we discussed this on the Semantic Web Interest Group, the conversation kept turning to how the formats could be fixed to make the use cases that I outlined happen. "Yeah, Sean's right, let's fix our languages!" But **it's not the languages which are broken,** except in as much as they are entirely broken: because **it's the *mentality* of their design which is broken.** You can't, it has turned out, make a metalanguage like RDF and then go looking for use cases. We thought you could, but you can't. It's taken eight years to realise. [38]

Developing digital infrastructure must be both bound to fulfilling immediate, incremental needs as well as guided by a long-range vision. The technical and social lessons run in parallel: We need software that solves problems people actually have, but can flexibly support an eventual form that allows new possibilities. We need a long-range vision to know what kind of tools we should build and which we shouldn't, and we need to keep it in a tight loop with the always-changing needs of the people it supports.

In short, to develop digital infrastructure we need to be *strategic.* To be strategic we need a *plan.* To have a plan we need to value planning as *work.* On the valuation of this kind of work, Ribes and Finholt are instructive:

> "On the one hand, I know we have to keep it all running, but on the other, LTER is about long-term data archiving. If we want to do that, we have to have the time to test and enact new approaches. But if we're working on the to-do lists, we aren't working on the tomorrow-list" (LTER workgroup discussion 10/05).

> The tension described here involves not only time management, but also the differing valuations placed on these kinds of work. The implicit hierarchy places scientific research first, followed by deployment of new analytic tools and resources, and trailed by maintenance work. [...] While in an ideal situation development could be tied to everyday maintenance, in practice, maintenance work is often invisible and undervalued. As Star notes, infrastructure becomes visible upon breakdown, and only then is attention directed at its everyday workings (1999). Scientists are said to be rewarded for producing new knowledge, developers for successfully implementing a novel technology, but the work of maintenance (while crucial) is often thankless, of low status, and difficult to track. *How can projects support the distribution of work across research, development, and maintenance?* [36]

### 2.2.4   "Neatness" vs "Scruffiness"

Closely related to the tension between "Now" and "Later" is the tension between "Neatness" and "Scruffiness." Lindsay Poirier traces its reflection in the semantic web community as the way that differences in "thought styles" result in different "design logics" [39]. On the question of how to develop technology for representing the ontology of the web – the system of terminology and structures with which everything should be named – there were (very roughly) two camps. The "neats" prioritized consistency, predictability, uniformity, and coherence – a logically complete and formally valid System of Everything. The "scruffies" prioritized local systems of knowledge, expressivity, "believing that ontologies will evolve organically as everyday webmasters figure out what schemas they need to describe and link their data. [39] "

This tension is as old as the internet, where amidst the dot-com bubble a telecom spokesperson lamented that the internet wasn't controllable enough to be profitable because "it was devised by a bunch of hippie anarchists." [40] The hippie anarchists probably agreed, famously rejecting "kings, presidents and voting" in favor of "rough consensus and running code" during an attempted ISO coup to replace TCP/IP with a proprietary protocol. Clearly, the difference in thought styles has an unsubtle relationship with beliefs about who should be able to exercise power and what ends a system should serve [41].

VIEWS OF THE FUTURE

The last force on us — us

The standards elephant of yesterday — OSI.

The standards elephant of today — its right here.

As the Internet and its community grows, how do we manage the process of change and growth?
• Open process — let all voices be heard.
• Closed process — make progress.
• Quick process — keep up with reality.
• Slow process — leave time to think
• Market driven process — the future is commercial.
• Scaling driven process — the future is the Internet.

We reject: kings, presidents and voting.
We believe in: rough consensus and running code.

SLIDE 19

**Figure 2.1:** A slide from David Clark's 1992 "Views of the Future"[42] that contrasts differing visions for the development process of the future of the internet. The struggle between engineered order and wild untamedness is summarized forcefully as "We reject: kings, presidents and voting. We believe in: rough consensus and running code"

Practically, the differences between these thought communities impact the tools they build. Aaron Swartz put the approach of the "neat" semantic web architects the way he did:

> Instead of the "let's just build something that works" attitude that made the Web (and the Internet) such a roaring success, they brought the formalizing mindset of mathematicians and the institutional structures of academics and defense contractors. They formed committees to form working groups to write drafts of ontologies that carefully listed (in 100-page Word documents) all possible things in the universe and the various properties they could have, and they spent hours in Talmudic debates over whether a washing machine was a kitchen appliance or a household cleaning device.
>
> With them has come academic research and government grants and corporate R&D and the whole apparatus of people and institutions that scream "pipedream." And instead of spending time building things, they've convinced people interested in these ideas that the first thing we need to do is write standards. (To engineers, this is absurd from the start—standards are things you write after you've got something working, not before!) [43]

The outcomes of this cultural rift are subtle, but the broad strokes are clear: the "scruffies" largely diverged into the linked data community, which has taken some of the core semantic web technology like RDF, OWL, and the like, and developed a broad range of downstream technologies that have found purchase across information sciences, library sciences, and other applied domains[4]. The linked data developers, starting by acknowledging that no one system can possibly capture everything, build tools that allow expression of local systems of meaning with the expectation and affordances for linking data between these systems as an ongoing social process.

The vision of a totalizing and logically consistent semantic web, however, has largely faded into obscurity. One developer involved with semantic web technologies (who requested not be named), captured the present situation in their description of a still-active developer mailing list:

> I think that some people are completely detached from practical applications of what they propose. [...] I could not follow half of the messages. these guys seem completely removed from our plane of existence and I have no clue what they are trying to solve.

This division in thought styles generalizes across domains of infrastructure, though outside of the linked data and similar worlds the dichotomy is more frequently between "neatness" and "people doing whatever" – with integration and interoperability becoming nearly synonymous with standardization. Calls for standardization without careful consideration and incorporation of existing practice have a familiar cycle: devise a standard that will solve everything, implement it, wonder why people aren't using it, funding and energy dissipates, rinse, repeat[5]. The difficulty of scaling an exacting vision of how data should be formatted, the tools researchers should use for their experiments, and so on is that they require dramatic and sometimes total changes to the way people do science. The alternative is not between standardization and chaos, but a potential third way is designing infrastructures that allow the diversity of approaches, tools, and techniques to be expressed in a common framework or protocol along with the community infrastructure to allow the continual negotiation of their relationship.

### 2.2.5 Taped-on Interfaces: Open-Loop User Testing

The point of most active competition in many domains of commercial software is the user interface and experience (UI/UX). To compete, software companies will exhaustively user-test and refine them with pixel precision to avoid any potential customer feeling even a thimbleful of frustration. Scientific software development is largely disconnected from usability testing, as what little support exists is rarely tied to it. This, combined with the preponderance of semi-amateurs and above incentives for developing new packages – and thus reduplicating the work of interface development – make it perhaps unsurprising that most scientific software is hard to use!

I intend the notion of "interface" in an expansive way: In addition to a graphical user interface (GUI) or set of functions and calling conventions exposed to the end-user, I am referring generally to all points of contact with users, developers, and other software. Interfaces are intrinsically social, and include the surrounding documentation and experience of use — part of using software is being able to figure out how to use it! The favored design idiom of scientific software is the black box: I implemented an algorithm of some kind, here are the two or three functions needed to use it, but beneath the surface there be dragons.

[4] This isn't a story of "good people" and "bad people," as a lot of the linked data technology also serves as the backbone for abusive technology monopolies like google's acquisition of Freebase [44] and the profusion of knowledge graph-based medical platforms.

[5] There is, of course, an XKCD for that to which we make obligatory reference: https://xkcd.com/927/

Ideally, software would be designed with developer interfaces and documentation at multiple scales of complexity to enable clean entrypoints for developers with differing levels of skill and investment to contribute. When this kind of design and documentation is underdeveloped, even widely used projects with excellent top-level interfaces like poetry struggle to respond to the pile of issues thousands deep as even users who have spent time reading the source have difficulty understanding what exactly needs to be fixed and maintainers have to spend their time triaging them and manually re-explaining the software hundreds of times[6].

Additionally, it would include interfaces for use and integration with other software — or APIs. While the term "API" most commonly refers to web APIs, the term generally refers to the means by which other programs can interact with a given program. All programs have some limit to their function, the question is how other programs are expected to handle them. One particularly successful approach to program interface design is the Unix philosophy as articulated by Doug McIlroy and colleagues [45] — which was originally designed to help build research software. Its first "make each program do one thing well" and second "expect the output of every program to become the input to another, as yet unknown, program" principles inspired a set of simple tools that can be composed together for complex tasks. When a program is monolithic and isn't designed to provide access to its component parts, it becomes difficult to reuse in downstream projects, potentially reskin with a more friendly user interface, and ultimately more likely to be a dead-end in a system of shared infrastructure.

Without care given to any of these types of interfaces, the barrier to use is likely to remain high, the community of co-developers is likely to remain small, and the labor they expend is less likely to be useful outside that single project. This, in turn, closes the loop with incentives to develop new packages and makes another vicious cycle reinforcing fragmentation[7].

## 2.2.6    Platforms, Industry Capture, and the Profit Motive

Publicly funded science is an always-irresistible golden goose for private industry. The fragmented interests of scientists and the historically light touch of funding agencies on encroaching privatization means that if some company manages to capture and privatize a corner of scientific practice they are likely to keep it. Industry capture has been thoroughly criticized in the context of the journal system (eg. recently, [24]), and that criticism should extend to the rest of our infrastructure as information companies seek to build a for-profit platform system that spans the scientific workflow (eg. [46]). The mode of privatization of scientific infrastructure follows the broader software market as a proliferation of software as a service (SaaS), from startups to international megacorporations, that rent access to some, typically proprietary software without selling the software itself.

While in isolation SaaS can make individual components of the infrastructural landscape easier to access — and even free!!* — the business model is fundamentally incompatible with integrated and accessible infrastructure. The SaaS model derives revenue from subscription or use costs, often operating as "freemium" models that make some subset of its services available for free. Even in freemium models, though, the business model requires that some functionality of the platform is enclosed and proprietary. To keep the particular domain of enclosure viable as a profit stream, the proprietor needs to actively defend against competitors as well as any technol-ogy that might fill the need for the proprietary technology[8] (See a more thorough treatment of platform capitalism in science in [4])

As isolated services, one can imagine the practice of science devolving along a similar path as the increasingly-fragmented streaming video market: to do my work I need to subscribe to a data storage service, a cloud computing service, a platform to host my experiments, etc. For larger platforms, however, vertical integration of multiple complementary services makes their impact on infrastructure more insidious. Locking users into more and more services makes for more and more revenue, which encourages platforms to be as mutually incompatible as they can get away with [48]. To encourage adoption, platforms that can offer multiple services may offer one of the services – say, data storage – for free, forcing the user to use the adjoining services – say, a cloud computing platform.

Since these platforms are often subsidiaries of information industry monopolists, scientists become complicit in their often profoundly unethical behavior of by funneling millions of dollars into them. Longterm, unconditional funding of wildly profitable journals has allowed conglomerates like Elsevier to become sprawling surveillance companies [49, 5] that are sucking as much data up as they can to market derivative products like algorithmic ranking of scientific productivity [13] and making data sharing agreements with ICE [17]. Or our reliance on AWS and the laundry list of human rights abuses by Amazon [50]. In addition to lock-in, dependence on a constellation of SaaS allows the opportunity for platform-holders to take advantage of their limitations and *sell us additional services to make up for what the other ones purposely lack* — for example Elsevier has taken advantage of our dependence on the journal system and its strategic disorganization to sell a tool for summarizing trending research areas for tailoring maximally-fundable grants [51].

Funding models and incentive structures in science are uniformly aligned towards the platformatization of scientific infrastructure. Aside from the corporate doublespeak "technology transfer" rhetoric that pervades the neoliberal university, the relative absence of major funding opportunities for scientific software developers competitive with the profit potential from "industry" often leaves it as the only viable career path. The preceding structural constraints on local infrastructural development strongly incentivize labs and researchers to rely on SaaS that provides a readymade solution to specific problems. Distressingly, rather than supporting infrastructural development that would avoid obligate payments to platform-holders, funding agencies seem all too happy to lean into them (emphases mine):

> NIH will **leverage what is available in the private sector,** either through strategic partnerships or procurement, to create a workable **Platform as a Service (PaaS)** environment. [...] NIH will partner with cloud-service providers for cloud storage, computational, and related infrastructure services needed to facilitate the deposit, storage, and access to large, high-value NIH datasets. [...]

> NIH's cloud-marketplace initiative will be the first step in a phased operational framework that **establishes a SaaS paradigm for NIH and its stakeholders.** (-NIH Strategic Plan for Data Science, 2018 [35])

The articulated plan being to pay platform holders to house data while also paying for the labor to maintain those databases veers into parody, haplessly building another triple-pay industry [52] into the economic system of science — one can hardly wait until they have the opportunity to rent their own data back with a monthly subscription. This isn't a metaphor: the STRIDES program, with the official subdomain cloud.nih.gov, has been authorized to pay $85 million to cloud providers since 2018. In exchange, NIH hasn't received any sort of new technology, but "extramural" scientists receive a maximum discount of 25% on cloud storage and "data egress" fees as well as plenty of training on how to give control of the scientific process to platform giants [53] [9]. Without exaggeration, we are paying them to let us pay for something that makes it so we need to pay them more later.

It is unclear to me whether this is the result of the cultural hegemony of platform capitalism narrowing the space of imaginable infrastructures, industry capture of the decision-making process, or both, but the effect is the same in any case.

### 2.2.7   Protection of Institutional and Economic Power

Aside from information industries, infrastructural deficits are certainly not without beneficiaries within science — those that have already accrued power and status.

Structurally, the adoption of SaaS on a wide scale necessarily sacrifices the goals of an integrated mass infrastructure as the practice of research is carved into small, marketable chunks within vertically integrated technology platforms. Worse, it stands to amplify, rather than reduce, inequities in science, as the labs and institutes that are able to afford the tolls between each of the weigh stations of infrastructure are able to operate more efficiently — one of many positive feedback loops of inequity.

More generally, incentives across infrastructures are often misaligned across strata of power and wealth. Those at the top of a power hierarchy have every incentive to maintain the fragmentation that prevents people from competing — hopefully mostly unconsciously via uncritically participating in the system rather than maliciously reinforcing it.

This poses an organizational problem: the kind of infrastructure that unwinds platform ownership is not only unprofitable, it's **anti-profitable** – making it impossible to profit from its domain of use. That makes it difficult to rally the kind of development and lobbying resources that profitable technology can, requiring organization based on ethical principles and a commitment to sacrifice control in order to serve a practical need.

The problem is not insurmountable, and there are strategic advantages to decentralized infrastructure and its development within science. Centralized technologies and companies might have more concerted power, but we have *numbers* and can make tools that let us combine small amounts of labor from many people. We often start (and end) our dreams of infrastructure with the belief that they will necessarily cost a lot of *money,* but that's propaganda. Of course development isn't *free,* but the cost of decentralized technologies is far smaller than the vast sums of money funneled into industry profits, labor hours spent compensating for the designed inefficiencies of the platform model, and the development of a fragmented tool ecosystem built around them.

Science, as one of few domains of non-economic labor, has the opportunity to be a seed for decentralized technologies that could broadly improve not only the health of scientific practice, but the broader information ecosystem. We can develop a plan and mobilize to make use of our collective expertise to build tools that have no business model and no means of development in commercial domains — we just need

[9] Their success stories tell the story of platform non-integration where scientists have to handbuild new tools to manage their data across multiple cloud environments: "We have been storing data in both cloud environments because we wanted the ecosystem we are creating to work on both clouds" [54]

to realize what's at stake and agree that the health of science is more important than the convenience of the cloud[10] or which journal our papers go into.

## 2.3   The Ivies, Institutes, and "The Rest of Us"

Given these constraints, who can build new digital infrastructure? Constraints, motivations, and strategies all depend on the circumstance of those doing the development. The undone work of infrastructure is being nibbled at around the edges[11] by several different kinds of organization already ranging in scale and structure. A short survey to give us some notion of how we should seek to organize infrastructure building:

### 2.3.1   Institutional Core Facilities

Centralized "core" facilities are maybe the most typical form of infrastructure development and resource sharing at the level of departments and institutions. These facilities can range from minimal to baroque extravagance depending on institutional resources and whatever complex web of local history brought them about.

A subproject within a PNI Systems Core grant echoes a lot of the thoughts here, particularly regarding effort duplication[12]:

> Creating an Optical Instrumentation Core will address the problem that much of the technical work required to innovate and maintain these instruments has shifted to students and postdocs, because it has exceeded the capacity of existing staff. This division of labor is a problem for four reasons: (1) lab personnel often do not have sufficient time or expertise to produce the best possible results, (2) the diffusion of responsibility leads people to duplicate one another's efforts, (3) researchers spend their time on technical work at the expense of doing science, and (4) expertise can be lost as students and postdocs move on. For all these reasons, we propose to standardize this function across projects to improve quality control and efficiency. Centralizing the design, construction, maintenance, and support of these instruments will increase the efficiency and rigor of our microscopy experiments, while freeing lab personnel to focus on designing experiments and collecting data.

While core facilities are an excellent way of expanding access, reducing redundancy, and standardizing tools *within* an institution, as commonly structured they can displace work spent on efforts that would be portable *outside* of the institution. Elite institutions can attract the researchers with the technical knowledge to develop the instrumentation of the core and infrastructure for maintaining it, but this development is only occasionally made usable by the broader public. The Princeton data science core is an excellent example of a core facility that does makes its software infrastructure development public[13], which they should be applauded for, but also illustrative of the problems with a core-focused infrastructure project. For an external user, the documentation and tutorials are incomplete – it's not clear to me how one would set this up for my institute, lab, or data, and there are several places of hard-coded Princeton-specific values that I am unsure how exactly to adapt[14]. I would consider this example a high-water mark, and the median openness of core infrastructure falls far below it. I was unable to find an example of a core facility that

[10] Though the system of engineered helpless that convinces us that we're incapable of managing our own web infrastructure is not actually as reliable and seamless as it claims, as the long history of dramatic outages at AWS can show us [55, 56]

[11] aka doing hard development work in sometimes adverse conditions.

[12] Thanks a lot to the one-and-only brilliant Dr. Eartha Mae Guthman for suggesting looking at the BRAIN initiative grants as a way of getting insight on core facilities.

[13] Project Summary: Core 2, Data Science [...] In addition, the Core will build a data science platform that stores behavior, neural activity, and neural connectivity in a relational database that is queried by the DataJoint language. [...] This data-science platform will facilitate collaborative analysis of datasets by multiple researchers within the project, and make the analyses reproducible and extensible by other researchers. [...] NIH RePORTER

[14] Though again, this project is exemplary, built by friends, and would be an excellent place to start extending towards global infrastructure.

maintained publicly-accessible documentation on the construction and operation of its experimental infrastructure or the management of its facility.

This might be unsurprising given the economic structure of most core facilities: an institution pays for a core to benefit the institution, and downstream public benefits are a nice plus but not high up in the list of concerns (if present at all). Core facilities are thus unlikely to serve as the source of mass infrastructure, but they do serve as a point of local coordination within institutions, and so given some larger means of coordination may still be useful.

### 2.3.2   Centralized Institutes

Outside of universities, the Allen Brain Institute is perhaps the most impactful reflection of centralization in neuroscience. The Allen Institute has, in an impressively short period of time, created several transformative tools and datasets, including its well-known atlases [57] and the first iteration of its Observatory project which makes a massive, high-quality calcium imaging dataset of visual cortical activity available for public use. They also develop and maintain software tools like their SDK and Brain Modeling Toolkit (BMTK), as well as a collection of hardware schematics used in their experiments. The contribution of the Allen Institute to basic neuroscientific infrastructure is so great that, anecdotally, when talking about scientific infrastructure it's not uncommon for me to hear something along the lines of "I thought the Allen was doing that."

Though the Allen Institute is an excellent model for scale at the level of a single organization, its centralized, hierarchical structure cannot (and does not attempt to) serve as the backbone for all neuroscientific infrastructure. Performing single (or a small number of, as in its also-admirable OpenScope Project) carefully controlled experiments a huge number of times is an important means of studying constrained problems, but is complementary with the diversity of research questions, model organisms, and methods present in the broader neuroscientific community.

Christof Koch, its director, describes the challenge of centrally organizing a large number of researchers:

> Our biggest institutional challenge is organizational: assembling, managing, enabling and motivating large teams of diverse scientists, engineers and technicians to operate in a highly synergistic manner in pursuit of a few basic science goals [58]

> These challenges grow as the size of the team grows. Our anecdotal evidence suggests that above a hundred members, group cohesion appears to become weaker with the appearance of semi-autonomous cliques and sub-groups. This may relate to the postulated limit on the number of meaningful social interactions humans can sustain given the size of their brain [59]

These institutes too are certainly helpful in building core technologies for the field, but they aren't necessarily organized for developing mass-scale infrastructure. They reflect the capabilities and needs of the institute itself, which are likely to be radically different than a small lab. They can build technologies on a background of expensive cloud storage and computation and rely on a team of engineers to implement and maintain them. So while the tools they make are certainly *useful* we shouldn't count on them to build the systems we need for scientists at large.

### 2.3.3   Meso-scale collaborations

Given the diminishing returns to scale for centralized organizations, many have called for smaller, "meso-scale" collaborations and consortia that combine the efforts of multiple labs [60]. The most successful consortium of this kind has been the International Brain Laboratory [61, 27], a group of 22 labs spread across six countries. They have been able to realize the promise of big team neuroscience, setting a new standard for performing reproducible experiments across many labs [62] and developing data management infrastructure to match [63] [15]. Their project thus serves as the benchmark for large-scale collaboration and a model from which all similar efforts should learn from.

Critical to the IBL's success was its adoption of a flat, non-hierarchical organizational structure, as described by Lauren E. Wool:

> IBL's virtual environment has grown to accommodate a diversity of scientific activity, and is supported by a flexible, 'flattened' hierarchy that emphasizes horizontal relationships over vertical management. [...] Small teams of IBL members collaborate on projects in Working Groups (WGs), which are defined around particular specializations and milestones and coordinated jointly by a chair and associate chair (typically a PI and researcher, respectively). All WG chairs sit on the Executive Board to propagate decisions across WGs, facilitate operational and financial support, and prepare proposals for voting by the General Assembly, which represents all PIs. [27]

They should also be credited with their adoption of a form of consensus decision-making, sociocracy, rather than a majority-vote or top-down decisionmaking structure. Consensus decision-making systems are derived from those developed by Quakers and some Native American nations, and emphasize collective consent rather than the will of the majority.

The infrastructure developed by the IBL is impressive, but its focus on a single experiment makes it difficult to expand and translate to widescale use. The hardware for the IBL experimental apparatus is exceptionally well-documented, with a complete and detailed build guide and library of CAD parts, but the documentation is not modularized such that it might facilitate use in other projects, remixed, or repurposed. The experimental software is similarly single-purpose, a chimeric combination of Bonsai [64] and PyBpod scripts. It unfortunately lacks the API-level documentation that would facilitate use and modification by other developers, so it is unclear to me, for example, how I would use the experimental apparatus in a different task with perhaps slightly different hardware, or how I would then contribute that back to the library. The experimental software, according to the PDF documentation, will also not work without a connection to an alyx database. While alyx was intended for use outside the IBL, it still has IBL-specific and task-specific values in its source-code, and makes community development difficult with a similar lack of API-level documentation and requirement that users edit the library itself, rather than temporary user files, in order to use it outside the IBL.

My intention is not to denigrate the excellent tools built by the IBL, nor their inspiring realization of meso-scale collaboration, but to illustrate a problem that I see as an extension of that discussed in the context of core facilities — designing infrastructure for one task, or one group in particular makes it much less likely to be portable to other tasks and groups. This argument is much more contingent on the specific circumstances of the consortium than the prior arguments about core facilities and institutes: when organized with mass-infrastructure in mind, collaborations between semi-autonomous groups across institutions could be a powerful mode of tool development.

[15] Seriously, don't miss their extremely impressive data portal.

It is also unclear how replicable these consortia are, and whether they challenge, rather than reinforce technical inequity in science. Participating in consortia systems like the IBL requires that labs have additional funding for labor hours spent on work for the consortium, and in the case of graduate students and postdocs, that time can conflict with work on their degrees or personal research which are still far more potent instruments of "remaining employed in science" than collaboration. In the case that only the most well-funded labs and institutions realize the benefits of big team science without explicit consideration given to scientific equity, mesoscale collaborations could have the unintended consequence of magnifying the skewed distribution of access to technical expertise and instrumentation.

The central lesson of the IBL, in my opinion, is that governance matters. Even if a consortium of labs were to form explicitly to build mass-scale digital infrastructure, without a formal system to ensure contributors felt heard and empowered to shape the project it would soon become unfocused or unsustainable. Even if this system is not perfect, with some labor still falling unequally on some researchers, it is a promising model for future collaborative consortia.

### 2.3.4   The rest of us...

Outside of ivies with rich core facilities, institutes like the Allen, or nascent multi-lab consortia, the rest of us are largely on our own, piecing together what we can from proprietary and open source technology. The world of open source scientific software has plenty of energy and lots of excellent work is always being done, though constrained by the circumstances of its development described briefly above. Anything else comes down to whatever we can afford with remaining grant money, scrape together from local knowledge, methods sections, begging, borrowing, and (hopefully not too much) stealing from neighboring labs.

The state of broader scientific deinfrastructuring is perhaps to be expected given our relationship to informational monopolies that in some part depend on it, but unlike many other industries or professions there is reason for hope in science. Science is packed with people with an enormous diversity of skills, resources, and perspectives. Publicly funded science is relatively unique as a labor system that does not strictly depend on profit. There is widespread discontent with the systems of scientific practice, and so the question becomes how we can organize our skill, labor, and energy to rebuild the systems that constrain us.

A third option from the standardization offered by centralization and the blooming, buzzing, beautiful chaos of disconnected open-source development is that of decentralized systems, and with them we might build the means by which the "rest of us" can mutually benefit by organizing our knowledge and labor.

We don't need to wait for permission from a memo from a funding body or the founding of some new organization. We do have to recognize that while we might have very different roles to play, we are all responsible for the state of digital scientific infrastructure. We should take courage and purpose in knowing that we are not alone, and that our problems are just one reflection of the model of digital en-closure and surveillance that defines the information economy. There is no need for distance or animosity with the other modes of organization described above, as if what we intend to build is truly useful to *everyone* except those that profit from its absence, then that certainly includes them. Shunting the vision of a better future onto some as-yet formed effort is precisely the trap we should avoid: our existing organizations *should* be a part of the work of rebuilding our infrastructure precisely because we should be reconsidering the ways that *we, ourselves* work. Seeing a subscription to this platform monopolist's cloud, or that knowledge baron's prestige hierarchy as not being a value-neutral decision begs an alternative from people, labs, and institutions alike. The diversity in what that means for different groups is a *strength,* not a weakness, but it does require some shared vision and notion of how to get there. The rest of the paper is an attempt to draft one.

# 3

# A Draft of Decentralized Scientific Infrastructure

What should we build?

The infrastructural systems I will describe here are similar to previous notions of "grass-roots" science articulated within systems neuroscience [60], "small tech" [65] or the anti software software club's manifesto [66] in the web development world , and shares some of the motivations of the Solid project [67], but ultimately draws from a set of ideas with broad and deep history in many domains of computing. My intention is to provide a more prescriptive scaffolding for their design and implementation as a way of painting a picture of what science could be like. This sketch is not intended to be final, but a starting point for further negotiation and refinement.

Throughout this section, when I am referring to any particular piece of software I want to be clear that I don't intend to be dogmatically advocating that software *in particular*, but software *like it* that *shares its qualities* — no snake oil is sold in this document. Similarly, when I describe limitations of existing tools, without exception I am describing a tool or platform I love, have learned from, and think is valuable — learning from something can mean drawing respectful contrast! Many of these technologies have long and torrid social histories, and so when invoked as examples I don't necessarily mean to import along with them all the unmentioned baggage that might accompany them[1].

## 3.1   Design Principles

I won't attempt to derive a definition of decentralized systems from first principles here, but from the constraints described above, some design principles that illustrate the idea emerge naturally. For the sake of concreteness, in some of these I will draw from the architectural principles of the internet protocols (specifically TCP/IP): the most successful decentralized digital technology project to date.

### 3.1.1   Protocols, not Platforms

Much of the basic technology of the internet was developed as protocols that describe the basic attributes and operations of a process. A simple and common example is email over SMTP (Simple Mail Transfer Protocol) [68]. SMTP describes a series of steps that email servers must follow to send a message: the sender initiates a connection to the recipient server, the recipient server acknowledges the connection, a few more handshake steps ensue to describe the senders and receivers of the message, and then the data of the message is transferred. Any software that implements the protocol can send emails to and from any other. The protocol basis of email is the reason why it is possible to send an email from a gmail account to a hotmail account (or any other hacky homebrew SMTP client) despite being wholly different pieces of software.

In contrast, *platforms* provide some service with a specific body of code usually with-out any pretense of generality. In contrast to email over SMTP, we have grown accustomed to not being able to send a message to someone using Telegram from WhatsApp, switching between multiple mutually incompatible apps that serve nearly identical purposes. Platforms, despite being *theoretically* more limited than associated protocols, are attractive for many reasons: they provide funding and administrative agencies a single point of contracting and liability, they typically provide a much more polished user interface, and so on. These benefits are short-lived, however, as the inevitable toll of lock-in and shadowy business models is realized.

[1] As one example, while I will write about linked data, I don't necessarily mean it in precisely the original instantiation as an irrevocable URI/RDF/SPARQL-only web, but do draw on its triplet link structure.

By virtue of being intended for use by many independent organizations rather than under the sole control of a platform-holder, protocols are a complicated political effort that embed and facilitate systems of belief and power (see re: TCP/IP [41], ActivityPub [69]). For example, in order to arrive at a version of TCP/IP that kept the intermediate relays relatively simple at the expense of reliability, the manufacturer of the "smart" relays had to be excluded from the group. TCP/IP's success was not inevitable: it was one of several protocols, becoming the default over proprietary competitors from telecommunication and network hardware companies because of some combination of timing, its relative absence of bureaucracy, and institutional adoption (depending on who does the accounting)[41].

Seemingly prosocial protocols can be used by industries to preempt an alternative that would undermine their profit model — a notable example for academics being the DOI system, created in order for publishers to preserve control over their intellectual property [70]. The STM association[2] hastily[3] threw its weight behind the DOI-X initiative at its 1999 meeting. The impending creation of PubMed Central by the National Library of Medicine (and see then-NIH Director Harold Varmus' and others self-described "radical" departure from publishers with what became PLoS [9, 8]) posed an existential threat to for-profit publishing. At the time there was no unified means of linking to scholarly work[4], and bilateral publisher-publisher linking deals threatened the smooth operation of business, so an NIH-owned platform might have made journals might lose their status as the obligate dissemination platform. According to Bob Campbell, STM chair at the time: "our consensus was that publishers should be the ones doing the linking." Unlike the anarchic URI/URL, The DOI system requires a registrar (denoted by the prefix before the slash, `doi:10.xxxx/yyyyy`) to create DOI names [72]. In the US, that means being an institution with an approved CrossRef membership, which requires members not to link to intellectual property infringing content, and to use DOIs as their default reference links to other works. Effectively, though it is an "open"[5] standard, the DOI system ensures that publishers remain in control of what counts as scholarly work [71].

When approaching protocols, we should do so with humility and caution: work in smaller teams with shared visions with the intention of rough consensus around multiple instances of working code. We should refuse participation by the wide range of industries and interest groups circling each domain of infrastructure, their protocols and standards are siren songs.

### 3.1.2 Integration, not Invention

At the advent of the internet protocols, several different institutions and universities had already developed existing network infrastructures, and so the "top level goal" of IP was to "develop an effective technique for multiplex utilization of exist-

[2] The global trade association of publishers that serves as its lobbying and propaganda arm.

[3] The description provided by the "official" CrossRef 10 year retrospective paints a picture of panicked executives making an announcement for something they didn't have a clear picture of yet, but it would be *something* to compete with pubmed:

"We decided to issue an announcement of a broad STM reference linking initiative. It was, of course, a strategic move only, since we had neither plan nor prototype.

A small group led by Arnoud de Kemp of Springer-Verlag met in an adjacent room immediately following the Board meeting to draft the announcement, which was distributed to all attendees of the STM annual meeting the following day and published in an STM membership publication.

Campbell recalled running into Bolman and Swanson (neither of whom was then on the STM Board) in the hotel lobby immediately after the drafting of the announcement. Their astonishment at hearing what had just transpired was matched by Campbell's own on learning what they had been working on. [...]

Bolman and Swanson chose to seize the moment, and called an ad hoc meeting the following evening, Tuesday, October 12, to announce their venture and assemble a coalition of publishers to launch it. [...]

The potential benefit of the service that would become CrossRef was immediately apparent. Organizations such as AIP and IOP (Institute of Physics) had begun to link to each other's publications, and the impossibility of replicating such one-off arrangements across the industry was obvious. As Tim Ingoldsby later put it, "All those linking agreements were going to kill us." [71]

[4] It is hard to appreciate in retrospect how radical URLs/URIs were at the time — it might seem trivial to us now to be able to arbitrarily link to different locations on the internet, but before the internet linking was a carefully controlled process within publishing, looking more like ISBN and ISSNs than hyperlinks.

[5] Reading the standard costs 88 Swiss Francs.

ing interconnected networks," and "come to grips with the problem of integrating a number of separately administered entities into a common utility" [73]. As a result, IP was developed as a 'common language' that could be implemented on any hardware, and upon which other, more complex tools could be built. This is also a cultural practice: when the system doesn't meet some need, one should try to extend it rather than building a new, separate system — and if a new system is needed, it should be interoperable with those that exist.

This point is practical as well as tactical: to compete, an emerging protocol should integrate or be capable of bridging with the technologies that currently fill its role. A new database protocol should be capable of reading and writing existing databases, a new format should be able to ingest and export to existing formats, and so on. The degree to which switching is seamless is the degree to which people will be willing to switch.

This principle runs directly contrary to the current incentives for novelty and fragmentation and the dominant economic model of software platforms, which must be counterbalanced by design choices elsewhere.

### 3.1.3    *Embrace Heterogeneity, Be Uncoercive*

In addition to integrating with existing systems, it must be straightforward for unanticipated future development to be integrated to accommodate unanticipated needs and practices. This idea is related to "the test of independent invention", summarized with the question "if someone else had already invented your system, would theirs work with yours?" [74]. Rather than attempting to *a priori* divine a single perfect universal protocol, we should design multiple with extensibility in mind (see this discussion of the extensibility models of ActivityPub to XMPP [75] and Christopher Yoo's description of the tradeoffs of the internet's layered protocols [76]) to leave open the opportunity for porting functionality between them.

This principle also has tactical elements. An uncoercive system allows users to gradually adopt it rather than needing to adopt all of its components in order for any one of them to be useful. We shouldn't rely on potential users making dramatic changes to their existing practices. For example, an experimental framework should not insist on a prescribed set of supported hardware and rigid formulation for describing experiments. Instead it should provide affordances that give a clear way for users to extend the system to fit their needs [77].There always needs to be a *benefit* to adopting further components of the system to encourage *voluntary* adoption, but it should never be *compulsory*. For example, again from experimental frameworks, it should be possible to use it to control experimental hardware without needing to use the rest of the experimental design, data storage, and interface system. To some degree this is accomplished with a modular system design where designers are mindful of keeping the individual modules independently useful.

A noncoercive architecture also prioritizes the ease of leaving. Though this is somewhat tautological to protocol-driven design, specific care must be taken to enable export and migration to new systems. Multiplicity of design and making leaving easy help ensure that early missteps in development of the system are not fatal, preventing lock-in to a component that becomes fixed and stagnant.

### 3.1.4   *Empower People, not Systems*

Because IP was initially developed as a military technology by DARPA, a primary design constraint was survivability in the face of failure. The model adopted by internet architects was to move as much functionality as possible from the network itself to the end-users of the network — rather than the network itself guaranteeing a packet is transmitted, the sending computer will do so by requiring a response from the recipient [73].

For infrastructure, we should make tools that don't require a central team of developers to maintain, a central server-farm to host data, or a small group of people to govern. Whenever possible, data, software, and hardware should be self-describing[6], so one needs minimal additional tools or resources to understand and use it. It should never be the case that funding drying up for one node in the system causes the entire system to fail.

Practically, this means that the tools of digital infrastructure should be deployable by individual people and be capable of recapitulating the function of the system without reference to any central authority. Researchers need to be given control over the function of infrastructure: from controlling sharing permissions for eg. clinically sensitive data to assurance that their tools aren't spying on them. Formats and standards must be negotiable by the users of a system rather than regulated by a central governance body.

### 3.1.5   *Infrastructure is Social*

The alternative to centralized governing and development bodies is to build the tools for community control over infrastructural components. This is perhaps the largest missing piece in current scientific tooling. On one side, decentralized governance is the means by which an infrastructure can be maintained to serve the ever-evolving needs of its users. On the other, a sense of community ownership is what drives people to not only adopt but contribute to the development of an infrastructure. In addition to being a source of all the warm fuzzies of socially affiliative "community-ness," any collaborative system needs a way of ensuring that the practice of maintaining, building, and using it is designed to *visibly and tangibly benefit* those that do, rather than be relegated to a cabal of invisible developers and maintainers [78, 79].

Governance and communication tools also make it possible to realize the infinite variation in application that infrastructures need while keeping them coherent: tools must be built with means of bringing the endless local conversations and modifications of use into a common space where they can become a cumulative sense of shared memory.

I will return to this idea in Archives Need Communities in the context of social dynamics of private bittorrent trackers, as well as propose a set of basic communication and governance tools in Rebuilding Scientific Communication.

### 3.1.6   *Usability Matters*

It is not enough to build a technically correct technology and assume it will be adopted or even useful, it must be developed embedded within communities of practice and *be useful for solving problems that people actually have.* We should learn from the struggles of the semantic web project. Rather than building a fully pre-scriptive and complete system first and deploying it later, we should develop tools whose usability is continuously improved *en route* to a (flexible) completed vision.

[6] AKA you shouldn't need to resort to some external source to understand it. Data should come packaged with clear metadata, software should have its own docs, etc.

The adage from RFC 1958[7] "nothing gets standardized until there are multiple instances of running code" [77] captures the dual nature of the constraint well. Work-able standards don't emerge until they have been extensively tested in the field, but development without an eye to an eventual protocol won't make one.

We should read the gobbling up of open protocols into proprietary platforms that defined "Web 2.0" as instructive[8][80]. *Why* did Slack outcompete IRC?[9] The answer is relatively simple: it was relatively simple to use. Using a contemporary example, to set up a Synapse server to communicate over Matrix one has to wade through dozens of shell commands, system-specific instructions, potential conflicts between dependent packages, set up an SQL server... and that's just the backend, we don't even have a frontend client yet! In contrast, to use Slack you download the app, give it your email, and you're off and running.

The control exerted by centralized systems over their system design does give certain structural advantages to their usability, and their for-profit model gives certain advantages to their development process. There is no reason, however, that decentralized systems *must* be intrinsically harder to use, we just need to focus on user experience to a degree comparable to centralized platforms: if it takes a college degree to turn the water on, that ain't infrastructure.

People are smart, they just get frustrated easily and have other things to do on a deadline. We have to raise our standards of design such that we don't expect users to have even a passing familiarity with programming, attempting to build tools that are truly general use. We can't just design a peer-to-peer system, we need to make the data ingestion and annotation process automatic, effortless, and expressive. We can't just build a system for credit assignment, it needs to happen as an automatic byproduct of using the system. We can't just make tools that *work,* they need to *feel good to use.*

Centralized systems also have intrinsic limitations that provide openings for decentralized systems, like cost, incompatibility with other systems, restrictions on independent extension, and opacity of function. The potential for decentralized systems to capture the independent development labor of all of its users, rather than just that of a core development team, is one means of competition. If a system is sufficiently easy to adopt, at least comparable to prior tooling, and gives people a satisfying means of having their work accepted and valued, the social and technical joy might be enough to outweigh the inertia of change and the convenience of centralized systems.

With these principles in mind, and drawing from other knowledge communities solving similar problems: internet infrastructure, library/information science, peer-to-peer networks, and radical organizing, I conceptualize a system of distributed infrastructure for (neuro)science as three objectives: **shared data**, **shared tools**, and **shared knowledge**.

## 3.2  Shared Data

### 3.2.1 Formats as Onramps

The shallowest onramp towards a generalized data infrastructure is to make use of existing discipline-specific standardized data formats. As will be discussed later, a truly universal pandisciplinary format is impossible and undesirable, but to arrive at the alternative we should first congeal the wild west of unstandardized data into a smaller number of established formats.

Data formats consist of some combination of an abstract specification, an implementation in a particular storage medium, and an API for interacting with the format. I won't dwell on the particular qualities that a particular format needs, assuming that most that would be adopted would abide by FAIR principles.

There are a dizzying number of scientific data formats [81], so a comprehensive treatment is impractical here and I will use Neurodata Without Borders:N (NWB) [82] as an example. NWB is the de facto standard for systems neuroscience, adopted by many institutes and labs, though far from universally. NWB consists of a specification language, a schema written in that language, a storage implementation in hdf5, and an API for interacting with the data. They have done an admirable job of engaging with community needs [83] and making a modular, extensible format ecosystem.

The major point of improvement for NWB, and I imagine many data standards, is the ease of conversion and use. The conversion API requires extensive programming, knowledge of the format, and navigation of several separate tutorial documents. This means that individual labs, if they are lucky enough to have some partially standardized format for the lab, typically need to write (or hire someone to write) their own software library for conversion.

Without being prescriptive about its form, substantial interface development is needed to make mass conversion possible. It's usually untrue that unstandardized data had *no structure,* and researchers are typically able to articulate it – "the filenames have the collection date followed by the subject id," and so on. Lowering the barriers to conversion mean designing tools that match the descriptive style of folk formats, for example by prompting them to describe where each of an available set of metadata fields are located in their data. It is not an impossible goal to imagine a piece of software that can be downloaded and with minimal recourse to reference documentation allow someone to convert their lab's data within an afternoon.

NWB also has an extension interface, which allows, for example, data from common hardware and software tools to be more easily described in the format. These are registered in an extensions catalogue, but at the time of writing it is relatively sparse. The preponderance of lab-specific conversion packages relative to extensions is indicative of an interface and community tools problem: presumably many people are facing similar conversion problems, but because there is not a place to share these techniques in a human-readable way, the effort is duplicated in dispersed codebases. We will return to some possible solutions for knowledge preservation and format extension when we discuss tools for shared knowledge.

For the sake of the rest of the argument, let us assume that some relatively trivial conversion process exists to subdomain-specific data formats and we reach some reasonable penetrance of standardization. The interactions with the other pieces of infrastructure that may induce and incentivize conversion will come later.

### 3.2.2   Peer-to-peer as a Backbone

We should adopt a *peer-to-peer* system for storing and sharing scientific data. There are, of course many existing databases for scientific data, ranging from domain-general like figshare and zenodo to the most laser-focused subdiscipline-specific. The notion of a database, like a data standard, is not monolithic. As a simplification, they consist of at least the hardware used for storage, the software implementation of read, write, and query operations, a formatting schema, some API for interacting with it, the rules and regulations that govern its use, and especially in scientific databases some frontend for visual interaction. For now we will focus on the storage software and read-write system, returning to the format, regulations, and interface later.

Centralized servers[10] are fundamentally constrained by their storage capacity and bandwidth, both of which cost money. In order to be free, database maintainers need to constantly raise money from donations or grants in order to pay for both. Funding can never be infinite, and so inevitably there must be some limit on the amount of data that someone can upload and the speed at which it can serve files[11]. Centralized servers are also intrinsically out of the control of their users, requiring them to abide whatever terms of use the server administrators set. Even if the database is carefully backed up, it serves as a single point of infrastructural failure, where if the project lapses then at worst data will be irreversibly lost, and at best a lot of labor needs to be expended to exfiltrate, reformat, and rehost the data. The same is true of isolated, local, institutional-level servers and related database platforms, with the additional problem of skewed funding allocations making them unaffordable for many researchers.

Peer-to-peer (p2p) systems solve many of these problems, and I argue are the only type of technology capable of making a database system that can handle the scale of all scientific data. They are also not new for science, used in projects like Academic-Torrents.com [84, 85] or the now defunct BioTorrents [86]. Whether we acknowledge it or not, most scientific work is already available on p2p networks via sci-hub and library genesis [87, 88, 89].

There is an enormous degree of variation between p2p systems[12], but they share a set of architectural advantages. The essential quality of any p2p system is that rather than each participant in a network interacting only with a single server that hosts all the data, everyone hosts data and interacts directly with each other.

For the sake of concreteness, we can consider a (simplified) description of Bittorrent [91], arguably the most successful p2p protocol. To share a collection of files, a user creates a `.torrent` file with their Bittorrent client which consists of a cryptographic hash, or a string that is unique to the collection of files being shared; and a list of "trackers." A tracker, appropriately, keeps track of the `.torrent` files that have been uploaded to it, and connects users that have or want the content referred to by the `.torrent` file. The uploader (or seeder) then leaves a torrent client open waiting for incoming connections. Someone who wants to download the files (a leecher) will then open the `.torrent` file in their client, which will then ask the tracker for the IP addresses of the other peers who are seeding the file, directly connect to them, and begin downloading. So far so similar to standard client-server systems, but say another person wants to download the same files before the first person has finished downloading it: rather than *only* downloading from the original seeder, the new leecher downloads from *both* the original seeder and the first leecher by requesting pieces of the file from each until they have the whole thing. Leechers are incentivized to share among each other to prevent the seeders from spending time reuploading the pieces that they already have, and once they have finished downloading they become seeders themselves.

[10] This applies to centrally managed traditional servers as well as rented space on larger CDNs like AWS, but in the case of the CDN the constraint is from their pricing model.

[11] As I am writing this, I am getting a (very unscientific sample of n=1) maximum speed of 5MB/s on the Open Science Framework

[12] Peer to peer systems are, maybe predictably, a whole academic subdiscipline. See [90] for reference.

From this very simple example, we can articulate a number of attractive qualities of p2p systems:

- First, p2p systems are extremely **inexpensive to maintain** since they take advantage of the existing bandwidth and storage space of the computers in the swarm. Near the height of its popularity in 2009, The Pirate Bay, a notorious bittorrent tracker [92], was estimated to cost $3,000 per month to maintain while serving approximately 20 million peers [93]. According to a database dump from 2013 [94], multiplying the size of each torrent by the number of seeders (ignoring any partial downloads from leechers), the approximate instantaneous amount of data stored by The Pirate Bay was ~26 Petabytes. The comparison to centralized services is not straightforward, since it is hard to evaluate the distributed costs of additional storage media (as well as the costs avoided by being able to take advantage of existing storage infrastructure within labs and institutes), but for the sake of illustration: hosting 26PB would cost $546,000/month with standard AWS S3 hosting ($0.021/GB/month). On AWS, downloads cost extra ($0.05/GB), so the much smaller academictorrents.com which has served nearly 18PB in 1.3m downloads since 2016 would have cost $900,000 in bandwidth costs alone — as opposed to the literally zero dollars it costs to operate.

- The **speed** of a bittorrent swarm *increases,* rather than decreases, the more people are using it since it is capable of using all of the available bandwidth in the system.

- The network is extremely **resilient** since the data is shared across many independent peers in the system. If our goal is to make a resilient and robust data architecture, we would benefit by paying attention to the tools used in the broader archival community, especially the archival communities that are frequent targets of governments and intellectual property holders [95]. Despite more than 15 years of concerted effort by governments and intellectual property holders, The Pirate Bay is still alive and kicking[13] [96]. This is because even if the entire infrastructure of the tracker is destroyed, as it was in 2006, the files are distributed across all of its users, the actual database of `.torrent` metadata is quite small, and the tracker software is extraordinarily simple to rehost [97] – The Pirate Bay was back online in 2 days. When another tracker, what.cd (which we will return to soon) was shut down, a series of successors popped up using the open source tools Gazelle and Ocelot that what.cd developers built. Within two weeks, one successor site had recovered and reindexed 200,000 of its torrents resubmitted by former users [98]. Bittorrent is also used by archival groups with little funding like Archive Team, who struggled – but eventually succeeded – to disseminate their geocities archive over a single "crappy cable modem" [99].

[13] knock on wood

- The network is extremely **scalable** since there is no cost to connecting new peers and the users of a system expand the storage capacity of the system depending on their needs. Rather than having one extremely fast data center, the model of p2p systems is to leverage many approachable peer/servers.

Peer-to-peer systems are not mutually exclusive with centralized servers: servers are peers too, after all. A properly implemented p2p system will always be *at least* as fast and have *at least* as much storage as any alternative centralized server because peers can use *both* the bandwidth of the server *and* that of any peers that have the file. In the bittorrent ecosystem large-bandwidth/storage peers are known as "seedboxes"[100] when they use the bittorrent protocol, and "web seeds"[101] when they use a protocol built on top of traditional HTTP. Archive.org has been distributing all of its materials with bittorrent by using its servers as web seeds since 2012 and makes this point explicitly: "BitTorrent is now the fastest way to download items from the Archive, because the Bittorrent client downloads simultaneously from two different Archive servers located in two different datacenters, and from other Archive users who have downloaded these Torrents already." [102]

p2p systems complement centralized servers in a number of ways beyond raw download speed, increasing the efficiency and performance of the network as a whole. Spotify began as a joint client/server and p2p system [103], where when a listener presses play the central server provides the data until the p2p system locates peers with a cached copy to download from. The central server is able to respond quickly and reliably, and is the server of last resort in the case of rare files that aren't being shared by anyone else in the network. The p2p system alleviates pressure on the central server, improving the performance of the network and reducing server costs.

A peer to peer system is a particularly natural fit for many of the common circumstances and practices in science, where centralized server architectures seem (and prove) awkward and inefficient. Most labs, institutes, or other organized bodies of science have some form of local or institutional storage systems. In the most frequent cases of sharing data within a lab or institute, sending it back and forth to some nationally-centralized server is like walking across the lab by going the long way around the Earth. That's the method invoked by a Dropbox or AWS link, which keeps a time-tested p2p system relevant: walking a flash drive across the lab. The system makes less sense when several people in the same place need to access the same data at the same time, as is frequently the case with multi-lab collaborations, or scientific conferences and workshops. Instead of needing to wait on the 300kb/s conference wifi bandwidth as it's cheese-gratered across every machine, we instead could directly beam it between all computers in range simultaneously, full blast through the decrepit network switch that won't have seen that much excitement in years.

If we take the suggestion of Andrey Andreev et al. and invest in server clusters within institutes [104, 105], their impact could be multiplied manyfold by fluidly combining them in a p2p swarm. While the NIH might be shy to start up another server farm for all scientific data and prefer to contract with AWS, the rest of us don't have to be. Nervous university administrators concerned about bandwidth costs should also favor p2p systems: instead of needing to serve entire datasets to each person who wants them, the load can be spread out across many institutes naturally based on the use of the file, and sharing the dataset internally would cost nothing at all.

So far I have relied on the Extraordinarily Simplified Bittorrent[14] depiction of a peer to peer system, but there are many improvements and variants that can address different needs for scientific data infrastructure.

One obvious need that bittorrent can't currently support is version control[15], but more recent p2p systems do. IPFS functions like "a single BitTorrent swarm, exchanging objects within one Git repository." [107] [16] Dat [108], specifically designed for data synchronization and versioning, handles versioning and more. A full description of IPFS is out of scope, and it has plenty of problems [109], but for now it suffices to say p2p systems can handle version control.

[14] TM

[15] Though the Bittorrent V2 protocol specification [106] adopts a Merkle tree data structure which could theoretically support versioned torrents, v2 torrents are still not widely supported.

[16] Git, briefly, is a version control system that keeps a history of changes of files (blobs) as a Merkle DAG: files can be updated, and different versions can be branched and reconciled.

Bittorrent swarms are vulnerable to data loss if all the peers seeding a file disconnect (though the tail is longer than typically assumed, see [110]), but this too can be addressed with updated p2p system design. A first-order solution to this problem is a variant of IPFS' notion of 'pinning.' Since backup to lab-level or institutional servers is already commonplace, one peer could be able to 'pin' another and automatically download all the data that they share. This concept could scale to institutes and national infrastructure as scientists can request the datasets they'd like to be saved permanently be pinned.

Another could be something akin to Freenet [111]. Peers could allocate a certain amount of their unused storage space to be used to automatically download, cache, and rehost shards of other datasets. Distributing chunks and encrypting them at rest so the rehoster can't inspect their contents would make it possible to maintain privacy and network availability for sensitive data (see, for example, ERIS). IPFS has an analogous concept – BitSwap – that is makes it into a barter system. Peers who seek to download will have to 'earn' it by finding some chunk of data that the other peers want, download, and share them, though it seems like an empirical question whether or not a barter system works or is necessary.

Solid is a project that almost exactly meets all these needs [112, 67, 113]. Solid allows people to share data in Pods, which let them control access and distribution across storage system with a unified identity system. It is implementation-agnostic, and so can support any peer-to-peer storage and transfer system that complies with its protocol specification.

There are a number of additional requirements for a peer to peer scientific data infrastructure, but even these seemingly very technical problems of versioning and distributed storage show the clear need to consider the structure of the surrounding social system. What control do we give to researchers over the version history of their data? Should people that aren't the originating researcher be able to issue new versions? What structure of distributed/centralized storage works? How should we incentivize sharing of excess storage and resources?

Even before considering additional social systems, a p2p structure in itself implies a different relationship to infrastructure. Scientists always unavoidably make their data available to at least one person: themselves; on at least one computer: theirs, and that computer is usually connected to the internet. With a p2p system that integrates metadata from domain-specific data formats, that's it, that's all, the data is already hosted by merely existing. Dust your palms off: open data achieved. A peer-to-peer backbone for scientific infrastructure realizes the unnecessarily radical notion that our infrastructure can be integrated into our daily practices, rather than existing exogenously as something "out there." It helps us internalize the slyly subversive notion that *we can build it ourselves* instead of renting something out of our control from someone else.

Scientists don't need to reinvent the notion of distributed, community curated data archives from scratch. In addition to scholarly work on the social systems of digital infrastructure, we can learn from communities of practice, and there has been no more important and impactful decentralized archival project than internet piracy.

### 3.2.3    Archives Need Communities

Why do hundreds of thousands of people, completely anonymously, with zero compensation, spend their time to do something that is as legally risky as curating pirated cultural archives?

Scholarly work, particularly from Economics, tends to focus on understanding piracy in order to prevent it [114, 115], taking the moral good of intellectual property markets as an *a priori* imperative and investigating why people behave *badly* and "rend [the] moral fabric associated with the respect of intellectual property." [115]. If we put the legality of piracy aside, we may find a wealth of wisdom and insight to draw from for building scientific infrastructure.

The world of digital piracy is massive, from entirely disorganized efforts of individual people on public sites to extraordinarily organized release groups [114], and so a full consideration is out of scope (see [116]), but many of the important lessons are taught by the structure of bittorrent trackers.

An underappreciated element of the BitTorrent protocol is the effect of the separation between the data transfer protocol and the "discovery" part of the system — the "overlay" — on the community structure of torrent trackers (for a more complete picture of the ecosystem, see [110]). Many peer to peer networks like KaZaA or the gnutella-based Limewire had searching for files integrated into the transfer interface. The need for torrent files to share .torrent files spawned a massive community of private torrent trackers that for decades have been iterating on cultures of archival, experimenting with different community structures and incentives that encourage people to share and annotate some of the world's largest, most organized libraries.

One of these private trackers was the site of one of the largest informational tragedies of the past decade: what.cd[17], which I will use as an example to describe some of these community systems.

What.cd was a bittorrent tracker that was arguably the largest collection of music that has ever existed. At the time of its destruction in 2016, it was host to just over one million unique releases, and approximately 3.5 million torrents[18] [117]. Every torrent was organized in a meticulous system of metadata communally curated by its roughly 200,000 global users. The collection was built by people who cared deeply about music, rather than commercial collections provided by record labels notorious for ceasing distribution of recordings that are not commercially viable — or just losing them in a fire [118]. Users would spend large amounts of money to find and digitize extremely rare recordings, many of which were unavailable anywhere else and are now unavailable anywhere, period. One former user describes one example:

> "I did sound design for a show about Ceaușescu's Romania, and was able to pull together all of this 70s dissident prog-rock and stuff that has never been released on CD, let alone outside of Romania" [119]

What.cd was a "private" bittorrent tracker, where unlike public trackers that anyone can access, membership was strictly limited to those who were personally invited or to those who passed an interview (for more on public and private trackers, see [120]). Invites were extremely rare, and the interview process was demanding to the point where extensive guides were written to prepare for them.

The what.cd incentive system was based on a required ratio of data uploaded vs. data downloaded [121]. Peer to peer systems need to overcome a free-rider problem

[17] for a detailed description of the site and community, see Ian Dunham's dissertation [117]

[18] Though spotify now boasts its library having 50 million tracks, back of the envelope calculations relating number of releases to number of tracks are fraught, given the long tail of track numbers on albums like classical music anthologies with several hundred tracks on a single "release."

where users might download a torrent ("leeching") and turn their computer off, rather than leaving their connection open to share it to others (or, "seeding"). In order to download additional music, then, one would have to upload more. Since downloading is highly restricted, and everyone is trying to upload as much as they can, torrents had a large number of "seeders," and even rare recordings would be sustained for years, a pattern common to private trackers [122].

The high seeder/leecher ratio made it so it was extremely difficult to acquire upload credit, so users were additionally incentivized to find and upload new recordings to the system. What.cd implemented a "bounty" system, where users with a large amount of excess upload credit would be able to offer some of it to whoever was able to upload the album they wanted. To "prime the pump" and keep the economy moving, highlight artists in an album of the week, or direct users to preserve rare recordings, moderators would also use a "freeleech" system, where users would be able to download a specified set of torrents without it counting against their download quantity [123, 124].

The other half of what.cd was the more explicitly social elements: its forums, comment sections, and moderation systems. The forum was home to roiling debates that lasted years about the structure of some tagging schema, whether one genre was just another with a different name, and so on. The structure of the community was an object of constant, public negotiation, and over time the metadata system evolved to be able to support a library of the entirety of human musical culture[19]. To support the good operation of the site, the forums were also home to a huge amount of technical knowledge, like guides on how to make a perfect copy of a CD or how to detect a fake upload, that eased new users into being able to use and contribute to the system.

A critical problem in maintaining coherent databases is correcting metadata errors and departures from schemas. Finding errors was rewarded. Users were able to discuss and ask questions of the uploader in a comment section below each upload, which would allow "polite" resolution of low-level errors like typos. More serious problems could be reported to the moderation team, which caused the upload to be visibly marked as under review, and the report could then be discussed either in the comment sections or the forum. The system wasn't perfect: being an anonymous, gray-area community, there was of course plenty of power to be abused. Rather than being a messy hodgepodge of fake, low-quality uploads, though, what.cd was always teetering just shy of perfection.

[19] Though music metadata might seem like a trivial problem (just look at the fields in an MP3 header), the number of edge cases are profound. How would you categorize an early Madlib cassette mixtape remastered and uploaded to his website where he is mumbling to himself while recording some live show performed by multiple artists, but on the b-side is one of his Beat Konducta collections that mix together studio recordings from a collection of other artists? Who is the artist? How would you even identify the unnamed artists in the live show? Is that a compilation or a bootleg? Is it a cassette rip, a remaster, or a web release?

These structural considerations do not capture the most elusive but indisputably important feature of what.cd's community infrastructure: *the sense of community*. The What.cd forums were the center of many user's relationships to music. Threads about all the finest scales of music nichery could last for years: it was a rare place people who probably cared a little bit too much about music could talk to people with the same condition. What made it more satisfying than other music forums was that no matter what music you were talking about, everyone else in the conversation would always have access to it if they wanted to hear it. Beyond any structural incentives, people spent so much time building and maintaining what.cd because it became a source of community and a sink of personal investment.

Structural norms supported by social systems converge as a sort of *reputational* incentive. Uploading a new album to fill a bounty both makes the network more functional and complete, but also *people respect you for it* because it's prominently displayed on your profile as well as in the bounty charts and that *feels good*. Becoming known on the forums for answering questions, writing guides, or even just having a good taste in music *feels good* and also contributes to the overall health of the system. Though there are plenty of databases, and even plenty of different communication venues for scientists, there aren't any databases (to my knowledge) with integrated community systems.

The tracker overlay model mirrors and extends some of the recommendations made by Benedikt Fecher and colleagues in their work on the reputational economy surrounding data sharing [125]. They give three policy recommendations: Increasing reputational benefits, reducing transaction costs, and "increasing market transparency by making open access to research data more visible to members of the research community." One way to accomplish implement them is to embed a data sharing system within a social system that is designed to reward communitarian behavior.

Many features of what.cd's structure are undesirable for scientific infrastructure, but they demonstrate that a robust archive is not only a matter of building a database with some frontend, but also building a community [126]. Of course, we need to be careful with building the structural incentives for a data sharing system: the very last thing we want is another coercive leaderboard that turns what should be a collaborative effort punitive. In contrast to what.cd, for infrastructure we want extremely low barriers to entry, and be agnostic to resources — researchers with access to huge server farms should not be unduly favored. We should think carefully about using downloading as the "cost," because downloading and analyzing huge amounts of data can be *good* and exactly what we *want* in some circumstances, but a threat to privacy and data governance in others.

This model has its own problems, including the lack of interoperability between different trackers, the need to recreate a new set of accounts and database for each new tracker, among others. It's also been tried before: sharing data in specific formats (as our running example, Neurodata Without Borders) on indexing systems like bit-torrent trackers amounts to something like BioTorrents [86] or AcademicTorrents [84]. Even with our extensions of version control and some model of automatic mir-roring of data across the network, we still have some work to do. To address these and several other remaining needs for scientific data infrastructure, we can take in-spiration from *federated systems*.

### 3.2.4   Linked Data or Surveillance Capitalism?

> Having become a dense and consistent historical reality, language forms the locus of tradition, of the unspoken habits of thought, of what lies hidden in a people's mind; it accumulates an ineluctable memory which does not even know itself as memory. Expressing their thoughts in words of which they are not the masters, enclosing them in verbal forms whose historical dimensions they are unaware of, men believe that their speech is their servant and do not realize that they are sub-mitting themselves to its demands.

Michel Foucault — *The Order of Things* [127]

There is no shortage of databases for scientific data, but their traditional structure chokes on the complexity of representing multi-domain data. Typical relational databases require some formal schema to structure the data they contain, which have varying reflections in the APIs used to access them and interfaces built atop them. This broadly polarizes database design into domain-specific and domain-general[20]. This design pattern results in a fragmented landscape of databases with limited interoperability. How shall we link the databases? In this section we'll con-sider the Icarian promise of creating the great unified database of everything as a way of motivating an alternative that blends *linked data* [128] with *federated systems* against our peer to peer backbone in the next section.

Domain-specific databases require data to be in one or a few specific formats, and usually provide richer tools for manipulating and querying by metadata, visualiza-tion, summarization, aggregation that are purpose-built for that type of data. For example, NIH's Gene tool has several visualization tools and cross-referencing tools for finding expression pathways, genetic interactions, and related sequences (Figure xx). This pattern of database design is reflected at several different scales, through in-stitutional databases and tools like the Allen brain atlases or observatory, to lab- and project-specific dashboards. This type of database is natural, expressive, and power-ful — for the researchers they are designed for. While some of these databases allow open data submission, they often require explicit moderation and approval to main-tain the guaranteed consistency of the database, which can hamper mass use.

General-purpose databases like figshare and zenodo[21] are useful for the mass aggre-gation of data, typically allowing uploads from most people with minimal barriers. Their general function limits the metadata, visualization, and other tools that are offered by domain-specific databases, however, and are essentially public, versioned, folders with a DOI. Most have fields for authorship, research groups, related publi-cations, and a single-dimension keyword or tags system, and so don't programmati-cally reflect the metadata present in a given dataset.

The dichotomy of fragmented, subdomain-specific databases and general-purpose databases makes combining information from across even extremely similar subdis-ciplines combinatorically complex and laborious. In the absence of a formal interoperability and indexing protocol between databases, even *finding* the correct subdomain-specific database often comes down to pure luck. It also puts researchers who want to be good data stewards in a difficult position: they can hunt down the appropriate subdomain specific database and risk general obscurity; use a domain-general database and make their work more difficult for themselves and their peers to use; or spend all the time it takes to upload to multiple databases with potentially conflicting demands on format.

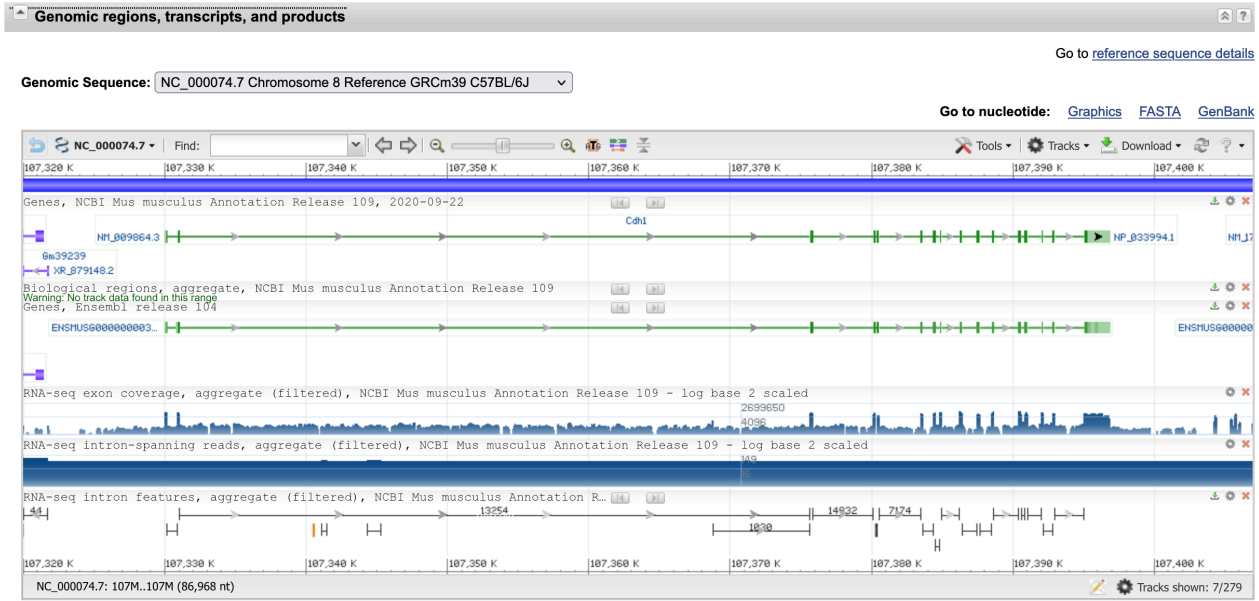

**Figure 3.2:** NIH's Gene tool includes many specific tools for visualizing, cross-referencing, and aggregating genetic data. Shown is the "genomic regions, transcripts, and product" plot for Mouse Cdh1, which gives useful, common summary descriptions of the gene, but is not useful for, say, visualizing reading proficiency data from educational research.

What can be done? There are a few naïve answers from standardizing different parts of the process: If we had a universal data format, then interoperability becomes trivial. Conversely, we could make a single ur-database that supports all possible formats and tools.

The notion of a universal database system almost immediately runs aground on the reality that organizing knowledge is intrinsically political. Every subdiscipline has conflicting *representational* needs, will develop different local terminology, allocate differing granularity and develop different groupings and hierarchies for the same phenomena. At their mildest, differences in representational systems can be incompatible, but at their worst they can reflect and reinforce prejudices and become the site of expression for intellectual and social power struggles [129, 130, 131, 132]. Every subdiscipline has conflicting *practical* needs, with infinite variation in privacy demands, different priorities between storage space, bandwidth, and computational power, and so on. In all cases the boundaries of our myopia are impossible to gauge: we might think we have arrived at a suitable schema for biology, chemistry, and physics... but what about the historians?

Matthew J Bietz and Charlotte P Lee articulate this tension in their ethnography of metagenomics databases:

"Participants describe the individual sequence database systems as if they were shadows, poor representations of a widely-agreed-upon ideal. We find, however, that by looking across the landscape of databases, a different picture emerges. Instead, **each decision about the implementation of a particular database system plants a stake for a community boundary. The databases are not so much imperfect copies of an ideal as they are arguments about what the ideal Database should be.** [...]

In the end, however, **the system was so tailored to a specific set of research questions that the collection of data, the set of tools, and even the social organization of the project had to be significantly changed.** New analysis tools were developed and old tools were discarded. Not only was the database ported to a different technology, the data itself was significantly restructured to fit the new tools and approaches. While the database development projects had begun by working together, in the end they were unable to collaborate. **The system that was supposed to tie these groups together could not be shielded from the controversies that formed the boundaries between the communities of practice.**" [133]

The pursuit of unified representation is an intimate part of the history of linked data, which relies on "ontologies" or controlled vocabularies that describe a set of objects (or classes) and the properties they can have. For example, schema.org maintains a widely used set of hierarchical vocabularies to describe the fundamental things that exist in the unfamiliar world in which a Person has a gender and net worth but lacks a race [39]. At one extreme in the world of ontology builders, the ideological nature of demarcating what is allowed to exist is as clear as a klaxon (emphasis in original):

An exception is the Open Biomedical Ontologies (OBO) Foundry initiative, which accepts under its label only those ontologies that adhere to the principles of ontological realism. [...] Ontologies, from this perspective, are representational artifacts, comprising a taxonomy as their central backbone, whose representational units are intended to designate *universals* (such as *human being* and *patient role*) or *classes defined in terms of universals* (such as *patient,* a class encompassing *human beings* in which there inheres a *patient role*) and certain relations between them. [...]

BFO is a realist ontology [15,16]. This means, most importantly, that representations faithful to BFO can acknowledge only those entities which exist in (for example, biological) reality; thus they must reject all those types of putative negative entities - lacks, absences, non-existents, possibilia, and the like [134]

In practice, because of the difficulty of changing the representation and encompassing database systems on a dime, using these ontologies to link disparate datasets tends to follow the pattern of metadata *overlays* where the structure of individual databases are mapped onto one "unifying" ontology to allow for aggregation and translation. This approach appears gentler than standardization at the level of individual databases, but has the same problems kicked up one level of abstraction.

To concretize the problems with a globally unified database or metadata overlay, the remainder of this section will trace the compromises and outcomes of the The NIH's "Biomedical Data Translator" project[22]. The Translator project was initially described in the 2016 Strategic Plan for Data Science as a means of translating between biomedical data formats:

> Through its Biomedical Data Translator program, the National Center for Advancing Translational Sciences (NCATS) is supporting research to develop ways to connect conventionally separated data types to one another to make them more useful for researchers and the public. [35]

[22] A conversation with someone close to the project just before the initial publication of this piece made clear several changes that need to be made to this section, but prevailing circumstances in academic publishing hastened its release before I could make them. I intend this section as good-faith criticism of a lot of hard work by researchers who I assume are working with the best intentions, and invite comment and clarification from anyone working within the Translator project for revisions in a future version.

The original funding statement from 2016 is similarly humble, and press releases through 2017 also speak mostly in terms of querying the data – though some ambition begins to creep in. By 2019, the vision for the project had veered sharply away from anything a basic researcher might recognize as a means of translating between data types. In their piece "Toward a Universal Biomedical Translator," then in a feasibility assessment phase, the members of the Translator Consortium assert that universal translation between biomedical data is impossible [135]. The impossibility they saw was not that of conflicting political demands on the structure of organization (as per [132]), but of the sheer numeracy of the data and vocabularies needed to describe them. The risk posed by a lack of a universal "language" was not being able to index all possible data, rather than inaccuracy or inequity.

Undaunted by their stated belief in the impossibility of a universalizing ontology, the Consortium created one in their biolink model [136, 137]. Biolink consists of a hierarchy of basic classes: eg. a BiologicalEntity like a Gene, or a ChemicalEntity like a Drug. Classes can then linked by any number of properties, or "Slots," like a therapeutic procedure that treats a disease.

The translator does not attempt to respond to the needs of researchers or labs who might want to link their raw data splayed out across flash drives and file structures whose chaos borders on whimsy. Instead, the Translator operates at the level of "knowledge," or "generally accepted, universal assertions derived from the accumulation of information" [138]. Rather than translating *between data types*, the meaning of "translation" shifted to meaning *"translating data into knowledge"* [135].

To feed the Translator, Biolink sits "on top of" a collection of database APIs that serve structured biomedical data, each called a "knowledge source." Individual APIs declare that they are able to provide data for a particular set of classes or slots, like drugs that affect genetic expression, and are then made browsable from the SmartAPI Knowledge Graph. Queries to individual APIs do not return "raw" data, but return assertions of fact in the parlance of the Biolink model: this procedure treats that disease, etc.

Because individual researchers do not typically represent their data in the form of factual assertions, knowledge sources are constrained to "highly curated biomedical databases" or other aggregated systems. The NIH RePORTER tool gives an overview of the way these knowledge sources are prepared when none already exist for a given Biolink class or predicate: automated text mining tools and a series of domain-specific data provider projects, rather than via tools provided to researchers.

The collection of knowledge sources, linked to nodes and edges in the Biolink model, are designed to be queried as a graph. To answer a query like "what drug treats this disease?" the translator considers the graph of entities linked to the disease: what symptoms does the disease have? what genes are linked to those symptoms? which drugs act on those genes? and so on [139]. The form of the Translator as a graph-based question answering machine bounds its application as a platform for re-searchers to guide their research and clinicians to guide their care [140], rather than a tool for linking data.

One primary example currently featured by NCATS is using the translator to propose novel treatments for drug-induced liver injury (DILI) [141] detailed in a 2021 conference paper [142]. To find a candidate drug, the researchers manually conducted three API queries: first they searched for phenotypes associated with DILI and selected "one of them"[23] — "red blood cell count". Then they queried for genes associated with red blood cell count to find telomerase reverse transcriptase (TERT), and then finally for drugs that affect TERT to find Zidovudine. The directionality of each of these relationships, high vs. low, increases vs. decreases, is unclear in each case. A more recent report on the Translator repeated this pattern of manual querying, arriving at a handful of different genes and drugs [138].

While the current examples are highly manual, providing an array of results for each query along with links to associated papers on pubmed, some algorithmic system for ranking results is necessary to make use of the information in the extended knowledge graph. Rather than just the first-order connections, it should be possible to make use of second, third, and n-th order connections to weight potential results. Algorithmic medical recommendation systems have been thoroughly problematized elsewhere (eg. [143, 144, 145, 146]). The primary ranking algorithm is developed by a defense contractor (CoVar) who has[24] named it ROBOKOP [147] [25]. Though ROBOKOP functions with a simple weighted graph metric based on citations and abstract text, the ranking system is intended to be extended with machine learning tools [147] that can be trained based on the way the provided answers are used [135]. Algorithmic recommendation platforms are in a regulatory gray area [148, 149], but would arguably need to have interpretable results with clear provenance to pass scrutiny. The DILI example uses a language model which explained the recommendation of Zidovudine with all the clarity of "one of 'DOWNREGU-LATOR,' 'INHIBITOR,' 'INDIRECT DOWNREGULATOR'."

The arrival at a biomedical question answering platform built atop an algorithmic ranking system for a knowledge graph that queries 200+ aggregated data sources has several qualities that should give us pause.

First, as with any machine-learning based system, the algorithm can only reflect the implicit structure of its creation, including the beliefs and values of its architects [150, 151], its training data and accompanying bias [152], and so on. The "mass of data" approach ML tools lend themselves to, in this case, querying hundreds of independently operated databases, makes dissecting the provenance of every entry from every data provider effectively impossible. For example, one of the providers, mydisease.info was more than happy to respond to a query for the outmoded definition of "transsexualism" as a disease [153] along with a list of genes and variants that supposedly "cause" it - see for yourself. At the time of the search, tracing the source of that entry first led to the disease ontology DOID:1234 which traced back into an entry in a graph aggregator Ontobee (Archive Link), which in turn listed the github repository **maintained by a single person** as its source[26]. This is, presumably, the fragility and inconsistency in input data that the machine learning layer is intended to putty over.

If the graph encodes being transgender as a disease, it is not farfetched to imagine the ranking system attempting to "cure" it. In a seemingly prerelease version of the translator's query engine, ARAX, it does just that: in a query for entities with a `biolink:treats` link to gender dysphoria[27], it ranks the standard therapeutics [154, 155] Testosterone and Estradiol 6th and 10th of 11, respectively — behind a recommendation for Lithium (4th) and Pimozide (5th) due to an automated text scrape of two conversion therapy papers.[28]. Queries to ARAX for treatments for gender identity disorder helpfully yielded "zinc" and "water," offering a paper from the translator group that describes automated drug recommendation as the only provenance [156]. A query for treatments for `DOID:1233` "transvestism" was predictably troubling.

Even if the curators do their best to prevent harmful queries and block searches for "cures" to being trans, the graph-based nature of the system means that any given entry will have unpredictable consequences on recommendations made from the surrounding network of objects like genes, treatment history, and so on. If the operation of the ranking algorithm is uninterpretable, as most are, or the algorithm it itself proprietary, harmful input data could have long-range influence on both the practice of medicine as well as the course of basic research *without anyone being able to tell.* The Consortium also describes a system whereby the algorithm is continuously updated based on usage of results in research or clinical practice [135], which stands to magnify the problem of algorithmic bias by uncritically treating harmful treatment and research practices as training data.

The approach creates a fundamental tradeoff between algorithmic interpretability and the system being useful at all. The paper cited in the 2021 DILI example as evidence that the system gives plausible results is for a specific subclass of liver injuries caused by anti-tuberculosis drugs [157], highlighting the danger of automated recommendations from noisy data, but also calling into question what novel contribution the Translator made if telomeres were already implicated in DILI. The 2022 report gives examples where the results were already expected by the researchers, or provided a series of papers that seems difficult to imagine being much more informative than a PubMed search. If the algorithmic recommendations are unexpected — ie. the system provides novel information — the process of confirming them appears to be near-identical to the usual process of reading abstracts and hopping citation trees.

Perhaps most worrisome is the eventual fate of the project in the hands of the broader ecosystem of orbiting information conglomerates. Centralized infrastructure projects can be an opportunity for for-profit companies to "dance until the music stops" and then scoop up any remaining technology when the funding dries up (so far roughly $81.6 million since 2016 for the Translator [158], and $84.7 million for the discontinued NIH Data Commons pilot which morphed into the STRIDES program). I have little doubt that the scientists and engineers working on the Translator are doing so with the best of intentions — the real question is what happens to it after it's finished.

Knowledge graphs in particular are promising targets for platform holders. Perhaps the most well known example is Google's 2010 acquisition of Freebase (via Metaweb) [159], a graph of structured data with a wealth of properties for common people, places and things. Google incorporated it into their Knowledge Graph [160] to populate its factboxes and make its search results more semantically aware in its Hummingbird upgrade in 2013, the largest overhaul of its search engine since 2001 [161], cementing its dominance as a search engine. The connection between swallowing up knowledge organization systems into search engines is not incidental, but reflective of the broader pattern of enclosing basic digital infrastructure be-

[27] To its credit, ARAX does transform the request for `DOID:10919` to `MONDO:0001153` - gender dysphoria.

[28] as well as a recommendation for "date allergenic extract" from a misinterpretation of "to date" in the abstract of a paper that reads "Cross-sex hormonal treatment (CHT) used for gender dysphoria (GD) could by itself affect well-being without the use of genital surgery; however, **to date,** there is a paucity of studies investigating the effects of CHT alone"

hind opaque platforms. Searching has a different set of cognitive expectations than browsing a database: we expect search results to be "best effort," not necessarily complete or accurate, where when browsing a database it's relatively clear when information is missing or inaccurate. For products packaged up into search platforms by for-profit companies, *it doesn't have to actually work* as long as it seems like it does.

The platformatization of the knowledge graph, along with carefully worded terms of service, is a clean means by which "good enough" results could be jackknifed into an expanded system of biomedical surveillance. Since the algorithm needs continual training, the translator has every incentive to suck up as much personal data as it can[29]. For-profit platform providers as a rule depend on developing elaborate personal profiles for targeted advertising algorithmically inferred from available data[30], that naturally includes diagnosed or inferred disease — a practice they explicitly describe in the patents for the targeting technology [163], have gone to court to defend [164, 165], and formed secretive joint projects with healthcare systems to pursue [166].

So while an algorithmic recommendation tool may have limited use for the basic researchers it was originally intended for, it is likely to be extremely useful for the booming business of "personalized medicine.[31]" Linking biomedical and patient data in a single platform is a natural route towards a multisided market where records management apps are sold to patients, treatment recommendation systems are sold to clinicians, research tools and advertising opportunities are sold to pharmaceutical companies, risk metrics are sold to insurance companies, and so on.

Multiple information conglomerates are poised to capitalize on the translator project. Amazon already has a broad home surveillance portfolio [167], and has been aggressively expanding into health technology [168] and even literally providing health care [169], which could be particularly dangerous with the uploading of all scientific and medical data onto AWS with entirely unenforceable promises of data privacy through NIH's STRIDES program [170].

RELX, parent of Elsevier, is as always the terrifying elephant in the room. In addition to distribution rights for a large proportion of scientific knowledge and a collection of research databases, it also sells a clinical reference platform in ClinicalKey, point of service products for planning patient care with ClinicalPath, medical education tools, and pharmaceutical advertisements designed to look like scientific papers [171], among others [172]. It also is explicitly expanding into "clinical decision support applications" [172] and recently embedded its medication management product into Apple's watchOS 9 [14]. Subsidiaries in RELX's "Risk" market segment sell risk profiles to insurance companies based on what they claim to be highly comprehensive profiles of harvested personal data. The Translator infrastructure is a perfect keystone to unify these products: after the NIH fronts the money to develop it and lends the credibility of basic research, RELX can cheaply expand its surveillance apparatus to enhanced medical risk profiles to insurers, priority placement in candidate drug rankings to pharmaceutical companies, and augment its ranking systems for funders and employers to include some proprietary metric of "promisingness" to encourage researchers to follow its research recommendations. This isn't speculative — it can just strap whatever clinical data Translator gains access to into its existing biomedical knowledge graph.

Even assuming the Translator works perfectly and has zero unanticipated conse-quences, the development strategy still reflects the inequities that pervade science rather than challenge them. Biopharmaceutical research, followed by broader biomedical research, being immediately and extremely profitable, attracts an enormous quantity of resources and develops state of the art infrastructure, while no similar infrastructure is built for the rest of science, academia, and society.

The eventual form of the Translator follows from a series of decisions centered around the intended universality of the system. From the funding statement in 2016, the system was conceptualized as an "informatics platform" intended to "bring together all biomedical and health data types." The surrounding background of cloud-based database storage imagined by the Strategic Plan for Data Science immediately constrained the design to consist of APIs that served small quantities of aggregated data, rather than potentially large quantities of raw data. Together with a platform, rather than tool-based approach, a system that allowed individual researchers to link and make sense of the subtlety of own their data was precluded from the start.

From these constraints, the form of the BioLink model comes into focus: high-level classes and logical relationships between them as asserted by a large number of separate knowledge sources. Since the data from each of these sources is heterogeneous, relatively uncurated, and potentially numerous for any given graph-based query, the need for a machine learning layer to make sense of it follows. The conceptualization of BioLink as a universal ontology seems to follow the lineage of the "neat" thought style [39] that emphasizes "deductive inference through logical rules" [137] or otherwise computing derived information from the structure of the knowledge graph rather than browsing the graph itself. Together, these constraints and design logics bring us to the form of the Translator as a graph-based query engine.

The Translator Consortium justifiably takes pride in its social organizing systems [173] — coordinating 200 researchers and engineers from dozens of institutions is no small feat. This system of social organization seems to have lent itself towards developing the individual components with an eye for them to be understood by the rest of the *consortium* rather than with the intention of inviting collaboration from the broader research community[32]. The very notion of a platform indicates that it is something that *they build* and *we use*: There is no explicit means for proposing changes to the BioLink model, to pick and choose how answers are ranked or queries are performed, etc. This is broadly true of platform-based scientific tools, especially databases, and contributes to how they *feel*: they feel disconnected with our work, don't necessarily help us do it more easily or more effectively, and contributing to them is a burdensome act of charity (if it is possible at all).

Given the real need for *some* means of combining heterogeneous data from disparate sources, what could have been done differently?

Problematizing the need for a system intended to link *all* or even *most* biomedical data in a single mutually coherent system opens the possibility for a very different data linking infrastructure. Perhaps paradoxically, any universal, logically complete schema intended to support algorithmic inference projects a relatively circumscribed group of people for whom it would be useful: nearly all of the publicly described use-cases are oriented around finding new drugs or targets to treat disease, presumably in part because that's what preoccupies the ontology. Rather than a set of generalizable *tools* for linking data, the need for universality strongly constrains the form of data that can be represented by the system, and its platform structure constrains its uses to only those imagined by the platform designers. Every infras-tructural model is an act of balancing constraints, and prioritizing "all data" seems to imply "for some people." Who is supposed to be able to upload data? change the ontology? inspect the machine learning model? Who is in charge of what? Who is a knowledge-graph query engine useful for?

[32] The descriptions of difficulty in interfacing the components of the project internally are littered throughout their public-facing documents: eg. "A lot of the work has been about defining standards, so that the components that each of the 15 teams are building can talk to each other" [139], "In part due to the speed with which the program has progressed, team members also have found it challenging to coordinate milestones and deliverables across teams and align the goals of the Translator program with the goals of their own nonTranslator research projects." [173]. These problems are, of course, completely reasonable. My comment here merely suggests that solving these problems, particularly on the self-described tight timeline of the Translator's development, may have edged out concerns for engagement with the broader research community.

Another conceptualization might be building systems for *all people* that can *embed with existing practices* and *help them do their work* which typically involves accessing *some data.* We can imagine a system designed to integrate data with schemas written in the *vernacular* of communities of knowledge work. Rather than the dichotomy of one singular database vs. many fragmented and incompatible databases, we can imagine a *pluralistic* system capable of supporting multiple overlapping and potentially conflicting representations, governable and malleable in local communities of practice. Taking seriously the notion of "translation," we could stand to learn from linguistics and translation studies: rather than attempting to project the dialects of each subdiscipline into some "true" meta-framework (a decidedly colonial project [174]), we could resist the urge for homogenization and preserve the multiplicity of representation, embracing the imperfection of mappings between heterogeneous representational systems at multiple scales without resigning ourselves to completely isolated incompatibility.

Maybe we don't *want* a universal system that presents itself with the authority of truth to be mined and spun off into derivative platforms by information conglomerates. We might abandon the techno-utopianism of a globally consistent schema that supports arbitrary logical inference by acknowledging that those inferences would always be colored by the decisions embedded in the structure of the system, unknowable beneath the shrouding weights of its ranking model.

Instead can we imagine a properly *human* data infrastructure? One that preserves the seams and imperfections in our representational systems, that is designed to represent precisely the contingency of representation itself? (eg. see [175, 176]). We might start with the propositional nature of links and mappings between formats — that rather than a divine received truth, the relationships between things are contextual and created. We could find grounding in *use,* that the schemas and mappings between them should arise from the need to link representations within the context of some problem, rather than to resolve their difference.

Picking up the thread of our peer to peer data sharing backbone, we might start to imagine the boistrous multiplicity of an infrastructure based around communication and expression, rather than platformatized perfection.

### 3.2.5   Folk Federation

Human language thrives when using the same term to mean somewhat different things, but automation does not. *Tim Berners-Lee (1999) The Semantic Web* [177]

Wittgenstein's contribution to communism was his robust proof of the proposition that there is no private language, but in our time, privatized languages are everywhere. And not just languages: Images, codes, algorithms, even genes can become private property, and in turn private property shapes what we imagine the limits and possibilities of this information to be. *McKenzie Wark (2021) Capital Is Dead: Is This Something Worse?* [7]

To structure our p2p data sharing system, we should use *Linked Data*. Linked data is at once exceptionally simple and deceptively complex, a set of technologies and social histories. In this section we will introduce the notion of linked data, extend it for a p2p context, and then add a twist from *federated systems*.[33] Our goal will be to articulate the foundation for a "protocol of protocols," a set of minimal operations by which individual people can create, extend, borrow, and collectively build a space of linked folk schemas and ontologies, or *folksonomies*.

When last we left it, we had developed the notion of a p2p system to the point where we had big torrentlike piles of files with a few additional features like versioning and sharded storage. We need to add an additional layer of *metadata* that exposes information about the contents of each of these file piles. But what is that metadata *made of*?

The core format of linked data is the Resource Document Format (RDF) [179] and its related syntaxes like Turtle [180]. Typical hyperlinks are *duplet* links — linking from the source to the target. The links of linked data are instead **triplet** links that consist of a **subject**, a **predicate** that *describes* the link, and an **object** that is linked to. Subjects and objects (generally, nodes) have particular types like a number, or a date, or something more elaborate like an Airline or Movie that have particular sets of predicates or properties: eg. a `Movie` has a `director` property which links to a `Person`. A `Person` has an `address` which links to a `PostalAddress`, and so on. Types and properties are themselves defined in **vocabularies** (or, seemingly interchangeably [181], ontologies and schemas) by a special subset of RDF schema modeling classes and properties [182]. Linked data thus consists of semantically annotated **graphs** of linked nodes[34].

Linked data representations are very general and encompass many others like relational [183] and object-oriented models, but have a few properties that might be less familiar. The first is that triplet links have the status of an utterance or a proposition: much like typical duplet hyperlinks, anyone can make whatever links they want to a particular object to say what they'd like about it. As opposed to object-oriented models where a class is defined beforehand and its attributes or data are stored "within" the object, RDF schemas are composed of links just like any other, and the link, object, and predicate can all be stored in separate places by different people [184]. For example:

> One person may define a `vehicle` as having a `number of wheels` and a `weight` and a `length`, but not foresee a `color`. This will not stop another person making the assertion that a given car is red, using the color vocabulary from elsewhere. [184]

Linked data has an ambivalent history of thought regarding the location and distribution of ontology building. Its initial formulation came fresh from the recent incendiary success of the internet, where without any system of organization "people were frightened of getting lost in it. You could follow links forever." [184] Linked data was conceptualized to be explicitly without authoritative ontologies, but intended to evolve like language with local cultures of meaning meshing and separating at multiple scales [177]. Perhaps one of the pieces that went missing when moving between writing about the semantic web and its realization in standards and protocols is that this language-like conception of links requires **quartet,** rather than triplet links: **author**, subject, object, predicate. The author is encoded implicitly in the source of the vocabulary: "Users are given [...] a single URI [...] for each persona they want to have," [22] so theoretically ontologies have the status of "schema.org says this." Without a first-class notion of author in the links themselves there is little means of "forking" a vocabulary, or having multiple versions of a term with the same name but different authors.

The dream of mass automaticity, however, with computational "agents" capable of seamlessly crawling consistent graphs of linked data to extract surplus meaning necessarily requires that the meaning of terms does not "mutate" between different uses. For many early linked data architects the resolution was more automation, to use additional semantic structure about the equivalence between different ontologies as a means of estimating how trustworthy a particular result was. This tension is sewn into one of its most well known ontologies, the Simple Knowledge Organization System (skos) [185], which is intended to represent relationships between terms and vocabularies [186].

The fluidity of the original vision for linked data never emerged, however, and is remembered instead as being monstrously overcomplicated [38, 187]. While HTML, CSS, and Javascript developed a rich ecosystem of abstractions that let people create websites without directly writing HTML, the same never materialized for RDF. While linked data entities are intended to be designated by the very general notion of a URI, in practice URIs are near-synonymous with URLs, and maintaining a set of URLs is hard. The initial vision for URI/URL-based linked data identifiers seems to have been, in part, based on a miscalculation of the centralizing effect of the DNS system, which makes them expensive and rarer than they need to be for each person to have their own[35]. In the absence of interfaces for manipulating linked data and the pain of hosting them, the dream of a distributed negotiation over language-like ontologies was largely confined to information scientists and what became corporate knowledge graphs. For those war-weary RDF vets, I will again clarify that we are describing the desirable *qualities* of RDF while trying to learn from its failures.

In our revival of this dream we are describing a system where heterogeneous data is indicated by its metadata, rather than representing all data in a uniform format — similarly to the mixture of RDF and non-RDF data in the linked data platform standard [189]. We want to handle a broad span of heterogeneity: data with different naming schemes, binary representations, sizes, nested structures, and so on. The first task is to describe some means of accessing this heterogeneous data in a reasonably standard way despite these differences.

While that may seem a tall order, researchers already do it, it's just mostly done manually whenever we want to use anyone else's data. One way of characterizing the task at hand is systematizing the idiosyncratic paths by which a researcher might dump out a .csv file from a sql database to load into MATLAB to save in the .mat format with the rest of their data. To do that we can draw from a parallel body of thought on *federated databases*.

Like our p2p system, federated systems consist of *distributed*, *heterogeneous*, and *autonomous* agents that implement some minimal agreed-upon standards for mutual communication and (co-)operation. Federated databases were proposed in the early 1980's [190] and have been developed and refined in the decades since as an alternative to either centralization or non-integration [191, 192, 193]. Their application to the dispersion of scientific data in local filesystems is not new [194, 195, 196], but their implementation is more challenging than imposing order with a centralized database or punting the question into the unknowable maw of machine learning.

[35] Tim Berners-Lee, in his 1999 "Weaving the Web," was already describing how the centralization of the DNS system compromised some of his loftier ambitions for the web: "It is essential that domain names be primarily owned by the people as a whole, and that they be governed in a fair and reasonable way by the people, for the people." and "[DNS Centralization] also shows how a technical decision to make a single point of reliance can be exploited politically for power and commercially for profit, breaking the technology's independence from these things, and weakening the Web as a universal space." [188] The Solid project instead uses OpenID for identification, rather than, eg. a FOAF record located at a URL.

Amit Sheth and James Larson, in their reference description of federated database systems, describe **design autonomy** as one critical dimension that characterizes them:

> Design autonomy refers to the ability of a component DBS to choose its own design with respect to any matter, including
>
> - (a) The **data** being managed (i.e., the Universe of Discourse),
>
> - (b) The **representation** (data model, query language) and the **naming** of the data elements,
>
> - (c) The conceptualization or **semantic interpretation** of the data (which greatly contributes to the problem of semantic heterogeneity),
>
> - (d) **Constraints** (e.g., semantic integrity constraints and the serializability criteria) used to manage the data,
>
> - (e) The **functionality** of the system (i.e., the operations supported by system),
>
> - (f) The **association and sharing with other systems**, and
>
> - (g) The **implementation** (e.g., record and file structures, concurrency control algorithms).

Susanne Busse and colleagues add an additional dimension of **evolvability,** or the ability of a particular system to adapt to inevitable changing uses and requirements [194].

In order to support such radical autonomy and evolvability, federated systems need some means of translating queries and representations between heterogeneous components. The typical conceptualization of federated databases have five layers that implement different parts of this reconciliation process [197] :

- A **local schema** is the representation of the data on local servers, including the means by which they are implemented in binary on the disk

- A **component schema** serves to translate the local schema to a format that is compatible with the larger, federated schema

- An **export schema** defines permissions, and what parts of the local database are made available to the federation of other servers

- The **federated schema** is the collection of export schemas, allowing a query to be broken apart and addressed to different export schemas. There can be multiple federated schemas to accomodate different combinations of export schemas.

- An **external schema** can further be used to make the federated schema better available to external users, but in this case since there is no notion of "external" it is less relevant.

This conceptualization provides a good starting framework and isolation of the different components of a database system, but a peer-to-peer database system has dif-ferent constraints and opportunities [198]. In the strictest, "tightly coupled" fed-
erated systems, all heterogeneity in individual components has to be mapped to a
single, unified federation-level schema. Loose federations don't assume a unified
schema, but settle for a uniform query language, and allow multiple translations
and views on data to coexist. A p2p system naturally lends itself to a looser fed-
eration, and also gives us some additional opportunities to give peers agency over
schemas while also preserving some coherence across the system. I will likely make
some database engineers cringe, but the emphasis for us will be more on building a
system to support distributed social control over the database, rather than guaran-
teeing consistency and transparency between the different components.

Let us take the notion of a loosely coupled systems to its extreme, and invert the
meaning of federation as it is used in other systems like ActivityPub: rather than a
server-first federation, where peers create accounts on servers that define their op-
eration and the other servers they federate with, ours will be peer-first federation.
In this system, individual peers will maintain their own vocabularies and be able
to make them available to other peers. Peers can directly connect to one another,
but can also federate into groups, which can federate into groups of groups, and so
on. A peer will implement the local, component, and export schema with a client
that handles requests for vocabularies and and datasets according to their scheme of
permissions. Translation from a metadata-based query to a particular binary repre-
sentation of a file, whether it be in a relational database, binary, file, or otherwise,
will also be supported by vocabularies that indicate the necessary code.

Clearly, we need some form of *identity* in the system so that a peer can have their
links unambiguously identified and discovered. This is a challenging problem that
we leave open here, but strategies ranging from URI-based resolution like `username@domain.com`,
to locally-held cryptographic key based identity, to decentralized systems like the
w3c's Decentralized Identifiers [199] would suffice. For the sake of example, let's
make identity simple and flat, denoted in pseudocode as `@username`. Someone would
then be able to use their `@namespace` as a root, under which they could refer to their
data, schemas, and so on, which will be denoted `@name:subobject` (see this no-
tion of personal namespaces for knowledge organization discussed in early wiki cul-
ture here [200]). Let us also assume that there is no categorical difference between
`@usernames` used by individual researchers, institutions, consortia, etc. — everyone
is on the same level.

To illustrate the system by example, we pick up where we left off earlier with a peer
who has their data in some discipline-specific format, which let us assume for the
sake of concreteness has a representation as an OWL schema.

That schema could be "owned" by the `@username` corresponding to the standard-
writing group — eg `@nwb` for neurodata without borders. In all the following exam-
ples, we will use a turtle-ish syntax that is *purposely pseudocode* with the intention of
demonstrating general qualities without being concerned with syntactic correctness
or indicating one syntax in particular. Our dataset might look like this:

```
@base @jonny

<#my-data>
  a @nwb:NWBFile
  @nwb:general:experimenter @jonny
```

```
@nwb:ElectricalSeries
  .electrodes [1, 2, 3]
  .rate 30000
  .data [ ... ]
```

Unpacking the pseudocode, this indicates:

- We declare a `@base` context underneath my identity, `@jonny`,

- Underneath the base, individual objects are declared with their name like `<#object-name>`, a shorthand for `<@base:object-name>`. In this case I have made a dataset identified as `@jonny:my-data`.

- I have identified the type of this object with the a token, in this case a `@nwb:NWBFile`

- Subsequent lines indicate particular properties of the indicated type and their value, specifically I have indicated that the `@nwb:general:experimenter` is me, `@jonny`, and that the dataset also contains a `@nwb:ElectricalSeries`. While my identity object might have additional links like an `@ORCID:ID`, we can assume some basic inference that resolves my identity to a string as specified in the NWB specification, or else specify it explicitly as `@jonny:name`

- Additional subproperties are assigned with a leading `.`, so `.electrodes` would resolve to `@nwb:ElectricalSeries:electrodes`.

How would my client know how to read and write the data to my disk so i can use and share it? In a system with heterogeneous data types and database implementations, we need some means of specifying different programs to use to read and write, different APIs, etc. This too can be part of the format specification. Suppose the HDF5 group (or anyone, really!) has a namespace `@hdf` that defines the properties of an `@hdf:HDF5` file, basic operations like `Read`, `Write`, or `Select`. NWB could specify that in their definition of `@nwb:NWBFile`:

```
<@nwb:NWBFile>
  a @hdf:HDF5
    .isVersion "x.y.z"
    .hasDependency "libhdf5"=="x.y.z"
  usesContainer @nwb:NWBContainer
```

So when I receive a request for the raw data of my electrical series, my client knows to use the particular methods from the HDF5 object type to index the data contained within the file.

I have some custom field for my data, though, which I extend the format specification to represent. Say I have invented some new kind of solar-powered electrophysiological device — the SolarPhys2000 — and want to annotate its specs alongside my data.

```
<#SolarEphys>
  extends @nwb:NWBContainer

  UsedWith @jonny:hw:SolarPhys2000
```

```
ManufactureDate
   a @schema:Date

InputWattageSeries
   extends @nwb:ElectricalSeries

sunIntensity
   a @nwb:TimeSeries
```

Here I create a new extension `@jonny:SolarEphys` that `extends` the `@nwb:NWBContainer` schema. We use `extends` rather than a because we are adding something new to the *description* of the container rather than *making* a container to store data. I declare that this container is `UsedWith` our SolarPhys2000 which we have defined elsewhere in our `hw` namespace using some hardware ontology. I then add two new fields, `ManufactureDate` and `InputWattageSeries`, declaring types from, for example `@schema:Date` and `@nwb`.

The abstraction around the file implementation makes it easier for others to consume my data, but it also makes it easier for *me* to use and contribute to the system. Making an extension to the schema wasn't some act of charity, it was the most direct way for me to use the tool to do what I wanted. Win-win: I get to use my fancy new instrument and store its data by extending some existing format standard. We are able to make my work part of a cumulative schema building effort by *aligning the modalities of use and contribution.*

For the moment our universe is limited only to other researchers using NWB. Conveniently, the folks at NWB have set up a federating group so that everyone who uses it can share their format extensions. In the same way that we can use schemas to refer to code as with our HDF5 files, we can use it to indicate the behavior of clients and federations. Say we want to make a federating peer that automatically `Accepts` request to `Join` and indexes any schema that inherits from their base `@nwb:NWBContainer`. Let's say `@fed` defines some basic properties of our federating system — it constitutes our federating "protocol" — and loosely use some terms from the ActivityStreams vocabulary as `@as`.

```
<@nwbFederation>
   a @fed:Federation
   onReceive
      @as:Join @as:Accept
   allowSchema
      extensionOf @nwb:NWBContainer
```

Now anyone that is a part of the `@nwbFederation` would be able to see the schemas we have submitted, sort of like a beefed up, semantically-aware version of the existing neurodata extensions catalog. In this system, many overlapping schemas could exist simultaneously under different namespaces, but wouldn't become a hopeless clutter because similar schemas could be compared and reconciled based on their semantic properties.

Now that I've got my schema extension written and submitted to the federation, time to submit my data! Since it's a p2p system, I don't need to manually upload it, but I do want to control who gets it. By default, I have all my NWB datasets set to be available to the `@nwbFederation` , and I list all my metadata on, say the Society for Neuroscience's `@sfnFederation`.

```
<#globalPermissions>
   a @fed:Permissions
   permissionsFor @jonny

   federatedWith
     name @nwbFederation
     @fed:shareData
        is @nwb:NWBFile

   federatedWith
     name @sfnFederation
     @fed:shareMetadata
```

Let's say this dataset in particular is a bit sensitive — say we apply a set of permission controls to be compliant with `@hhs.HIPAA` — but we do want to make use of some public server space run by our Institution, so we let it serve an encrypted copy that those I've shared it with can decrypt.

```
<#datasetPermissions>
   a @fed:Permissions
   permissionsFor @jonny:my-data

   accessRuleset @hhs:HIPAA
      .authorizedRecipient <#hash-of-patient-ids>

   federatedWith
     name @institutionalCloud
     @fed:shareEncrypted
```

Now I want to make use of some of my colleagues data. Say I am doing an experiment with a transgenic dragonfly and collaborating with a chemist down the hall. This transgene, known colloquially in our discipline as `@neuro:superstar6` (which the chemists call `@chem:SUPER6`) fluoresces when the dragonfly is feeling bashful, and we have plenty of photometry data stored as `@nwb:Fluorescence` objects. We think that its fluorescence is caused by the temperature-dependent conformational change from blushing. They've gathered NMR and Emission spectroscopy data in their chemistry-specific format, say `@acs:NMR` and `@acs:Spectroscopy`.

We get tired of having our data separated and needing to maintain a bunch of pesky scripts and folders, so we decide to make a bridge between our datasets. We need to indicate that our different names for the gene are actually the same thing and relate the spectroscopy data.

Let's make the link explicit, say we use an already-existing vocabulary like the "simple knowledge organization system" for describing logical relationships between concepts: `@skos`?

```
<#links:super6>
   @neuro:superstar6
     @skos:exactMatch @chem:SUPER6
```

Our `@nwb:Fluorescence` data has the emission wavelength in its `@nwb:Fluorescence:excitation_lambda` property[36], which is the value of their `@acs:Spectroscopy` data at a particular value of its `wavelength`. Unfortunately, `wavelength` isn't metadata for our friend, but does exist as a column in the `@acs:Spectroscopy:readings` table, so where we typically have a singular value they have a set of measurements. Since the same information has a structurally different meaning across disciplines, we dont expect there to be an automated 1:1 mapping between them, but presumably their data format also specifies some means of reading the data akin to the HDF5 methods indicated by our NWB data format so we can add an additional translation later like `@math:mean` and pick it up in our analysis tools.

[36] not really where it would be in the standard, but again, for the sake of example...

```
<#links:lambda>
   @acs:Spectroscopy:readings:wavelength
     @skos:narrowMatch @nwb:Fluorescence:excitation_lambda
       @skos:note
         "Multiple spectrographic readings are
         aggregated to a single excitation lambda"
       @translate:aggregate @math:mean
```

This makes it much easier for us to index our data against each other and solves a few real practical problems we were facing in our collaboration. We don't need to do as much cleaning when it's time to publish the data since it can be released as a single linked entity.

Though this example is relatively abstract (which metadata from spectroscopy readings would need to match which in a fluorescence series to compare wavelengths to lambda?), it serves as an example in its own right of the quasi-inversion of reasoning that we can make use of in our particular version of linked data with code. We refer to the general notion of taking a `@math:mean`, but don't specify a particular implementation of it. Other package maintainers could indicate that their function implements it, so we could be prompted to choose one when resolving the link. Alternatively, if we specified our aggregation used `@numpy:mean`, we could trace it backwards to find which general operation it implements and choose a different one. Since the objects of any triplet link have their own type, we can use the *context* of the link to infer how to use it.

Rinse and repeat our sharing and federating process from our previous schema extension, add a little bit of extra federation with the `@acs` namespace, and in the normal course of our doing our research we've contributed to the graph structure linking two common data formats. Our link is one of many, and is a proposition that other researchers can evaluate in the context of our project rather than as an authoritative reference link. We might not have followed the exact rules, but we have also changed the nature of rules — rather than logical coherence guaranteed *a priori* by adherence to a specification language, much like language the only rules that matter are those of *use*. We may have only made a few links rather than a single authorative mapping, but if someone is interested in compiling one down the line they'll start off a hell of a lot further than if we hadn't contributed it! Rather than this for-mat translation happening ad-hoc across a thousand lab-specific analysis libraries, we have created a space of *discourse* where our translation can be contextually compared to others and negotiated by the many people concerned, rather than handed down by a standards body.

Queries across what amounts to the federated schema, in the federated database parlance, are by design less seamless than they would be with centrally governed schema — which is a feature, not a bug. While this example deals with relatively dry fluorescence and spectrographic data, if this system were to expand to clinical, cultural, and personal data, the surveillance economy that emerged subsequent to they heyday of the semantic web has made it abundantly clear that *we don't necessarily want* arbitrary actors to be able to index across all available data. It is much more valuable to have low-barrier, vernacular expression usable by collections of subdisciplines and communities of people than a set of high-barrier, fixed, logically correct schemas. Researchers and people alike typically are only concerned with using the information within or a few hops outside of their local systems of meaning, so who is a totalizing database of everything *for?* This framing of linked data, by rejecting the goal of global inference altogether, could be considered beyond even Lindsay Poirier's conception of "scruffiness" to something we might properly call *vulgar linked data.*

The act of translation is always an act of creation, and by centering the multiplicity of links between extensible schemas we center the dialogic reality of that creation: *who says* those things are equivalent? Since the act of using translating links between schemas itself creates links — ie. I link to `@<user>`'s link to link my dataset and another — we are both able to assess the status of consensus around which links are used, as well as bring a currently invisible form of knowledge work into a system of credit. As we will develop in the following two sections, this multiplicity also naturally lends itself to a fluid space of tools that implement translations and analyses, as well as a means of discussing and contextualizing the results.

We have been intentionally vague about the technical implementation here, but there are many possible strategies and technologies for each of the components.

For making our peers and the links within their namespace discoverable we could use a distributed hash table, or **DHT**, like bittorrent, which distributes references to information across a network of peers (eg. [201]). We could use a strategy like the **Matrix** messaging protocol, where peers could federate with "relay" servers. Each server is responsible for keeping a synchronized copy of the messages sent on the servers and rooms it's federated with, and each server is capable of continuing communication if any of the others failed. We could use **ActivityPub** (AP) [202], a publisher-subscriber model where users affiliated with a server post messages to their 'outbox' and are sent to listening servers (or made available to HTTP GET requests). AP uses JSON-LD [203], so is already capable of representing linked data, and the related ActivityStreams vocabulary [204] also has plenty of relevant action types for creating, discussing, and negotiating over links (also see cpub). We could use a strategy like IPFS where peers will voluntarily rehost each other's data in order to gain trust with one another. To preserve interoperability with existing systems, we will want to make links referenceable from a URI (as IPFS does) as well as be able to resolve multiple protocols, but beyond that the space of possible technologies is broad.

Indexing and querying metadata across federated peers could make use of the SPARQL query language [205] as has been proposed for biology many times before [206, 195, 196]. The distinction between metadata and data is largely practical — a query shouldn't require transferring and translating terabytes of data — so we will need some means of resolving references to data from metadata as per the linked data platform specification [189]. A mutable/changeable/human-readable name and metadata system that points to a system of unique content addressed identifiers has been one need that has hobbled IPFS, and is the direction pointed to by DataLad[37] [178]. A parallel set of conversations has been happening in the broader linked data community with regard to using ActivityPub as a way to index data on Solid.

The design of federations of peers is intended to resolve several of the problems of prior p2p protocols. Rather than a separate swarm for every dataset per bittorrent, or a single global swarm per IPFS, this system would be composed of peers that can voluntarily associate and share metadata structure at multiple scales. Bittorrent requires trackers to aggregate and structure metadata, but they become single points of failure and often function as means of gatekeeping by the beloved petty tyrants who host them. IPFS has turned to filecoin to incentivize donating storage space among quasi-anonymous peers, a common design pattern among the radical zero-trust design of many cryptocurrencies and cryptocurrency-like systems.

Voluntary federations are instead explicitly social systems that can describe and organize their own needs: peers in a federation can organize tracker or serverlike rehosting of their data for performance, discoverability, guaranteed longevity. A federation can institute a cooperative storage model akin to private bittorrent trackers that requires a certain amount of rehosted data per data shared. A small handful of researchers can form a small federation to share data while collaborating on a project in the same way that a massive international consortioum could. Without enumerating their many forms, federations can be a way to realize the evolvable community structure needed for sustained archives. As may become clearer as we discuss systems for communication, in the context of science they might be a way of reconceptualizing scientific societies as something that supports the practice of science beyond their current role as ostensibly nonprofit journal publishers and event hosts.

So far we have described a system for sharing data with a p2p system integrated with linked data. We have given a few brief examples of how linked data can be used for standardized and vernacular metadata, integrating with heterogeneous local storage systems, and to perform actions like creating and joining federations of peers. As described, though, the system would still be decidedly unapproachable for most scientists and doesn't offer the kind of strong incentives that would create a broad base of use. We clearly need one or several *interfaces* to make the creation and use metadata easy. We will return to those in Shared Knowledge and also describe a set of communication and governance systems sorely needed in science. To get there, we will first turn to a means of integrating our shared data system with analytical and experimental tools to make each combinatorically more useful than if considered alone.

## 3.3   Shared Tools

Straddling our system for sharing data are the tools to gather and analyze it — combining tools to address the general need for *storage* with *computational resources*. Considering them together presents us with new opportunities only possible with cross-domain interoperability. In particular, we can ask how a more broadly in-tegrated system makes each of the isolated components more powerful, enables a kind of deep provenance from experiment to results, and further builds us towards reimagine the form of the community and communication tools for science. Where the previous section focused on integrating linked metadata with data, here our focus is how to make linked data *do things* by integrating it with code.

[37] DataLad [207, 178] and its application in Neuroscience as DANDI are two projects that are *very close* to what I have been describing here — developing a p2p backend for datalad might even be a promising development path towards it.

This section will be relatively short compared to shared data. We have already introduced, motivated, and exemplified many of the design practices of the broader infrastructural system. There is much less to argue against or "undo" in the spaces of analytical and experimental tools because so much more work has been done, and so much more power has been accrued in the domain of data systems. Distributed computing does have a dense history, with huge numbers of people working on the problem, but its dominant form is much closer to the system articulated below than centralized servers are to federated semantic p2p systems. I also have written extensively about experimental frameworks before [208], and develop one of them so I will be brief at risk of repeating myself or appearing self-serving.

Integrated scientific workflows have been written about many times before, typically in the context of the "open science" movement. One of the founders of the Center for Open Science, Jeffrey Spies, described a similar ethic of toolbuilding as I have in a 2017 presentation:

> Open Workflow: 1. Meet users where they are 2. Respect current incentives 3. Respect current workflow
>
> - We could... demonstrate that it makes research more efficient, of higher quality, and more accessible.
>
> - Better, we could... demonstrate that researchers will get published more often.
>
> - Even better, we could... make it easy.
>
> - Best, we could... make it automatic [209]

Similar to the impossibility of a single unified data format, it is unlikely that we will develop one tool to rule them all. We will take the same tactic of thinking about *frameworks* to integrate tools and make them easier to build, rather than building any tool in particular.

### 3.3.1   Analytical Frameworks

The first natural companion of shared data infrastructure is a shared analytical framework. A major driver for the need for everyone to write their own analysis code largely from scratch is that it needs to account for the idiosyncratic structure of everyone's data. Most scientists are (blessedly) not trained programmers, so code for loading and loading data is often intertwined with the code used to analyze and plot it. As a result it is often difficult to repurpose code for other contexts, so the same analysis function is rewritten in each lab's local analysis repository. Since sharing raw data and code is still a (difficult) novelty, on a broad scale this makes results in scientific literature as reliable as we imagine all the private or semi-private analysis code to be.

Analytical tools (anecdotally) make up the bulk of open source scientific software, and range from foundational and general-purpose tools like numpy [210] and scipy [211], through tools that implement a class of analysis like DeepLabCut [30] and scikit-learn [212], to tools for a specific technique like MoSeq [213] and DeepSqueak [214]. The pattern of their use is then to build them into a custom analysis system that can then in turn range in sophistication from a handful of flash-drive-versioned scripts to automated pipelines.

Having tools like these of course puts researchers miles ahead of where they would be without them, and the developers of the mentioned tools have put in a tremendous amount of work to build sensible interfaces and make them easier to use. No matter how much good work might be done, inevitable differences between APIs is a relatively sizable technical challenge for researchers — a problem compounded by the incentives for fragmentation described previously. For toolbuilders, many parts of any given tool from architecture to interface have to be redesigned each time with varying degrees of success. For science at large, with few exceptions of well-annotated and packaged code, most results are only replicable with great effort.

Discontinuity between the behavior and interface of different pieces of software is, of course, the overwhelming norm. Negotiating boundaries between (and even within) software and information structures is an elemental part of computing. The only time it becomes a conceivable problem to "solve" interoperability is when the problem domain coalesces to the point where it is possible to articulate its abstract structure as a protocol, and the incentives are great enough to adopt it. That's what we're trying to do here.

It's unlikely that we will solve the problem of data analysis being complicated, time consuming, and error prone by teaching every scientist to be a good programmer, but we can build experimental frameworks that make analysis tools easier to build and use.

Specifically, a shared analytical framework should be

- **Modular** - Rather than implementing an entire analysis pipeline as a monolith, the system should be broken into minimal, composable modules. The threshold of what constitutes "minimal" is of course to some degree a matter of taste, but the framework doesn't need to make normative decisions like that. The system should support modularity by providing a clear set of hooks that tools can provide: eg. a clear place for a given tool to accept some input, parameters, and so on. Since data analysis can often be broken up into a series of relatively independent stages, a straightforward (and common) system for modularity is to build hooks to make a directed acyclic graph (DAG) of data transformation operations. This structure naturally lends itself to many common problems: caching intermediate results, splitting and joining multiple inputs and outputs, distributing computation over many machines, among others. Modularity is also needed within the different parts of the system itself – eg. running an analysis chain shouldn't require a GUI, but one should be available, etc.

- **Pluggable** - The framework needs to provide a clear way of incorporating external analysis packages, handling their dependencies, and exposing their parameters to the user. Development should ideally not be limited to a single body of code with a single mode of governance, but should instead be relatively conservative about requirements for integrating code, and liberal with the types of functionality that can be modified with a plugin. Supporting plugins means supporting people developing tools for the framework, so it needs to make some part of the toolbuilding process easier or otherwise empower them relative to an independent package. This includes building a visible and expressive system for submitting and indexing plugins so they can be discovered and credit can be given to the developers. Reciprocal to supporting plugins is being interoperable with existing and future systems, which the reader may have assumed was a given by now.

- **Deployable** - For wide use, the framework needs to be easy to install and deploy locally and on computing clusters. A primary obstacle is dependency management, or making sure that the computer has everything needed to run the program. Some care needs to be taken here, as there are multiple emphases in deployability that can be in conflict. Deployable for who? A system that can be relatively challenging to use for routine exploratory data analysis but can distribute analysis across 10,000 GPUs has a very circumscribed set of people it is useful for. This is a matter of balancing design constraints, but we should prioritize broad access, minimal assumptions of technological access, and ease of use over being able to perform the most computationally demanding analyses possible when in conflict. Containerization is a common, and the most likely strategy here, but the interface to containers may need a lot of care to make accessible compared to opening a fresh .py file.

- **Reproducible** - The framework should separate the *parameterization* of a pipeline, the specific options set by the user, and its *implementation*, the code that constitutes it. The parameterization of a pipeline or analysis DAG should be portable such that it, for example, can be published in the supplementary materials of a paper and reproduced exactly by anyone using the system. The isolation of parameters from implementation is complementary to the separation of metadata from data and if implemented with semantic triplets would facilitate a continuous interface from our data to analysis system. This will be explored further below and in shared knowledge

Thankfully a number of existing projects that are very similar to this description are actively being built. One example is DataJoint [215], which recently expanded its facility for modularity with its recent Elements project [216]. Datajoint is a system for creating analysis pipelines built from a graph of processing stages (among other features). It is designed around a refinement on traditional relational data models, which is reflected throughout the system as most operations being expressed in its particular schema, data manipulation, and query languages. This is useful for operations that are expressed in the system, but makes it harder to integrate external tools with their dependencies — at the moment it appears that spike sorting (with Kilosort [217]) has to happen outside of the extracellular electrophysiology elements pipeline.

Kilosort is an excellent and incredibly useful tool, but its idiomatic architecture designed for standalone use is illustrative of the challenge of making a general-purpose analytic framework that can integrate a broad array of existing tools. It is built in MATLAB, which requires a paid license, making arbitrary deployment difficult, and MATLAB's flat path system requires careful and usual manual orchestration of potentially conflicting names in different packages. Its parameterization and use are combined in a "main" script in the repository root that creates a MATLAB struct and runs a series of functions — requiring some means for a wrapping framework to translate between input parameters and the representation expected by the tool. Its preprocessing script combines I/O, preprocessing, and plotting, and requires data to be loaded from disk rather than passed as arguments to preserve memory — making chaining in a pipeline difficult.

This is not a criticism of Datajoint or Kilosort, which were both designed for different uses and with different philosophies (that are of course, also valid). I mean this as a brief illustration of the design challenges and tradeoffs of these systems.

We can start getting a better picture for the way a decentralized analysis framework might work by considering the separation between the metadata and code modules, hinting at a protocol as in the federated systems sketch above. In the time since the heydey of the semantic web there has been a revolution in containerization and dependency management that makes it possible to imagine extending the notion of linked data to being able to not only indicate binary data but also *executable code*. Software dependencies form a graph structure, with one top level package specifying a version range from a 1st-order dependent, which in turn has its own set of 2nd-order packages and versions, and so on. Most contemporary dependency managers (like Python's poetry, Javascript's yarn, Rust's cargo, Ruby's Bundler, etc.) compute an explicit dependency graph from each package's version ranges to create a 'lockfile' containing the exact versions of each package, and usually the repositories where they're located and the content hashes to verify them. More general purpose package managers like spack [218], or nix [219] can also specify system-level software outside of an individual programming language, and containerization tools like docker can create environments that include entire operating systems.

Since we're considering modular analysis elements, each module would need some elemental properties like the parameters that define it, its inputs, outputs, as well as some additional metadata about its implementation (eg. this one takes *numpy arrays* and this one takes *matlab structs*). The precise implementation of a modular protocol also depends on the graph structure of the analysis pipelining system. We invoked DAGs before, but analysis graph structure of course has its own body of researchers refining them into eg. Petri nets which are graphs whose nodes necessarily alternate between "places" (eg. intermediate data) and "transitions" (eg. an analysis operation), and their related workflow markup languages (eg. WDL or [220]). In that scheme, a framework could provide tools for converting data between types, caching intermediate data, etc. between analysis steps, as an example of how different graph structures might influence its implementation.

The graph structure of our linked data system could flexibly extend to be continuous with these dependency pipeline graphs. With some means for a client to resolve the dependencies of a given analysis node, it would be possible to reconstruct the environment needed to run it. By example, how might a system like this work?

Say we use @analysis as the namespace for our specifying each analysis node's properties, and someone has provided bindings to objects in numpy (we'll give an example of how these bindings might work below, but for now assume they work analogously to the module structure of numpy, ie. @numpy:ndarray = numpy.ndarray). We can assume they are provided by the package maintainers, but that's not necessary: this is my node and it takes what I want it to!

In pseudocode, I could define some analysis node for, say, converting an RGB image to grayscale under my namespace as @jonny:bin-spikes like this:

```
<#bin-spikes>
  a @analysis:node
    Version " ⩾ 1.0.0"

  hasDescription
    "Convert an RGB Image to a grayscale image"

  inputType
    @numpy:ndarray
      # ... some spec of shape, dtype ...

  outputType
    @numpy:ndarray
      # ... some spec of shape, dtype ...

  params
    bin_width int
      default 10
```

I have abbreviated the specification of shape and datatype to not overcomplicate the pseudocode example, but say we successfully specify a 3 dimensional (width x height x channels) array with 3 channels as input, and a a 2 dimensional (width x height) array as output. An optional `bin_width` parameter with default "10" can also be provided.

The code doesn't run on nothing! We need to specify our node's dependencies. Say in this case we need to specify an operating system image `ubuntu`, a version of `python`, a system-level package `opencv`, and a few python packages on `pip`. We are pinning specific versions with semantic versioning, but the syntax isn't terribly important. Then we just need to specify where the code for the node itself comes from:

```
dependsOn
  @ubuntu:"^20.*":x64
  @python:"3.8"
  @apt:opencv:"^4.*.*"
  @pip:opencv-python:"^4.*.*"
    .extraSource "https://pywheels.org/"
  @pip:numpy:"^14.*.*"

providedBy
  @git:repository
    .url "https://mygitserver.com/binspikes/fast-binspikes.git"
    .hash "fj9wbkl"
  @python:class "/main-module/binspikes.py:Bin_Spikes"
    method "run"
```

Here we can see the practical advantage of the "inverted" link-based system rather than an object-oriented-like approach. `@ubuntu` refers to a specific software image that would have a specific `providedBy` value, but both `@apt` and `@pip` can have different repositories that they pull packages from, and for a given version and repository there will be multiple possible software binaries for different CPU architec-tures, python versions, etc. Rather than needing to specify a generalized specification format, each of these different types of links could specify their own means of resolving dependencies: a `@pip` dependency requires some `@python` version to be specified. Both require some operating system and architecture. If we hadn't provided the `.extraSource` of pywheels (for ARM architectures), someone who had defined some link between a given architecture and `@pip` could be proposed as a way of finding the package.

Our `@analysis.node` protocol gives us several slots to connect different tools together, each in turn presumably provides some minimal functionality expected by that slot: eg. `inputType` can expect `@numpy:ndarray` to specify its own dependencies, the programming language it is written in, shape, data type, and so on. Coercing data between chained nodes then becomes a matter of mapping between the `@numpy` and, say a `@nwb` namespace of another format. In the same way that there can be multiple, potentially overlapping between data schemas, it would then be possible for people to implement mappings between intermediate data formats as needed. This gives us an opportunity to build pipelines that use tools from multiple languages, a problem typically solved by manually saving, loading, and cleaning intermediate data.

This node also becomes available to extend, say someone wanted to add an additional input format to my node:

```
<@friend#bin-spikes>
    extends @jonny:bin-spikes

    inputType
      @pandas:DataFrame

    providedBy
      ...
```

They don't have to interact with my potentially messy codebase at all, but it is automatically linked to my work so I am credited. One could imagine a particular analysis framework implementation that would then search through extensions of a particular node for a version that supports the input/output combinations appropriate for their analysis pipeline, so the work is cumulative. This functions as a dramatic decrease in the size of a unit of work that can be shared.

This also gives us healthy abstraction over implementation. Since the functionality is provided by different, mutable namespaces, we're not locked into any particular piece of software — even our `@analysis` namespace that gives the `inputType` etc. slots could be forked. We could implement the dependency resolution system as, eg. a docker container, but it also could be just a check on the local environment if someone is just looking to run a small analysis on their laptop with those packages already installed.

The relative complexity required to define an analysis node, as well as the multiple instances of automatically computed metadata like dependency graphs hints that we should be thinking about tools that avoid needing to write it manually. We could use an `Example_Framework` that provides a set of classes and methods to implement the different parts of the node (a la luigi). Our `Bin` class inherits from `Node`, and we implement the logic of the function by overriding its `run` method and specify an

`output` file to store intermediate data (if requested by the pipeline) with an `output` method. Our class is within a typical python package that specifies its dependencies, which the framework can detect. We also specify a `bin_width` as a `Parameter` for our node, as an example of how a lightweight protocol could be bidirectionally specified as an <span style="color:red">interface</span> to the linked data format: we could receive a parameterization from our pseudocode metadata specification, or we could write a framework with a `Bin.export_schema()` that constructs the pseudocode metadata specification from code.

```python
from Example_Framework import Node, Param, Target

class Bin(Node):
  bin_width = Param(dtype=int, default=10)

  def output(self) → Target:
    return Target('temporary_data.pck')

  def run(self, input:'numpy.ndarray') → 'numpy.ndarray':
    # do some stuff
    return output
```

Now that we have a handful of processing nodes, we could then describe some `@workflow`, taking some `@nwb:NWBFile` as input, as inferred by the `inputType` of the `bin-spikes` node, and then returning some output as a `:my-analysis:processed` child beneath the input. We'll only make a linear pipeline with two stages, but there's no reason more complex branching and merging couldn't be described as well.

```
<#my-analysis>
    a @analysis:workflow

    inputType
        @jonny:bin-spikes:inputType

    outputName
        input:my-analysis:processed

    step Step1 @jonny:bin-spikes
    step Step2 @someone-else:another-step
        input Step1:output
```

Since the parameters are linked from the analysis nodes, we can specify them here (or in the workflow). Assuming literally zero abstraction and using the tried-and-true "hardcoded dataset list" pattern, something like:

```
<#my-project>
    a @analysis:project

    hasDescription
        "I gathered some data, and it is great!"
```

```
researchTopic
    @neuro:systems:auditory:speech-processing
    @linguistics:phonetics:perception:auditory-only

inPaper
    @doi:10.1121:1.5091776

workflow Analysis1 @jonny:my-analysis
  globalParams
      .Step1:params:bin_width 10

  datasets
      @jonny.mydata1:v0.1.0:raw
      @jonny.mydata2:^0.2.*:raw
      @jonny.mydata3:≥0.1.1:raw
```

And there we are! The missing parameters like `outputName` from our workflow can be filled in from the defaults. Our project is an abstract representation of the analysis to be performed and where its output will be found - in this case as `:processed` beneath each dataset link. From this very general pseudocode example it's possible to imagine executing the code locally or on some remote server, pulling the data from our p2p client, installing the environment, and duplicating the resulting data to the clients configured to mirror our namespace. This system would work similar to the combination of configuration and lockfiles from package managers: we would give some abstract specification for a project's analysis, but then running it would create a new set of links with the exact dependency graph, links to intermediate products, and so on. We get some inkling of where we're going later by also being able to specify the paper this data is associated with, as well as some broad categories of research topics so that our data as well as the results of the analysis can be found.

From here we could imagine how existing tools might be integrated without needing to be dramatically rewritten. In addition to wrapping their parameters, functions, and classes with the above `Node` class, we could imagine our analysis linking framework providing some function to let us indicate code within a package and prompt us for any missing pieces like dependency specification from, for example, old style python packages that don't require it. For packages that don't have an explicit declarative parameterization, but rely on programmatically created configuration files, we could imagine a tool ingestion function being able to extract default fields and then refer to them with a `fromConfig @yaml` link. A single tool need not be confined to a single analysis node: for example a tool that requires some kind of user interaction could specify that with an `@analysis:interactive` node type that feeds its output into a subsequent analysis node. There are infinitely more variations to be accounted for — but adapting to them is the task of an extensible linking system.

As soon as we extend our relatively static protocol to the realm of arbitrary code we immediately face the question of security. Executing arbitrary code from many sources is inherently dangerous and worthy of careful thought, but any integrative framework becomes a common point where security practices could be designed into the system as opposed to the *relative absence of security practices of any kind* in most usages of scientific software. There is no reason to believe that this system is intrinsically more dangerous than running uninspected packages from PyPI, which, for example, have been known to steal AWS keys and environment variables [221] [38].

[38] They took advantage of being able to run arbitrary code in legacy `setup.py` scripts to run a separate shell command, an illustration of the urgency with which we need to deprecate that horrible system.

Having analysis code and its dependency graph specified publicly presents opportunities for being able to check for identified vulnerabilities at the time of execution — a role currently filled by platform tools like GitHub's dependabot or npm's audit. Running code by default in containers or virtual environments could be a way towards making code secure by default.

So that's useful, but comparable to some existing pipelining technologies. The important part is in the way this hypothetical analysis framework and markup interact with our data system — it's worth unpacking a few points of interaction.

A dataset linked to an analysis pipeline and result effectively constitutes a "unit of analysis." If I make my data publicly available, I would be able to see all the results and pipelines that have been linked to it. Within a single pipeline, comparing the results across a grid of possible parameterizations gives us a "multiverse analysis [222]" for estimating the effects of each parameterization for free. Conversely, "rules of thumb" for parameter selection can be replaced by an evaluation of parameters and results across prior applications of the pipeline. Since some parameters like model weights in neural networks are not trivial to reproduce, and their use is linked to the metadata of the dataset they are applied to, all analyses contribute to a collection of models like the DeepLabCut model zoo decreasing the need for fine tuning on individual datasets and facilitating metalearning across datasets.

Across multiple pipelines, a dataset need no longer be dead on publication, but can instead its meaning and interpretation can continuously evolve along with the state of our tools and statistical practices. Since pipelines themselves are subject to the same kind of metadata descriptions as datasets are, it becomes to find multiple analysis nodes that implement the same operation, or to find multiple pipelines that perform similar operations despite using different sets of nodes. Families of pipelines that are applied to semantically related datasets would then become the substrate for a field's state of the art, currently buried within disorganized private code repositories and barely-descriptive methods sections. Instead of a 1:1 relationship where one dataset is interpreted once, we could have a many-to-many relationship where a cumulative body of data is subject to an evolving negotiation of interpretation over time — ostensibly how science is *"supposed to"* work.

This system also allows the work of scientific software developers to be credited according to use, instead of according to the incredibly leaky process of individual authors remembering to search for all the citations for all the packages they may have used in their analysis. Properly crediting the work of software developers is important not only for equity, but also for the reliability of scientific results as a whole. A common admonishment in cryptography is to "never roll your own crypto," but that's how most homebrew analysis code works, and the broader state of open source scientific code is not much better without incentives for maintenance. Bugs in analysis code that produce inaccurate results are inevitable and rampant [223, 224, 225, 226], but impossible to diagnose when every paper writes its own pipeline. A common analysis framework would be a single point of inspection for bugs and means of providing credit to people who fix them. When a bug is found, rather than irreparably damaging collective confidence in a field, it would then be trivial to re-run all the analyses that were impacted and evaluate how their results were changed.

Finally, much like how we are building towards the social systems to support federations for sharing data, integrating analysis pipelines into a distributed network of servers is a means of realizing a generalized Folding@Home-style distributed computing grid [227, 228]. Existing projects like F@H and the Pacific Research Platform [229] show the promise of these distributed computing systems for solving previously-intractable problems, but they require large amounts of coordination and are typically centrally administered towards a small number of specific projects with specific programming requirements. With some additional community systems for governance, resource management, and access, they become tantalizingly in-reach from the system we are describing here. We will return to that possibility after discussing experimental tools.

### 3.3.2   Experimental Frameworks

Across from the tools to analyze data are those to collect it, and tools to integrate the diversity of experimental practice are a different challenge altogether: *everyone needs completely different things!* Imagine the different stages of research as a cone of complexity: at the apex we can imagine the relatively few statistical outcomes from a family of tests and models. For every test statistic we can imagine a thousand analysis scripts, for every analysis script we might expect a thousand data formats, and so the complexity of the thousand experimental tools used to collect each type of data feels ... different.

Beyond a narrow focus of the software for performing experiments itself, the surrounding contextual knowledge work largely lacks a means of communication and organization. Methods sections have been increasingly marginalized, abbreviated, pushed to the end, and relegated to the supplement. The large body of work that is not immediately germane to experimental results, like animal care, engineering instruments, lab management, etc. have effectively no formal means of communication — and so little formal means of credit assignment.

Extending our ecosystem to include experimental tools has a few immediate benefits: bridging the gap between collection and sharing of data would resolve the need for format conversion as a prerequisite for inclusion in the linked system, allowing the expression of data to be a fluid part of the experiment itself. It would also serve as a means of building a body of cumulative contextual knowledge in a creditable system.

I have previously written about the design of a generalizable, distributed experimental framework [208], so to avoid repeating myself, and since many of the ideas from the section on analysis tools apply here as well, I will be relatively brief.

We don't have the luxury of a natural formalism like a DAG to structure our experimental tools. Some design constraints on experimental frameworks might help explain why:

- They need to support a wide variety of instrumentation, from **off-the-shelf parts,** to **proprietary instruments** as are common in eg. microscopy, to **custom, idiosyncratic designs** that might make up the existing infrastructure in a lab. Writing and testing embedded code that controls external hardware is a wholly different kind of difficulty than writing analysis tools.

- To be supportive, rather than constraining, they need to be able to **flexibly perform many kinds of experiments** in a way that is **familiar to patterns of existing practice.** That effectively means being able to coordinate heterogeneous instruments in some "task" with a flexible syntax.

- They need to be **inexpensive to implement,** in terms of both money and labor, so it can't require buying a whole new set of hardware or dramatically restructuring existing research practices.

- They need to be **accessible and extensible,** with many different points of control with different expectations of expertise and commitment to the framework. It needs to be useful for someone who doesn't want to learn it to its depths, but also have a comprehensible codebase at multiple scales so that reasearchers can **easily extend** it when needed.

- They need to be designed to support **reproducibility and provenance,** which is a significant challenge given the heterogeneity inherent in the system. On one hand, being able to produce *data that is clean at the time of acquisition* simplifies automated provenance, but enabling experimental replication requires multiple layers of abstraction to keep the idiosyncrasies of an experiment separable from its implementation: it shouldn't require building *exactly* the same apparatus with *exactly* the same parts connected in *exactly* the same way to replicate an experiment.

- Ideally, they need to support **cumulative labor and knowledge organization,** so an additional concern with designing abstractions between system components is allowing work to be made portable and combinable with others. The barriers to contribution should be extremely minimal, not requiring someone to be a professional programmer to make a pull request to a central library, and contributions should come in many modes — code is not the only form of knowing and it's far from the only thing needed to perform an experiment.

Here, as in the domains of data and analysis, the temptation to universalize is strong, and the parts of the problem that are emphasized influence the tools that are produced. A common design tactic for experimental tools is to design them as state machines, a system of states and transitions not unlike the analysis DAGs above. One such nascent project is BEADL [230] from a Neurodata Without Borders working group. BEADL is an XML-based markup for standardizing a behavioral task as an abstraction of finite state machines called statecharts. Experiments are fully abstract from their hardware implementation, and can be formally validated in simulations. The working group also describes creating a standardized ontology and metadata schema for declaring all the many variable parameters for experiments, like reward sizes, stimuli, and responses [231]. This group, largely composed of members from the Neurodata Without Borders team, understandably emphasize systematic description and uniform metadata as a primary design principle.

Personally, I *like* statecharts. The problem is that it's not necessarily natural to express things as statecharts as you would want to, or in the way that your existing, long-developed local experimental code does. There are only a few syntactical features needed to understand the following statechart: blocks are states, they can be inside each other. Arrows move between blocks depending on some condition. Entering and exiting blocks can make things happen. Short little arrows from filled spots are where you start in a block, and when you get to the end of the chart you go back to the first one. See the following example of a statechart for controlling a light, described in the introductory documentation and summarized in the figure caption:

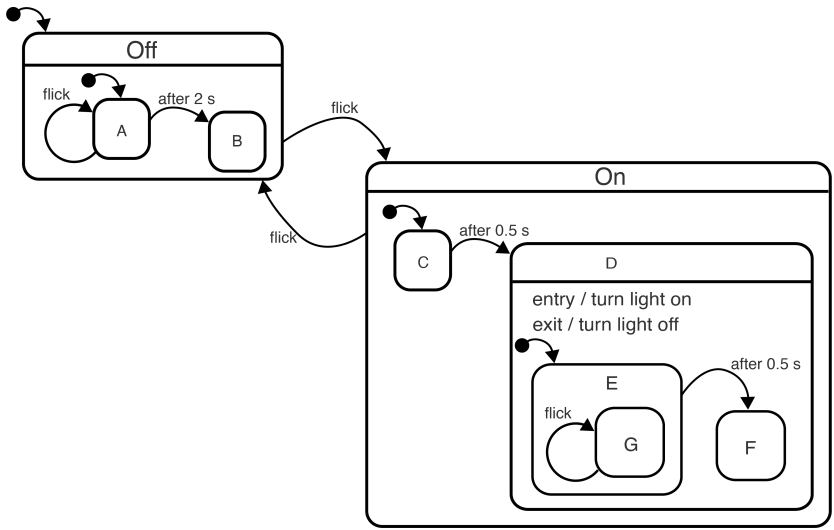

They have an extensive set of documents that defend the consistency and readability of statecharts on their homepage, and my point here is not to disagree with them. My point is instead that tools that aspire to the status of generalized infrastructure can't ask people to dramatically change the way they think about and do science. There are many possible realizations of any given experiment, and each is more or less natural to every person.

The problem here is really one of emphasis, BEADL seeks to solve problems with inconsistencies in terminology by standardizing them, and in order to do that seeks to standardize the syntax for specifying experiments.

This means of standardization has many attractive qualities and is being led by very capable researchers, but I think the project is illustrative of how the differing structures of problems constrain the possible space of tooling. Analysis tasks are often asynchronous, where the precise timing of each node's completion is less important than the path dependencies between different nodes. Analysis tasks often have a clearly defined set of start, end, and intermediate cache points, rather than branching or cyclical decision paths that change over multiple timescales. Statecharts are a hierarchical abstraction of finite state machines, the primary advantage of which is that they are better able to incorporate continuous and history-dependent behavior, which cause state explosion in traditional finite-state machines.

The difficulty of a controlled ontology for experimental frameworks is perhaps better illustrated by considering a full experiment. In Autopilot, a full experiment can be parameterized by the `.json` files that define the task itself and the system-specific configuration of the hardware. An example task from our lab consists of 7 behavioral shaping stages of increasing difficulty that introduce the animal to different features of a fairly typical auditory categorization task. Each stage includes the parameters for at most 12 different stimuli per stage, probabilities for presenting lasers, bias correction, reinforcement, criteria for advancing to the next stage, etc. So just for one relatively straightforward experiment, in one lab, in one subdiscipline, there are **268 parameters** – excluding all the default parameters encoded in the software.

How might we approach this problem differently, to accommodate diversity of thought styles and to be complementary to our data and analysis systems? The primary things we need from our experimental frameworks are a) to be able to link a particular realization of an experiment with the metadata that describes it, and b) to be able to produce similarly metadata-rich data. Rather than linked data indicating code as in our analysis frameworks, we might invert our strategy and think about code that draws from linked data.

As an example, Autopilot [208] approaches the problem by avoiding standardizing *experiments* themselves, instead providing smaller building blocks of experimental tools like hardware drivers, data transformations, etc. and emphasizing understanding their use in *context*. This approach sacrifices some of the qualities of a standardized system like being a logically complete or guaranteeing a standardized vocabulary in order to better support integrating with existing work patterns and making work cumulative. Because we can't possibly predict the needs and limitations of a totalizing system, we split the problem along a different set of concerns, those of the elements of experimental practice, and give facility for describing how they are used together.

For concrete example, we might imagine the lightswitch in an autopilot-like framework like this:

```python
from autopilot.hardware.gpio import Digital_Out
from time import sleep
from threading import Lock

class Lightswitch(Digital_Out):
  def __init__(self,
    off_debounce: float = 2,
    on_delay:     float = 0.5,
    on_debounce:  float = 0.5):
    """
    Args:
      off_debounce (float):
        Time (s) before light can be turned back on
      on_delay (float):
        Time (s) before light is turned on
      on_debounce (float):
        Time (s) after turning on that light can't be turned off
    """
    self.off_debounce = off_debounce
    self.on_delay     = on_delay
    self.on_debounce  = on_debounce

    self.on = False
    self.lock = Lock()

  def switch(self):
    # use a lock to make sure if
    # called while waiting, we ignore it
    if not self.lock.acquire():
      return
```

```python
    # if already on, switch off
    if self.on:
        self.on = False
        sleep(self.off_debounce)

    # otherwise switch on
    else:
        sleep(self.on_delay)
        self.on = True
        sleep(self.on_debounce)

    self.lock.release()
```

The class `Lightswitch` inherits from the `Digital_Out` class, which in turn inherits from `GPIO` and eventually `Hardware`. This hierarchy of inheritance carries with it a progressive refinement of meaning about what this class does. The terms `off_debounce`, `on_delay`, and `on_debounce` are certainly not part of a controlled ontology, but the context of their use bounds their meaning. Rather than being bound by, for example, the abstract `Latency` term from interlex, we have defined terms that we need to make a hardware object do what we need it to. These terms don't have too much meaning on their own — there isn't even much in this class to uniquely identify it as a "lightswitch" beyond its name, it is just a timed digital output. What makes them meaningful is how they are used.

The way Autopilot handles various parameters are part of set of layers of abstraction that separate idiosyncratic logic from the generic form of a particular `Task` or `Hardware` class. The general structure of a two-alternative forced choice task is shared across a number of experiments, but they may have different stimuli, different hardware, and so on. Autopilot `Tasks` use abstract references to classes of hardware components that are required to run them, but separates their implementation as a system-specific configuration so that it's not necessary to have *exactly the same* components plugged into *exactly the same* GPIO pins, etc. Task parameters like stimuli, reward timings, etc. are similarly split into a separate task parameterization that both allow `Tasks` to be generic and make provenance and experimental history easier to track. `Task` classes can be subclasses to add or modify logic while being able to reuse much of the structure and maintain the link between the root task and its derivatives — for example one task we use that starts a continuous background sound but otherwise is the same as the root `Nafc` class. The result of these points of abstraction is to allow exact experimental replication on inexactly replicated experimental apparatuses.

This separation of the different components of an experiment is a balance between reusable code and clear metadata: we might allow freedom of terminology for each individual class, but by designing the system to encourage reuse of flexible classes we reduce the number of times unique terms need to be redefined. For example, we can imagine a trivial use of our lightswitch inside a task measuring an experimental subject's estimation of time intervals: we toggle the switch once some analog sensor reaches a certain threshold, and then the subject tries to press a button at the same time as the light turns on after a fixed delay. While this is very similar to how Autopilot currently works, note that we are using pseudocode to indicate how it might extend the system we're describing.

```python
from autopilot import Task
from autopilot.data.modeling import Field
from datetime import datetime, timedelta

class Controlled_Switch(Task):
    """
    A [[Discipline::Psychophysics]] experiment
    to measure [[Research Topic::Interval Estimation]].
    """

    class Params(Task.Param_Spec):
        on_delay: '@si:seconds' = Field(
            description="Delay (s) before turning light on",
            parameterizes="@jonny:hardware:Lightswitch")
        threshold: float = Field(
            description="Flick switch above this value",
            is_a="@interlex:Threshold")

    class TrialData(Task.TrialData_Spec):
        switch_time: datetime = Field(
            description="Time the switch was flicked")
        target_time: datetime = Field(
            description="Time the subject should respond")
        response_time: datetime = Field(
            description="Time the subject did respond")
        error: timedelta = Field(
            description="Difference between target and response",
            is_a="@psychophys:ReactionTime")

    HARDWARE = {
        'sensor': 'Analog_In',
        'button': 'Digital_In',
        'lightswitch': '@jonny:hardware:Lightswitch'
    }

    def __init__(self,
        on_delay:float,
        threshold:float):
        self.on_delay = on_delay
        self.threshold = threshold

        super(Controlled_Switch, self).__init__()
        self.poll()

    def poll(self):
        while self.running:
            if self.hardware['sensor'].value > self.threshold:
                self.hardware['lightswitch'].switch()
                switch_time = datetime.now()
```

```
        target_time = switch_time + self.on_delay

        # Wait for the subject to press the button
        response_time = self.hardware['button'].wait()

        # Send the data for storage
        self.node.send(key="DATA", value={
            'switch_time': switch_time,
            'target_time': target_time,
            'response_time': response_time,
            'error': target_time - response_time
        })
```

In this example, we first define a data model (see section 3.2 - Data in [232]) for the Tasks `Params`, the data that the task produces as `TrialData`, and the `HARDWARE` that the task uses. Our `Params` each have a type hint indicating what type of data they are, as well as a `Field` that gives further detail about them. Specifically, we have exposed the Lightswitch's `on_delay` parameter, indicated that it will be in seconds by referring to some namespace that defines SI units `@si` and that it parameterizes the lightswitch object that we defined above. The `TrialData` is similarly annotated, and by default Autopilot will use this specification to create an hdf5 table to store the values. The `HARDWARE` dictionary makes abstract references the hardware objects that will be made available in the task, each of which would have its configuration — which GPIO pin they are plugged into, the polarity of the signal, etc. — using some local system configuration. Finally, the single `poll()` method continuously compares the value of the sensor to the threshold, switches the lightswitch when the threshold is crossed, records the time the button was pressed, and sends it for storage with its network node.

As before, we are using our experimental framework as an interface to our linked data system. Currently, Autopilot uses a semantic wiki to organize technical knowledge and to share plugins - https://wiki.auto-pi-lot.com. In this case, I would write my task and hardware classes inside a git repository and then add them to Autopilot's plugin registry, which uses a form to fill in semantic properties and allows further annotation in free text and semantic markup.

We could instead imagine being able to document the task in its docstring, including describing the relevant subdiscipline, research topic, and any other relevant metadata. Rather than manually entering it in the wiki, then, we might export the triplet annotations directly from the class and make them available from my `@jonny` namespace and mirroring that to the wiki. Since the plugin specifies its dependencies using standard Python tools, it would then be possible for other researchers to use its task and hardware objects by referring to them as above.

In our pseudocode, the (abbreviated) exported metadata for this task might look like this:

```
<#tasks:Controlled_Switch>
    a @autopilot:Task

    hasDescription
        "A Psychophysics experiment
```

```
    to measure Interval Estimation."

Discipline "Psychophysics"
Research_Topic "Interval Estimation"

Params
   on_delay @si:seconds
      hasDescription " ... "
      parameterizes @jonny:hardware:Lightswitch
   ...

TrialData
   switch_time @python:datetime
   ...

usesHardware
   @autopilot:hardware:Analog_In
      hasID "sensor"
   @autopilot:hardware:Digital_In
      hasID "button"
   @jonny:hardware:Lightswitch
      hasID "lightswitch"
```

and we might combine it with metadata that describes our particular use of it like this, where we combine that task with a series of other `levels` that shape the behavior, make it more challenging, or measure something else entirely:

```
<#projects:my-project>
   a @autopilot:protocol
   experimenter @jonny
   ...

   level @jonny:tasks:Controlled_Switch
      on_delay 2
      threshold 0.5
      graduation @autopilot:graduation:ntrials
         n_trials 200

   level @jonny:tasks:Another_Task
      ...

   hardwareConfig
      button @autopilot:hardware:Digital_In
         gpioPin 17
         polarity 1
      sensor @autopilot:hardware:Analog_In
         usesPart @apwiki:parts:<Part_Number>
         ...
```

On the other side, our output data can be automatically exported to NWB[39]. Our experimental framework knows that data contained within a `TrialData` model is a

[39] Recall that we're using NWB for the sake of concreteness, but this argument applies to any standardized data format.

`@nwb:behavior:BehavioralEvents` object, and can combine it with the metadata in our task docstring and system configuration. If we needed more specific data export - say we wanted to record the timeseries of the analog sensor - we could use the same `is_a` parameter to declare it as a `@nwb:TimeSeries` and create an extension to store the metadata about the sensor alongside it[40].

So while our code is mildly annotated and uses a mixture of standard and nonstandard terminology, we make use of the structure of the experimental framework to generate rich provenance to understand our data and task in context. It's worth pausing to consider what this means for our infrastructural system as a whole

To start, we have a means of integrating our task with the knowledge that precedes it in the hardware and system configuration that runs it. In addition to documenting plugins, among others, the Autopilot wiki also has schema for custom built and off-the-shelf hardware alike like sensors and sound cards. These correspond to local hardware configuration entries that link them to the hardware classes required to use them[41]. That link can be used bidirectionally: metadata about the hardware used to perform an experiment can be used in the experiment and be included with the produced data data, but the data from experiments can also be used to document the hardware. That means that usage data like calibrations and part longevity can be automatically collected and contributed to the wiki and then used to automatically configure hardware in future uses. This makes using the experimental framework more powerful, but also makes building a communal library of technical knowledge a normal part of doing experiments. Though the wiki is a transitional medium towards what we will discuss in the next section, since contributions are tracked and versioned that allows a currently undervalued class of knowledge work to be creditable.

This gives us a different model of designing and engineering experiments than we typically follow. Rather than designing most of it from scratch or decoding cryptic methods sections, researchers could start with a question and basic class of experiment, browse through various implementations based on different sets of tools, see which hardware they and analogous experiments use, which is then linked to the code needed to run it. From some basic information researchers would then be most of the way to performing an experiment: clone the task, download the necessary system configuration information to set up the hardware, make incremental modifications to make the experiment match what they had designed, all the while contributing and being credited for their work.

Much of this is possible because of the way that Autopilot isolates different components of an experiment: hardware is defined separately from tasks, both are separate from their local configuration. In addition to thinking about how to design tools for our infrastructural system, we can also think of the way it might augment existing tools. Another widely used and extremely capable tool, Bonsai [233, 234], is based on XML documents that combine the pattern of nodes that constitute an experiment with specific parameters like a crop bounding box. That makes sharing and reusing tasks difficult without exactly matching the original hardware configuration, but we could use our metadata system to *generate* code for Bonsai in addition to consuming data from it. Given some schematic pattern of nodes that describes the operation of the experiment, we could combine that with the same notion of separable parameterization and hardware configurations as we might use in Autopilot to generate the XML for a bonsai workflow. As with analytical tools, our infrastructural system could be used to make a wide array of experimental tools interoperable with an evolving set of vernacular metadata schema.

Together, our data, experimental, and analytical infrastructures would dramatically reshape what is possible in science. What we've described is a complete provenance chain that can be traced from analyzed results back through to the code and hardware used to perform the experiment. Trivially, this makes the elusive workflow where experimental data is automatically scooped up and analyzed as soon as it is collected that is typically a hard-won engineering battle within a single lab the normal mode of using the system. Developing tools that give researchers control over the mode of exported data renders the act of cleaning data effectively obsolete. The role of our experimental tool is to be able to make use of collected technical knowledge, but also to lower the barriers to using the rest of the system by integrating it with normal experimental practice.

The effects on collaboration and metascience are deeper though. Most scientific communication describes collecting and analyzing a single dataset. Making sense of many experiments is only possible qualitatively as a review or quantitatively as meta-analysis. Even if we have a means of linking many datasets and analysis pipelines as in the previous section, the subtle details in how a particular experiment is performed matter: things as small as milliseconds of variation in valve timings through larger differences in training sequences or task design powerfully influence the collected data. This makes comparing data from even very similar experiments — to say nothing of a class of results from a range of different experiments — a noisy and labor-intensive statistical process, to the degree that it's possible at all. This system extends the horizon of meta-analysis to the experiment itself and turns experimental heterogeneity into a strength rather than a weakness. Is some result a byproduct of some unreported parameter in the experimental code? Is a result only visible when comparing across these different conditions? Individual experiments only allow a relatively limited set of interpretations and inferences to be drawn, but being able to look across the variation in experimental design would allow phenomena to be described in the full richness supported by available observations.

This would also effectively dissolve the "file drawer problem." [235, 236] Though malice is not uncommon in science, I think it's reasonably fair to assume that most researchers do not withhold data given a null result in order to "lie" about an effect, but because there is no reward for a potentially laborious cleaning and publication process. Collecting data that is clean and semantically annotated at the time of acquisition resolves the problem. Even without the analysis or paper, being able to index across experiments of a particular kind would make it possible to have a much fairer look at a landscape distorted by the publication process and prevent us from repeating the same experiments because no one has bothered to publish the null. This would also open new avenues for collaboration as we will explore in the next section.

To review:

We have described a system of three component modalities: **data, analytical tools, and experimental tools** connected by a **linked data** layer. We started by describing the need for a **peer-to-peer** data system that makes use of **data standards** as an onramp to linked metadata. To interact with the system, we described an identity-based linked data system that lets individual people declare linked data resources and properties that link to **content addressed** resources in the p2p system, as well as **federate** into multiple larger organizations. We described the requirements for **DAG-based analytical frameworks** that allow people to declare individual nodes for a processing chain linked to code, combine them into workflows, and apply them to data. Finally, we described a design strategy for **component-based experimental frameworks** that lets people specify experimental metadata, tools, and output data.

This system as described is a two-layer system, with a few different domains linked by a flexible metadata linking layer. The metadata system as described is not merely *inert* metadata, but metadata linked to code that can *do something* — eg. specify access permissions, translate between data formats, execute analysis workflows, parameterize experiments, etc. Put another way, we have been attempting to describe a system that *embeds the act of sharing and curation in the practice of science*. Rather than a thankless post-hoc process, the system attempts to provide a means for aligning the daily work of scientists so that it can be cumulative and collaborative. To do this, we have tried to avoid rigid specifications of system structure, and instead described a system that allows researchers to pluralistically define the structure themselves.

## 3.4   Shared Knowledge

> The Web is more a social creation than a technical one. I designed it for a social effect — to help people work together — and not as a technical toy. [...] We clump into families, associations, and companies. We develop trust across the miles and distrust around the corner. What we believe, endorse, agree with, and depend on is representable and, increasingly, represented on the Web. We all have to ensure that the society we build with the Web is of the sort we intend.
>
> Tim Berners-Lee (1999) *Weaving the Web* [188]

The remaining set of problems implied by the infrastructural system sketched so far are the *communication* and *organization* systems that make up the interfaces to maintain and use it. We can finally return to some of the breadcrumbs laid before: the need for negotiating over distributed and conflicting data schema, for incentivizing and organizing collective labor, and for communicating information within and without academia.

The communication systems that are needed double as *knowledge organization* systems. Knowledge organization has the rosy hue of something that might be uncontroversial and apolitical — surely everyone involved in scientific communication wants knowledge to be organized, right? The reality of scientific practice might give a hint at our naïveté. Despite being, in some sense, itself an effort to organize knowledge, *scientific results effectively have no system of explicit organization*. There is no means of, say, "finding all the papers about a research question."[42] The problem is so fundamental it seems natural: the usual methods of using search engines, asking around on Twitter, and chasing citation trees are flex tape slapped over the central absence of a system for formally relating our work as a shared body of knowledge.

Information capitalism, in its terrifying splendor, here too pits private profit against public good. Analogously to the necessary functional limitations of SaaS platforms, artificially limiting knowledge organization opens space for new products and profit opportunities. In their 2020 shareholder report, RELX, the parent of Elsevier, lists increasing the number of journals and papers as a primary means of increasing revenue [49]. This represents a shift in their business model from subscriptions to deals like open access, which according to RELX CEO Erik Nils Engström "is where revenue is priced per article on a more explicit basis" [238].

[42] Also see Eve Marder's recent short and characteristically refreshing piece which in part discusses the problem of keeping up with scientific literature the context of maintaining the joy of discovery [237].

In the next breath, they describe how "in databases & tools and electronic reference, representing over a third of divisional[43] revenue, we continued to drive good growth through content development and enhanced machine learning [ML] and natural language processing [NLP] based functionality."

[43] RELX is a huge information conglomerate, and scientific publication is just one division.

What ML and NLP systems are they referring to? The 2019 report is a bit more revealing (emphases mine):

> Elsevier looks to enhance quality by building on its premium brands and **grow article volume** through **new journal launches,** the expansion of open access journals and growth from emerging markets; and add value to core platforms by implementing capabilities such as **advanced recommendations on ScienceDirect and social collaboration through reference manager and collaboration tool Mendeley.**
>
> **In every market, Elsevier is applying advanced ML and NLP techniques** to help researchers, engineers and clinicians perform their work better. For example, in research, ScienceDirect Topics, a free layer of content that enhances the user experience, uses **ML and NLP techniques to classify scientific content and organise it thematically,** enabling users to get faster access to relevant results and related scientific topics. The feature, launched in 2017, is proving popular, generating 15% of monthly unique visitors to ScienceDirect via a topic page. **Elsevier also applies advanced ML techniques that detect trending topics per domain,** helping researchers make more informed decisions about their research. **Coupled with the automated profiling and extraction of funding body information from scientific articles,** this process supports the whole researcher journey; from planning, to execution and funding. [239]

Reading between the lines, it's clear that the difficulty of finding research is a feature, not a bug of their system. Their explicit business model is to increase the number of publications and sell organization back to us with recommendation services. The recommendation system might be free[44], but the business is to maintain the self-reinforcing system of prestige where researchers compete for placement in highly visible journals to stand out among a wash of papers, in the process reifying the mythology [240] of the "great journals." With semantic structure to locate papers, it becomes much more difficult to sell high citation count as a product — people can find what they need, rather than needing to pay attention to a few high-profile journals. Without it, which papers might a paper discovery system created by a publisher recommend? The transition from a strictly journal-based discovery system to a machine learning powered search and feed model mirrors the strategic displacement of explicit organization by search in the rest of the digital economy, and presents similar opportunities for profit. Every algorithmically curated feed is an opportunity to sell ad placement[45] — which they proudly describe as looking very similar to their research content [243, 171].

[44] "free"

[45] a strategy that the reprehensible digital marketing disciplines call "native advertising" [241, 242]

The extended universe of profitmaking from knowledge disorganization gets more sinister: Elsevier sells multiple products to recommend 'trending' research areas likely to win grants, rank scientists, etc., algorithmically filling a need created by knowledge disorganization. The branding varies by audience, but the products are the same. For pharmaceutical companies "scientific opportunity analysis" promises custom reports that answer questions like "Which targets are currently being studied?" "Which experts are not collaborating with a competitor?" and "How much funding is dedicated to a particular area of research, and how much progress has been made?" [244]. For academics, "Topic Prominence in Science" offers university administrators tools to "enrich strategic research planning with portfolio overviews of their own and peer institutions." Researchers get tools to "identify experts and potential cross-sector collaborators in specific Topics to strengthen their project teams and funding bids and identify Topics which are likely to be well funded." [245] This reflects RELX's transition "from electronic reference, information reference tools, databases to [...] analytics and decision tools." [238] Publishing is old news, the real money is in tools for extending control through the rest of the process of research.

These tools are, of course, designed for a race to the bottom — if my colleague is getting an algorithmic leg up, how can I afford not to? Naturally only those labs that *can* afford them and the costs of rapidly pivoting research topics will benefit from them, making yet another mechanism that reentrenches scientific inequity for profit. Knowledge disorganization, coupled with a little surveillance capitalism that monitors the activity of colleagues and rivals [24, 246], has given publishers powerful control over the course of science, and they are more than happy to ride algorithmically amplified scientific hype cycles in fragmented research bubbles all the way to the bank.

One more turn of the screw: the ability of the (former) publishers to effectively invent the metrics that operationalize "prestige" in the absence of knowledge organization systems gives them broad leverage with governments and funding agencies. In an environment of continuously dwindling budgets and legislative scrutiny, seemingly mutually beneficial platform contracts offer the sort of glossy comfort that only predictive analytics can. In 2020 the National Research Foundation of Korea (NRF) and Elsevier published a joint report that used a measurement derived from citation counts - "Field-weighted citation impact", or FWCI - to argue for the underrated research prestige of South Korea [247]. While I don't dispute the value of South Korea's research program, the apparent bargain that was struck was chilling. South Korea gets a very fancy report arguing that more scientists in other countries should work with theirs, and Elsevier gets to cement itself into the basic operation of science. Elsevier controls the journals that can guarantee high citation counts *and* the metrics built on top of them. The Brain Korea program Phase II report [46] [248], issued just before the 2009 formation of the NRF argued that rankings and funding should be dependent on citation counts. The NRF now relies on SciVal and their FWCI measurement as a primary means of ranking researchers and determining funding, built into the Brain Korea 21 funding system [249, 250]. Without exaggeration, scientific disorganization and reliance on citation counts allowed Elsevier to buy control over the course of research in South Korea.

The consequences for science are hard to overstate. In addition to literature search being an unnecessarily huge sink of time and labor, science operates as a wash of tail-chasing results that only rarely seem to cumulatively build on one another. The need to constantly reinforce the norm that purposeful failure to cite prior work is research misconduct is itself a symptom of how engaging with a larger body of work is both extremely labor intensive and *strictly optional* in the communication regime of journal publication. The combination of more publications translating into more profit and the strategic disorganization of science contributes to conditions for sci-

[46] the result of another corporate collaboration with the Rand corporation.

entific fraud. An entirely fraudulent paper can be undetectable even by domain experts. Since papers can effectively be islands — given legitimacy by placement in a journal strongly incentivized to accept all comers — and there is no good means of evaluating them in context with their immediate semantic neighbors, investigating fraud is extremely time consuming and almost entirely without reward. And since traditional peer review happens once, rather than as a continual public process, the only recourse outside of posting on PubPeer is to wait on journal editorial boards to self-police by reviewing each individual complaint. Forensic peer-reviewers have been ringing the alarm bell, saying that there is "no net" to bad research [251], and brave and highly-skilled investigators like Elisabeth Bik have found thousands of papers with evidence of purposeful manipulation [252, 253]. The economic structure of for-profit journals pits their profit model against their function as providing a venue for peer review — the one function most scientists are still sympathetic to. Trust in science is critical for addressing our most dire problems from global pandemics to climate change [254], but attitudes towards scientists are lukewarm at best [255]. Even when it isn't fake news, why would anyone trust us when it's *effectively impossible* to find or assess the quality of scientific information? [256] Not even scientists can: despite the profusion of papers, by some measures progress in science has slowed to a crawl [257].

While Chu and Evans [257] correctly diagnose *symptoms* of knowledge disorganization like the need to "resort to heuristics to make continued sense of the field" and reliance on canonical papers, by treating the journal model as a natural phenomenon and citation as the only means of ordering research, they misattribute root *causes*. The problem is not people publishing *too many papers,* or a *breakdown of traditional publication hierarchies,* but the *staggering profitability of knowledge disorganization.* Knowledge disorganization is precisely the precondition of information-as-capital and the outcome of its concentration by our century's robber barons (see [258]). Their prescription for "a clearer hierarchy of journals" misses the role of organizing scientific work in journals ranked by prestige, rather than by the content of the work, as a potentially major driver of extremely skewed citation distributions. It also misses the publisher's stated goals of *publishing more papers* within an ecosystem of algorithmic recommendations, and there is nothing recommendation algorithms love recommending more than things that are already popular. Without diagnosing knowledge disorganization as a core part of the business model of scientific publishers, we can be led to prescriptions that would make the problem worse.

It's hard to imagine an alternative to journals that doesn't look like, well, journals. While a full treatment of the journal system is outside the scope of this paper, the system we describe here renders them *effectively irrelevant* by making papers as we know them *unnecessary.* Rather than facing the massive collective action problem of asking everyone to change their publication practices on a dime, by reconsidering the way we organize the surrounding infrastructure of science we can flank journals and replace them "from below" with something qualitatively more useful.

Beyond journals, the other technologies of communication that have been adopted out of need, though not necessarily design, serve as desire paths that trace other needs for scientific communication. As a rough sample: Researchers often prepare their manuscripts using platforms like Google Drive, indicating a need for collaborative tools in preparation of an idea. When working in teams, we often use tools like Slack to plan our work. Scientific conferences reflect the need for federated communication within subdisciplines, and we have adopted Twitter as a de facto platform for socializing and sharing our work to a broader audience. We use a handful of blogs and other sites like OpenBehavior [259], Open Neuroscience, and many others to index technical knowledge and tools. Last but not finally, we use sites like PubPeer and ResearchGate for comment and criticism.

These technologies point to a few overlapping and not altogether binary axes of communication systems.

- **Durable vs Ephemeral** - journals seek to represent information as permanent, archival-grade material, but scientific communication also necessarily exists as contextual, temporally specific snapshots.

- **Structured vs Chronological** - scientific communication both needs to present itself as a structured basis of information with formal semantic linking, but also needs the chronological structure that ties ideas to their context. This axis is a gradient from formally structured references, through intermediate systems like forums with hierarchical topic structure that embeds a feed, to the purely chronological feed-based social media systems.

- **Messaging vs Publishing** - Communication can be person-to-person, person-to-group with defined senders and recipients, or person-to-all statement to an undefined public. This ranges from private DMs through domain-specific tool indexes like OpenBehavior through the uniform indexing of Wikipedia.

- **Public vs. Private** - Who gets to read, who gets to contribute? Communication can be composed of entirely private notes to self, through communication in a lab, collaboration group, discipline, and landing in the entirely public realm of global communication.

- **Formal vs. Informal** - Journal articles and encyclopedia-bound writing that conforms to a particular modality of expression vs. a vernacular style intended to communicate with people outside the jargon culture.

- **Push vs. Pull** - Do you go to get information from a reference location, or does information come to you as an alert or message? Or, generally, where is the information "located," is an annotation pushed and overlaid on a document, or stored elsewhere requiring the audience to explicitly pull it?

"Peer reviewed vs. unrefereed" is purposely excluded as an axis of communication tools, as the ability to review and annotate multiple versions of a document — subject to the context of the medium — should be a basic part of any communication system. Fear over losing the at once immutable but also paradoxically fragile ecosystem of journal-led peer review is one of the first strawmen that stops consideration of radically reorganizing scientific communication[47]. The belief that peer review as we know it is an intrinsic part of science is ahistorical (eg. [261]), and the belief that journal-led peer review is somehow a unique venue for evaluating scientific work ignores the immense quantity of criticism and discussion that happens in almost every communicative context, scientific and otherwise. The notion that the body of scientific knowledge is best curated by passing each paper through a gauntlet of three anonymous reviewers, after which it becomes Fact is ridiculous on its face. Focusing on preserving peer review is a red herring that unnecessarily constrains the possible forms of scientific communication. Instead we will try and sketch systems

[47] For a recent example, see the responses to Dan Goodman's argument why he has stopped doing pre-publication peer review altogether [260] ]

that address the needs for communication and knowledge organization left unmet precisely because of the primacy of peer reviewed journal publications.

Clearly a variety of different types of communication tools are needed, but there is no reason that each of them should be isolated and inoperable with the others. We have already seen several of the ideas that help bring an alternative into focus. Piracy communities demonstrate ways to build social systems that can sustain distributed infrastructure. Federated and protocol-based systems show us that we don't need to choose between a single monolithic system or many disconnected ones, but can have a heterogeneous space of tools linked by a basic protocol. The semantic web gives us the unfilled promise of triplet links as a very general means of structuring data and building interfaces for disparate systems. We can bridge these lessons with some from wiki culture to get a more practical sense of distributed governance and organization. Together with our sketches of data, analytical, and experimental tools we can start imagining a system for coordinating them — as well as displacing some of the more intractable systems that misstructure the practice of science.

### 3.4.1   The Wiki Way

If we take radical collaboration as our core, then it becomes clear that extending Wikipedia's success doesn't simply mean installing more copies of wiki software for different tasks. It means figuring out the key principles that make radical collaboration work. What kinds of projects is it good for? How do you get them started? How do you keep them growing? What rules do you put in place? What software do you use? [262]

So that's it — insecure but reliable, indiscriminate and subtle, user hostile yet easy to use, slow but up to date, and full of difficult, nit-picking people who exhibit a remarkable community camaraderie. Confused? Any other online community would count each of these "negatives" as a terrible flaw, and the contradictions as impossible to reconcile. Perhaps wiki works because the other online communities don't. [263] and in WhyWikiWorks

Aside from maybe the internet itself, there is no larger public digital knowledge organization effort than Wikipedia. While there are many lessons to be learned from Wikipedia itself, it emerged from a prior base of thought and experimentation in radically permissive, self-structuring read/write — sometimes called "peer production" [264] — communities. Wikis are now quasi-ubiquitous[49], perhaps largely thanks to Wikipedia, but its specific history and intent to be an *encyclopedia* entwines it with a very particular technological and social system that obscures some of the broader dreams of early wikis.

Aaron Swartz recounts a quote from Jimmy Wales, co-founder of Wikipedia:

"I'm not a wiki person who happened to go into encyclopedias," Wales told the crowd at Oxford. "I'm an encyclopedia person who happened to use a wiki." [265]

And further describes how this origin and mission differentiates it from other internet communities:

[48] Interestingly, this quote is almost, but not exactly the same as that on Ward's wiki: "So that's it - insecure, indiscriminate, user-hostile, slow, full of difficult, nit-picking people, and frivolous. Any other online community would count each of these strengths as a terrible flaw. Perhaps wiki works because the other online communities do not." I can't tell if Ward Cunningham wrote the original entry in the wiki, but in any case seems to have found a bit of optimism in the book.

[49] though their corporate manifestations would probably be unrecognizable to the project early wiki users imagined.

But Wikipedia isn't even a typical community. Usually Internet communities are groups of people who come together to discuss something, like cryptography or the writing of a technical specification. Perhaps they meet in an IRC channel, a web forum, a newsgroup, or on a mailing list, but the focus is always something "out there", something outside the discussion itself.

But **with Wikipedia, the goal is building Wikipedia.** It's not a community set up to make some other thing, it's a community set up to make itself. And since Wikipedia was one of the first sites to do it, we know hardly anything about building communities like that. [262]

We know a lot more now than in 2006, of course, but Wikipedia still has outsized structuring influence on our beliefs about what Wikis can be. Wikipedia has since spawned a large number of technologies and projects like Wikidata and Wikimedia Commons, each with their own long and occasionally torrid histories. I won't dwell on the obvious and massive feat of collective organization that the greater Wikipedia project represents — we should build on and interoperate with its projects and respect the amount of work the foundation and its editors have put in to preserve free access to information, but learning from its imperfections is more useful to us here, especially for things that aren't encyclopedias. The dream of a centralized, but mass-edited "encyclopedia of everything" seems to be waning, and its slow retreat from wild openness has run parallel to a long decline in contributors [264, 266]. Throughout that time, there has been a separate (and largely skeptical) set of wiki communities holding court on what a radically open web can be like, inventing their worlds in real time. These communities have histories that are continuous with one another, and in their mutual reaction and inspiration sometimes teach similar lessons from across the divides of their very different structure.

The first wiki was launched in 1995[50] (it's still up) and came to be known as Ward's wiki after its author WardCunningham. Technically, it was extremely simple: a handful of TextFormattingRules and use of WikiCase where if you JoinCapitalizedWords you create a link to a (potentially new) WikiPage — and the ability for anyone to edit any page. These very simple WikiDesignPrinciples led to a sprawling and continuous conversation that spanned more than a decade and thousands[51] of pages that, because of the nature of the medium, is left fully preserved in amber. Those conversations are a history of thought on what makes wiki communities work (eg. WhyWikiWorks, WhyWikiWorksNot), and what is needed to sustain them.

One tension that emerged early without satisfying resolution is the balance between "DocumentMode" writing that serves as linearly-readable reference material, similar to that of Wikipedia, and "ThreadMode" writing that is a nonlinear representation of a conversation. Order vs spontaneity is a fundamental challenge of inventing culture in plaintext. The purpose of using a wiki as opposed to other technologies that existed at the time like bulletin boards, newsgroups, IRC, etc. was that it provided a means of fluid structure[52]. The parallel need to communicate and attribute work made it a seeming inevitability that even if you went out of your way to restructure a lot of writing into a sensible DocumentMode page, someone would soon after create a new horizontal divider and start a fresh ThreadMode section.

Ward Cunningham and other more organizationally-oriented contributors opposed ThreadMode (eg. ThreadModeConsideredHarmful, InFavorOfDissertation) for a number of reasons, largely due to the ThreadMess and WikiChaos it had the potential of creating.

[50] it's complicated: WardsWikiTenthAnniversary

[51] 23,244 unique page names according to the edit history, but the edit history was also purposely pruned from time to time.

[52] Giving a means of organizing the writing of the Portland Pattern Repository was the reason for creating Ward's Wiki in the first place.

> I occasionally suggest how this site should be used. My GoodStyle suggestions
> have been here since the beginning and are linked from the edit page should
> anyone forget. I have done my best to discourage dialog InFavorOfDissertation
> which offers a better fit to this medium. I've been overruled. I will continue to
> make small edits to pages for the sake of brevity. – WardCunningham [267]

Most pages are thus a combination of both, usually with some DocumentMode
text at the top with ThreadMode conversations interspersed throughout without
necessarily having any clean delineation between the two. Far from just being raw
disorder, this mixed mode of writing gave it a peculiar character of being *both* a folk
reference for a library of concepts *as well as* a history of discussion that made the
contingency of that reference material plain. Beka Valentine put it well:

> c2wiki is an exercise in dialogical methods. of laying bare the fact that knowledge
> and ideas are not some truth delivered from On High, but rather a social process,
> a conversation, a dialectic, between various views and interests [268]

This tension and its surrounding discussions point to the need for multiple repre-
sentations of a single idea: that both the social and reference representations of a
concept are valuable, but aren't necessarily best served by being represented in the
same place. There was relatively common understanding that the intended order of
things was to have many ThreadMode conversations that would gradually be con-
verted to DocumentMode in a process of BrainStormFirstCleanLater. Many pro-
posed solutions orbit around making parallel pages with similar names (like <page-
name>Discussion) to clean up a document while preserving the threads (though
there were plenty of interesting alternatives, eg. DialecticMode)[53].

Wikipedia, in its attendant WikiEngine MediaWiki, cut the Gordian Knot by split-
ting each page into a separate *Article* and *Talk* pages, with the talk page in its own
**Namespace** – eg. Gordian_Knot vs Talk:Gordian_Knot. Talk pages resemble a lot
of the energy of early wikis: disorganized, sometimes silly, sometimes angry, and
usually charmingly pedantic. Namespaces extend the traditional "everything is a
page" notion encoded in the WikiCase link system by giving different pages differ-
ent roles. In addition to having parallel conversations on articles and talk pages, it is
possible to have template pages that can be included on wiki pages with `{{double
curly bracket}}` syntax – eg. Template:Citation_Needed renders `{{Citation needed}}`
as [citation needed]. Talk pages have their own **functional differentiation,** with
features for threading and annotating discussions that aren't present on the main
article pages (see Wikipedia:Flow [269]). Generalized beyond the context of wikis,
functional differentiation of a single item into its multiple representations is rela-
tively common in computing: eg. this document exists as a git repository, the ren-
dered page, a pdf, hypothes.is annotations, etc.

The complete segregation of discussion to Talk pages is driven by Wikipedia's ex-
clusivity as an encyclopedia, with reminders that it is the "sole purpose" peppered
throughout the rules and guidelines. The presence of messy subjective discussions
would of course be discordant with the very austere and "neutral" articles of an en-
cyclopedia. There are no visible indications that the talk pages even exist in the main
text, and so even deeply controversial topics have no references to the conversations
in talk pages that contextualize them — despite this being a requested feature by
both administrators and editors [270].

Talk pages serve as one of the primary points of coordination and conflict resolu-tion on Wikipedia, and also provide a low-barrier entrypoint for questions posed to a space they perceive to be "an approachable community of experts" [271]. The separation of Talk pages and the labyrinthine rules governing their use function to obscure the dialogical and collective production of knowledge at the heart of wikis and Wikipedia. The body of thought that structures Wikipedia, most of which is in its Wikipedia:* namespace, is immense and extremely valuable, but is largely hidden except from those who care to look for it. Since Wikipedia is "always already there" often without trace of its massively collective nature, relatively few people ever contribute to it. Reciprocally, since acknowledging personal contribution is or point of view is explicitly against some of its core policies and traditions, there is little public credit outside the Wikipedia community itself for the labor of maintaining it.

[53] Contemporary wikis have continued this
conversation, see DocumentsVsMessages on
communitywiki.org

The forking of Wards Wikis into the first SisterSites teaches a parallel strain of lessons. Ward's Wiki started as a means of organizing knowledge for the Portland Pattern Repository[54], a programming community (referred to as DesignPatterns below), and in 1998 they were overwhelmed with proponents of ExtremeProgramming (or XP), which caused the first fissure in the wiki:

> XP advocates seemed to be talking about XP at every possible opportunity and seemingly on every page with content the least bit related to software development. This annoyed a number people who were here to discuss patterns, leading to the tag XpFreeZone, as a request not to talk about ExtremeProgramming on that page.

> It was difficult to pick out the DesignPatterns discussion on RecentChanges[55], because most of the activity was related to ExtremeProgramming. Eventually, most of the DesignPatterns people left, to discuss patterns in a "quieter" environment, and people started referring to this site as WardsWiki instead of the PortlandPatternRepository [267]

One of the first and most influential Sister Sites was Meatball Wiki, described on Wards Wiki:

> SunirShah founded MeatballWiki to absorb and enlarge the discussion of what wiki and wiki like sites might be. That discussion still simmers here. But here it can take on a negative tone sounding more like complaining. On meatball, under Sunir's careful leadership, the ideas, wild or not, stay amazingly upbeat. - SisterSites

MeatballWiki became the spiritual successor to Ward's Wiki, which at that point had its own momentum of culture less interested in being the repository of wiki thought[56]. Meatball has its own prolific history of thought, including reflections on its very existence as a SisterSite. These were a series of discussions that melded thoughts from open source computing to social systems; in part: RightToFork, RightToLeave, EnlargeSpace, and TransClusion.

What can be done when the internal divisions in a wiki community and the weight of its history make healthy contribution impossible? The simplest is to exercise the RightToLeave, as it is almost always possible to just stop being part of a digital community. This approach is clearly the most destructive, as it involves abandoning the emotional bonds of a community, prior work (see the WikiMindWipe where a user left and took all their contributions with them), and doesn't necessarily provide an alternative that alleviates the cause of the tension. The next idea is to *fork* the community, where its body — in the case of wikis the pages and history — can be duplicated so that it can proceed along two parallel tracks. Exercising the right to fork

---

[54] The initial motivations are actually stunningly close to the kinds of communication and knowledge organization problems we are still solving today (even in this piece)

"Cunningham had developed a database to collect the contributions of the listserv members. He had noticed that the content of the listserv tended to get buried, and therefore the most recent post might be under-informed about posts which came before it. The way around this problem was to collect ideas in a database, and then edit those ideas rather than begin anew with each listserv posting. Cunningham's post states that "The plan is to have interested parties write web pages about the People, Projects and Patterns that have changed the way they program. Short stories that hint at patterns are welcome too." As to the rhetorical expectations, Cunningham added "The writing style is casual, like email or netnews, but doesn't have to be so repetitive since the things being discussed don't disappear. Think of it as a moderated list where anyone can be moderator and everything is archived. It's not quite a chat, still, conversation is possible." - [272] is, according to Meatball, "people exercising their RightToLeave whilst maintaining their emotional stake" [273].

[55] Recent Changes was the dominant, if not controversial means of keeping track with recent wiki traffic, see RecentChangesJunkie

[56] There seems to have been an overriding belief that theoretical ideas about wikis and wiki culture belong on Meatball Wiki, from WikiWikiWebFaq: > Q: Do two separate wikis ever merge together to create one new wiki? Has this happened before? Keep in mind that I don't just mean two different pages within a wiki. (And for that matter, where is an appropriate page where I can post questions about the history of all wikis, not just this one?) > > A1: I don't know of any such wiki merge, nor of any discussion of the history of all wikis. Such a discussion should probably reside (if created) on MeatballWiki.

The discussion around the Right to Fork on Meatball is far from uniformly positive, and is certainly colored by the strong presence of its BenevolentDictator Sunir Shah who viewed it as a last resort after all attempts at ConflictResolution have failed. They point to the potentially damaging effects of a fork, like bitterness, disputes over content ownership (see MeatballIsNotFree), and potentially an avoidance of conflict resolution that is a normal and healthy part of any community. Others place it more in the realm of a radical *political* action rather than a strictly social action. Writing about the fork of OpenOffice to LibreOffice, Terry Hancock writes:

> [In] proprietary software [a] political executive decision can kill a project, regardless of developer or user interest. But with free software, the power lies with the people who make it and use it, and the freedom to fork is the guarantee of that power. [...] The freedom to fork a free software project is [a] "tool of revolution" intended to safeguard the real freedoms in free software. [274]

Forking digital communities can be much less acrimonious than physically-based communities because of the ability to EnlargeSpace given by the medium:

> In order to preserve GlobalResources, create more public space. This reduces limited resource tension. Unlike the RealWorld, land is cheap online. In effect, this nullifies the TragedyOfTheCommons by removing the resource pressure that created the "tragedy" in the first place. **You can't overgraze the infinity.** - [275]

Enlarging space has the natural potential to make the broader social scene bewildering with a geyser of pages and communities, but can be made less damaging by having mechanisms to link histories, trace their divergence, and potentially resolve a fork as is common in open source software development. Forking is then a natural process of community regeneration, allowing people to regroup to make healthier spaces when needed, where the fork is itself part of the history of the community rather than an unfathomable rift.

Forking communities is not the same as forking community resources: "you can't fork a community [...] what you can do is fork the content and to *split* the community" [276]. As described so far, a fork divides people into unreconciled and separate communities. In some cases this makes forking difficult, in others it makes it impossible: the prime example, again, is Wikipedia. It is simply too large and too culturally dominant to fork. Even though it is technically possible to fork Wikipedia, if you succeeded, then what? Who would come with you to build it, and who would that be useful for? This is partly a product of its totalizing effort to be an encyclopedia of everything (what good would *another* encyclopedia of everything be?) but also the weight of history: you won't get enough long-encultured Wikipedians to join you, and you probably won't be able to recruit a new generation of them on your own.

The last major effort to fork Wikipedia was in 2002 with an effort led by Edgar Enyedy to move the Spanish Wikipedia to The Enciclopedia Libre Universal en Español [277, 278]. Though it was brief and unsuccessful, Enyedy claims that because Jimmy Wales was worried about other non-English communities following their lead, he and the other admins capitulated to the demands for no advertising and a transfer to a .org domain, among others[57]. Even a politically symbolic fork is de-pendent on the perceived threat to the original project, and that window seems to have been closed after 2002.

[57] Jimmy Wales, naturally, disputes this characterization of events.

The cultural tensions and difficulties that lead other wikis and projects to fork have taken their toll on the editorship and culture of Wikipedia. The community is drawn into dozens of conflicting philosophical camps: the Deletionists[58] vs. the Inclusionists, Eventualists vs. Immediatists, Mergists vs. Separatists, and yes even a stub page for Wikisecessionism. Editorship has steadily declined from a peak in 2007. Its relatively invisible community systems make it mostly a matter of chance or ideology that new contributors are attracted in the first place. In its calcification of norms, largely to protect against legitimate challenges to the integrity of the encyclopedia, any newcomers that do find their way into editing now have little chance to catch a foothold in the culture before they are frustrated by (sometimes algorithmic) rejection [264, 266].

Arguably all internet communities have some kind of life cycle, so the question becomes how to design systems that support healthy forking without replicating the current situation of fragmentation. Wikis, including Meatball and MediaWiki, as well as other projects like Xanadu often turn to **transclusion** — or being able to reference and include the content of one wiki (or wiki page) in another. Rather than copying and pasting, the remote content is kept updated with any changes made to it.

Transclusion naturally brings with it a set of additional challenges: Who can transclude my work? Whose work can I transclude? Can my edits be propagated back to their work? What can be transcluded, at what level of granularity, and how? While before we had characterized splitting communities as an intrinsic part of a fork, that need not be the case in a system built for transclusion. Instead relationships post-fork are then made an *explicit social process* within the system, where even if a community wants to work as separate subgroups, it is possible for them to arrive at some agreement over what they want to share and what they want to keep separate. This kind of decentralized work system resembles radical organizing tactics like affinity groups where many autonomous groups fluidly work together or separately on an array of shared projects without aspiring to create "one big movement" [279]. Murray Bookchin describes:

> The groups proliferate on a molecular level and they have their own "Brownian movement." Whether they link together or separate is determined by living situations, not by bureaucratic fiat from a distant center. [...]
>
> [N]othing prevents affinity groups from working together closely on any scale required by a living situation. They can easily federate by means of local, regional or national assemblies to formulate common policies and they can create temporary action committees (like those of the French students and workers in 1968) to coordinate specific tasks. [...] As a result of their autonomy and localism, the groups can retain a sensitive appreciation of new possibilities. Intensely experimental and variegated in lifestyles, they act as a stimulus on each other as well as on the popular movement. [280]

To cherrypick a few lessons from more than 25 years of thought from tens of thousands of people: The differing models of document vs. thread modes and separate article vs. talk pages show us that using **namespaces** is an effective way to bridge multimodal expression on the same topic across perceived timescales or other conflict-ing communicative needs. This is especially true when the namespaces have **functional differentiation**[59] like the tools for threading conversations on Wikipedia Talk pages and the parsing and code generation tools of Templates. These namespaces need to be **visibly crosslinked** both to preserve the social character of knowledge work, but also to provide a means of credit assignment and tool development between namespaces. Any communication system needs to be designed to **prioritize ease of leaving** and **ease of forking** such that a person can take their work and represent it on some new system or start a new group to encourage experimentation in governance models and technologies. One way of accomplishing these goals might be to build a system around **social transclusion** such that work across many systems and domains can be linked into a larger body of work without needing to create a system that becomes too large to fork. The need for communication across namespaces and systems, coupled with transclusion further implies the need for **bidirectional transclusion** so that in addition to being able to transclude something in a document, there is visible representation on the original work being transcluded (eg. commented on, used in an analysis, etc.) by allowed peers and federations.

These lessons, coupled with those from private bittorrent trackers, linked data communities, and the p2p federated system we have sketched so far give us some guidelines and motivating examples to build a varied space of communication tools to communicate our work, govern the system, and grow a shared, cumulative body of knowledge.

### 3.4.2  Rebuilding Scientific Communication

It's time to start thinking about interfaces. We have sketched our system in turtle-like pseudocode, but directly interacting with our linking syntax would be labor intensive and technically challenging. Instead we can start thinking about tools for interacting with it in an abstract way. Beneath every good interface we're familiar with, a data model lies in wait. A .docx file is just a zipped archive full of xml, so a blank word document that contains the single word "melon" is actually represented (after some preamble) like:

```
<w:body>
   <w:p>
      w14:paraId="0667868A"
      w14:textId="50600F77"
      w:rsidR="002B7ADC"
      w:rsidRDefault="00A776E4">
      <w:r>
          <w:t>melon</w:t>
      </w:r>
   </w:p>
</w:body>
```

Same thing with jupyter notebooks, where a block of code:

```
>>> rating = 100
>>> print(f'I rate this dream {rating}')
```

```
'I rate this dream 100'
```

is represented as JSON (simplified for brevity):

```
{
  "cell_type": "code",
  "id": "thousand-vermont",
  "outputs": [{
    "name": "stdout",
    "output_type": "stream",
    "text": [
      "I rate this dream 100\n"
    ]
  }],
  "source": [
    "rating = 100\n",
    "print(f'I rate this dream {rating}')"
  ]
}
```

So we are already used to working with interfaces to data models, we just need to think about what kind of interfaces we need for a scientific communication system.

Let's pick up where we left off with our linked data and tools. Recall that we had a `project` named `#my-project` that linked an experiment, a few datasets that it produced, and an analysis pipeline that we ran on it. We *could* just ship the raw numbers from the analysis, wash our hands of it, and walk straight into the ocean without looking back, but usually scientists like to take a few additional steps to visualize the data and write about what it means.

To explore the communicative tools that might be useful, we can start by considering traditional documents, and attempt to generalize them by separating their form as "units" or "cells" of information with accompanying metadata from their representation in interfaces for interacting and communicating about them.

**Documents & Notebooks**

Say we have reached the stage where we are writing a brief summary of our experiment and analysis, but not yet at the stage of writing a "formal" scientific paper. We might do so in a notebook-like [281] environment with different kinds of "cells," specifically cells that execute *code* and cells that render *markdown*. We want to plot some of the results of our analysis, so to do that we might load the data and use matplotlib [282] to make our point (Fig. 3.4)

# Mathematical foundations of smiling

In my experiment, I have demonstrated that smiling is a simple matter of making the mouth into an arc, having eyes, and optionally a nose.

**Figure 3.4:** An example notebook where we use some data loading framework that embeds the linked dataset in the loading cell's metadata field and then plot it!

```python
from p2p_system import get_data
import matplotlib.pyplot as plt
```

```python
x, y, sizes = get_data('@jonny:my-project:Analysis1')
```

```
Downloading dataset @jonny:my-project:Analysis1
--------------------
  0%|          | 0/100 [00:00<?, ?it/s]
```

```python
plt.scatter(x, y, s=sizes)
```

```
<matplotlib.collections.PathCollection at 0x12826dee0>
```

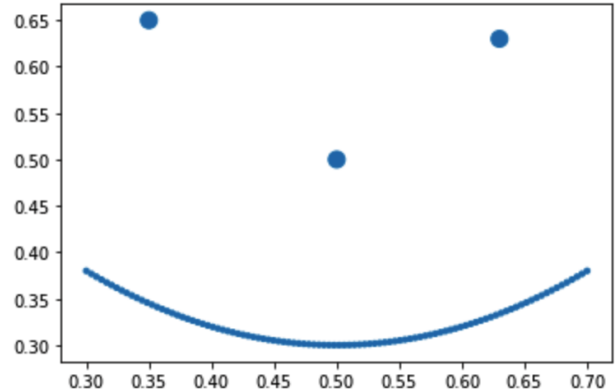

QED.

Our notebook file would then include an array of JSON objects that describe the contents of its cells. For example, our data loading cell would look something like this:

```json
{
    "cell_type": "code",
    "execution_count": 2,
    "id": "rapid-information",
    "metadata": {
     "scrolled": true
    },
    "outputs": [
     " ... "
    ],
    "source": [
     "x, y, sizes = get_data('@jonny:my-project:Analysis1')"
    ]
}
```

The "outputs" description has been abbreviated above, but it describes to the jupyter notebook server how to display it. Regular text piped through stdout is represented like this:

```json
{
    "name": "stdout",
    "output_type": "stream",
    "text": [
     "Downloading dataset @jonny:my-dataset\n",
     "--------------------\n"
    ]
}
```

And multiple output types can be combined in a single cell, for example a widget like our loading progress bar is described like this:

```json
{
    "data": {
     "application/vnd.jupyter.widget-view+json": {
      "model_id": "5799ac2959084a4596ffbad3f9940f48",
      "version_major": 2,
      "version_minor": 0
     },
     "text/plain": [
      "  0%|          | 0/100 [00:00<?, ?it/s]"
     ]
    },
    "metadata": {},
    "output_type": "display_data"
}
```

where the model_id, version_major, and version_minor describe which rendering code to use for the cell, similarly to the "metadata that indicates code" that we discussed in analytical frameworks.

Notice that there is already a metadata field! In order to link our notebook to our analysis — and thus to our extended graph of data, experiment, etc. — we could do it manually, but since we're thinking about interfaces we can also imagine that our `p2p_framework` is capable of filling it in for us. We don't need to invent a new metadata protocol for JSON, JSON-LD is already quite similar to the syntax we've been using already. For simplicity, say we use a `@comms` ontology to denote various features of our communication system. Our data loading function might then populate a field in our cell like this:

```
"metadata": {
    "scrolled": true,
    "@comms:usesData": "@jonny:my-project:Analysis1"
}
```

Other frameworks might make their own metadata annotations, like an indication that we're plotting some feature of the data, or performing some statistical analysis on the data. These annotations might be responsive to the parameterization of the function call or its results, but if we emphasize a design process that makes interfaces at multiple levels we could also imagine using something like iPython "magic commands" to declare metadata for our cell. For example, each cell is automatically assigned a random combination of words as an ID, but if we wanted to be able to specifically refer to a cell we could give it an explicit one:

```
%%meta @comms:cellID smilePlot
plt.scatter(x, y, s=sizes)
```

We're familiar with two types of cells, code and markdown, but we can extend our thinking to arbitrary cell types. What is a cell? A cell has a a) **type** that indicates its capabilities and representation, b) **metadata** that describes it, we can also generalize that to include *arguments* that parameterize it, and c) the content, or information contained by the cell. The jupyter document model more or less reflects this already, but in its base model only has code, markdown, and raw cell types, and the metadata field is unstructured. Its extension system allows for additional cell types as well as restructuring the program more generally, but since we're focused on self-contained documents we'll limit our discussion to additional cell types.

From this it's relatively trivial to imagine additional cell types that serve common needs in academic writing: a citation cell type[60] that takes a BibTeX object (or its fields) as arguments and then preserves the full metadata as well as renders it in a chosen style. A figure cell type that takes an image or plot and a caption. A contributor cell type that takes an author's name, affiliation, ORCID, email, and so on. Currently jupyter extensions use the NPM registry, but we could imagine being able to use other people's cell types directly by referring to them like `@jonny:celltypes:citation`.

Notebooks have multiple levels of metadata, so we can also specify document-level metadata that describe the type of our document (like a `@schema:ScholarlyArticle`), its `creativeWorkStatus` as a `Draft`, our authorship information, permissions, and whatever else we'd like. But what is a document? In the case of our jupyter notebook, it's a series of cell descriptions in a JSON array. Trivially, a document is a cell that contains other cells. What about in the other direction? The contents of our cells are *also* a cell-like system. The very notion of a programming language is a means of mapping structured syntax to machine instructions, and to do that code (in some languages) is interpreted or compiled by parsing it into an abstract syntax tree that relates data and its structuring metadata. Markdown can also be thought of as a series of subcells, where using a `#` header indicates how the text is to be represented as compared to `*italic*` text or `[links](https://link.com)`. The use of a programming language or markup syntax is represented by the `cell_type` field, which the notebook server knows to translate `"code"` to mean Python and `"markdown"` to mean its particular flavor of markdown (of which there are several).

[60] The original Jupyter Notebook paper describes the need for this near the end [281].

This points towards a model of **recursive** cells that can contain other cells. An editor could, for example, draw from templating engines like liquid, where an abstract representation of the content of a cell could include a `{{ content }}` marker that indicates that additional cells can be included inside of it. Recursive models, coupled with structuring metadata that indicates the relationship between a parent and child cell could then be used to model compound concepts. Another simple example using citation might be to have a cell with one child cell containing a reference to another work that ours `@cito:disagrees_with` [283], and another child cell that in turn contains some writing in markdown and a plot. Recursive cells also naturally lend themselves to **transclusion** by making each of the individual subcomponents of a document referenceable with full granularity. We will expand on both compound concepts and transclusion in a moment in talking about the extension of our cellular system to wikis.

Before we go beyond a document system that would be unrecognizable to most scientists, and thus yet another nice pipedream, it's important to pause on the continuity with existing document systems. Microsoft Word, or Word-like WYSIWYG editors like LibreOffice or Google docs are the dominant mode of preparing academic documents. Word-like editors are *already* create recursive cell-like documents, though their interface obscures them. They support semantic markup like heading styles (though their use compared to manual formatting is far from universal [284]), and every paragraph can be considered a cell, with the default paragraph styling as its metadata and additional styled elements like bolded words as sub-cells. It should then be possible to import existing word documents into a cellular document system. Care should also be taken to smooth the cognitive transition from word-like editors: Jupyter currently treats cells as being strictly separate, and new cells need to be created manually. Instead it should be possible for cells to "recede into the background" and be created with common gestures like a double return to make a new paragraph. The "insert" menu used to create things like tables or images is already a familiar tool in word-like editors, so the notion of adding elaborated types like citations shouldn't be that big of a lift.

The other major document preparation tool modalities are markup syntaxes and their associated builders like LaTeX. Though TeX-like tools have an exceedingly opinionated and obscure design history [285], they have three major affordances: 1) document-level structure provided by document classes, packages, and the options they provide, 2) environments that enclose some section of text between `\begin{}` and `\end{}` and provide some specific functionality or formatting like lists, and 3) commands that accept arguments and modify some smaller unit of text like creating a link with `\href{https://url.com}{link text}`. Each of these maps onto a cellular document system, with document-level metadata or the templates commonly used to render markdown, and cells that take arguments to approximate environ-ments and commands. Markdown extensions like MyST [286] make this translation even more straightforward with direct analogies to LaTeX commands and environments and their "role" and "directive" counterparts in reStructuredText. Since the goal should be a 1:1 relationship between source code and visual editor, the difference between representing a cell visually versus in markup should be left as a matter of author preference.

Bidirectional translation from a WYSIWYG editor to its markup is not a trivial task — the mediawiki team started writing theirs in 2011 and rolled it out as a default feature in 2020 [287]. It's a careful balance between ease of use, power of syntax, and accomodation of historical usage patterns. Markdown is on one extreme of ease with only a handful of markup elements to learn, but has a relatively steep learning curve to do anything more complex. On the other end is the wonderful dokieli [288] (and see Sarven's masterpiece [289], spiritual cousin to this document), which does essentially everything that we want our linked documents to do, but requires authors to write their documents in HTML and manually manage the semantic markup. Extending Notebooks to use recursive cells with reusable types sacrifices some of the ability to directly edit the source of a document as a potential way to balance familiarity and expressiveness.

Notebooks, with some architectural and interfaces then become a straightforward way of breaking up the scientific paper as a singular unit of knowldge work when embedded in a linked data system. Their use in scholarly publishing has been proposed many times before, but our linking system lets us resolve some of the largest outstanding limitations [290] : dependency management [291], archiving [292], and discovery, among others. The same gradient of access control rules we discussed in controlling access to sensitive data would support a process of gradual publication of smaller units of work, from a private demo in our lab meeting to a public part of scientific discourse.

What happens when we invite other people to respond?

**Forums & Feeds**

What if we think of our documents as "threads" and their cells as "posts?" What makes a cellular document a document is some (relatively arbitrary) notion of a 'root' cell that contains the others — ie. for notebooks a JSON array of cells. That could be trivially reformulated as cells with metadata indicating that they are `PartOf` a document, each indicating their `position` or linked to the cells they are before and after. If we also allow cells to be `inReplyTo` each other, we have the basis of a threaded communication system continuous with documents. Where cells in a linear document have at most one preceding and succeeding cell, multiple replies allow a tree structure that maps onto the patterns of most contemporary social media. Metadata that describes category and content extends this to include the structure of forums, and could be the basis of a rich continuum of media spanning order and chaos, permanence and ephemerality, between the *magnum opus* and the shitpost: media absent but sorely needed in academic communication.

Traditional forums like phpBB and contemporary social media operate from a single host with a fixed interface and representation of posts. What would a communication system that decouples hosting, identity, interface, and format look like? We can draw inspiration from the "fediverse," a collection of interoperable software platforms and protocols. The fediverse makes it possible to communicate across rad-ically different interfaces: someone using Funkwhale, which resembles music soft-ware like spotify, can communicate with people on PeerTube, a p2p video stream-ing program like YouTube, and Mastodon, a microblogging medium like Twitter. Rather than a single host, instances of each of these programs are hosted indepen-dently and can choose to federate with other instances to enable communication between them. Most of these programs use the ActivityPub [202] protocol, which defines a standard set of capabilities for client-server and server-server communica-tion.

Mastodon posts (or "toots") already resemble the kind of document-interoperable medium hinted at above. For example this post is represented in (abbreviated) JSON[61]:

```
{
  "to":[
    "https://www.w3.org/ns/activitystreams#Public"
  ],
  "cc":[
    "https://social.coop/users/jonny/followers"
  ],
  "id": "107328829457619549",
  "created_at": "2021-11-23T22:52:49.044Z",
  "in_reply_to_id": "107328825611826508",
  "in_reply_to_account_id": "274647",
  "visibility": "public",
  "url": "https://social.coop/@jonny/107328829457619549",
  "content": "<p>and making a reply to the post to show the in_reply_to and context fields</p>",
  "account":
  {
    "id": "274647",
    "username": "jonny",
    "fields":
    [ ... ]
  },
  "media_attachments": [],
  "mentions": [],
  "tags": [],
}
```

[61] Leaving in this string escaping its box because I think it's sort of cute

As described previously, ActivityPub supports linked data with JSON-LD – a re-markable feat despite the justifiable angst with the protocol [293, 75] given the his-torical grudges between linked data and indieweb communities (See this retrospec-tive by one of its authors, Christine Lemmer-Webber [69]). So we could imagine that post using a reference to a document or one of its cells in its `in_reply_to` field.

Mastodon might be a good transitional medium, but we can extend it to make use of our linked p2p system. The fediverse decouples the network from a single platform, but instances still bundle together the underlying data of a post with an interface, host, and account (but see hubzilla). p2p helps us decouple accounts from hosts (see this discussion on a p2p ActivityPub [294]), but we would also like to decouple interfaces from the underlying data so that we have a continuous communication medium where different interfaces are just *views* on the data. To do that we would want to start by replacing Mastodon's flat "`content`" field with the kind of typed cells in our documents that indicate what kind of message they are. For example a simple text-based message might use the ActivityStreams `Note` type:

```
{
  "@context": "https://www.w3.org/ns/activitystreams",
  "type": "Note",
  "name": "My Message",
  "content": "A note I send to you!"
}
```

But we might equivalently send a `@jupyter:Notebook` as a message, or some compound object like a `Collection`:

```
{
  "@context": "https://www.w3.org/ns/activitystreams",
  "summary": "A Compound Message!",
  "type": "Collection",
  "totalItems": 2,
  "items": [
    {
      "type": "Note",
      "name": "Hey how ya doin here's a notebook"
    }
    {
      "@context": "https://jupyter.com/",
      "type": "Notebook",
      "content": " ... "
    },
  ]
}
```

So the *existence* of a particular type of message is not bound to the ability of any given program's ability to render it. Our notebook program might not be able to understand what it means to have people responding to and making threads about its cells, but we would still be able to receive them and open them with an interface that does, and we could further imagine the ability for a type to recommend a program to us for rendering it as we did with the ability for analysis nodes to specify the code to execute them. We will set aside for a moment the issues of moderation and permission for which messages can link to our work and the practicalities of sending, receiving, storing, and serving messages and return to them in the context of <span style="color:red">annotations</span> and <span style="color:red">trackers</span>, respectively.

*Where* do our posts go? For concreteness, we can start with a forum called "NeuroChat." `@neurochat` is a peer like any other, and it supports some of the basic ActivityStreams vocabulary. We can request to join it by sending a `@as:Join` request, which gives it permission to index our public posts and issue links on our behalf through its web interface. It has a few broad categories like "Neuromodulation" and "Sensory Neuroscience," within which are collections of threads full of chronologically-sorted posts. Threads are objects that indicates a category like

`@neurochat:categories:Neuromod`, and when we post in them we create links that
are `@as:attributedTo` us with the `@as:context` of the thread we're posting in and
any `@as:inReplyTo` links to preceding or quoted posts.

We want to announce and describe some recent results in our document `@jonny:my-project:Writeup`.
This kind of post is common in `@neurochat`, and so instead of a generic citation we
use a `@neurochat:AnnouncesResult` link to indicate the relevant document. In our
forum pseudocode we'll use a `#prefix` macro to give a short name to our project
and semantic wikilinks with a `[[predicate :: object]]` syntax for the purpose of
demonstration, though ideally these would be part of the forum's interface. We
think we really have something that challenges some widely held previous results:

```
#prefix project @jonny:my-project
#prefix nc @neurochat

Hi everyone, happy to present my new work
[[nc:AnnouncesResult :: project:Writeup]].

I think it raises a number of interesting questions,
in particular @rival's longstanding argument
[[@cito:disputes :: @rival:TheBrainIsInTheLiver]].

I also wonder what this means about the conversation
we've been more generally about
[[@cito:discusses :: @discipline:whereAreTheOrgans]].

Anyway, write back soon, xoxo.
```

Our rival takes the criticism in stride but wants to run their own analysis. They
follow the links back to find our data, and reanalyze it. Their analysis framework
has already issued a link indicating that it reanalyzes our data, and rather than do an
independent writeup our rival returns to the thread to continue the discussion.

```
Interesting result, you old scoundrel.

That indeed [[disputes :: @doi:<id>]],
in particular its section [[.:results:main]]
and my re-analysis adds another wrinkle to the problem!
Take a look:

[[nc:embed :: @rival:reanalysis]]

This really complicated another project of mine,
[[@rival:projects:NeuronsCanSwim]]
```

Our forum's `embed` link knows how to embed the notebook our rival used to do
their reanalysis and in the underlying message indicates the the current version so
if they update it in the future the message will still be comprehensible. Our rival
doesn't use a predicate for their link to their side-project and our forum uses its
default `Mentions` predicate. It's still more informative than a duplet link because
the context of being a discussion in our forum the links in the surrounding posts.

We could imagine additional capabilities we give to our forum, like the ability to automatically trigger a re-analysis by someone mentioning a different pipeline for a given dataset, but we'll leave those as an exercise to the reader.

This example is a relatively trivial instance of scientific communication: sharing results, relating them to previous findings, and thinking about the broader implications on the field. However in our current regime of scientific communication, even in the most progressive publication venues that allow communication directly on a work, this kind of communication is *entirely invisible* to the broader state of our understanding. With our system of linked communication, however, the entire provenance chain from our experiment through its analysis and contextualizing discussion is related to immediately related work as well as the standing questions in our field. Our work is enriched by the additional analysis from our rival, and their work is continuously contextualized as the state of our understanding develops. We were capable of making incremental refinements to our shared understanding using units of work that were much smaller than the traditional scientific paper. It would be possible for someone entirely outside our field to browse through the general links from basic research questions to relevant work and its surrounding discussion. If they were to ask questions, our answers would represent the latent diffusion of understanding to other disciplines based on the graph context of our respective work — and we could be credited the time we spent doing so! In short, scientific communication could actually be *cumulative.*

Forums are just one point in a continuous space of threaded media. If we were to take forum threads out of their categories, pour them into our water supply, and drink whatever came our way like a dog drinking out of an algorithmic fire hydrant, we would have Twitter. Remove the algorithm and arrange them strictly chronologically and we have Mastodon. In both, the "category" that organizes threads is the author of the initial post. Algorithmic, rather than purposefully organized threaded systems have their own sort of tachycardic charm. They are effective at what they aim to do, presenting us whatever maximizes the amount of time we spend looking at them in a sort of hallucinatory timeless now of infinite disorganization — at the expense of desirable features of a communication system like a sense of stable, autonomously chosen community, perspective on broader conversation, and cumulative collective memory.

Nevertheless the emergence of a recognizable "Science Twitter" points towards a need for relatively informal all-to-all communication. Serendipitously being able to discover unlikely collaborators or ideas is a beautiful dream, if one ill-served by the for-profit attention economy. Our formulation of the `@neurochat` forum was as an equal peer that mirrored, collected, and organized posts that otherwise are issued from other peers such as ourselves. In the same way that we might use the ActivityStreams `Join` action to have our posts mirrored by it, we might also use `@as:Follow` to receive posts from any peer, and in the case of a federation that might include posts from its members sent to the federation. Notice in the example mastodon post above how it uses JSON-LD and the activitystreams ontology: a "me to the world" tweetlike message is addressed to `activitystreams#Public` and cc'd to the URL that corresponds symbolically to the list of `@jonny`'s followers.

We can take advantage of the graph structure and rich metadata of our social network in ways that are impossible in corporate social media networks that require the expectation of disorder to be able to sell "native" ad placement. The instance-to-instance federation model of the fediverse, and the accompanying absence of any "global" scope of all posts, results in the need for multiple views on the network: in Mastodon, a "local" timeline that shows only posts from within the host instance, and a "federated" timeline that shows posts from all instances that the host instance has federated with. Since our network allows peer-to-peer, federation-to-federation, and peer-to-federation interaction, we can extend that further. We can construct views of the network based on granular control over graph depth: instead of seeing just the posts from the peers that we follow, we can request to see n-depth posts, from the peers that our peers follow, and so on. This could be done at the level of a "view" or at the level of the follow link itself — since I know this person well, I want to see a graph depth of 2 from them, and a depth of 1 from others. At the federation level, we might imagine that @neurochat is federated with another @linguisticsChat group and the two mirror and rehost each other's posts. We could then make use of our extended social graph and prioritize posts from people who are part of overlapping subsets of the federations we are a part of. The peer-based nature of our social network serves as the basis for a system of fluid scoping and filtering of the kind of communication we are looking for at any given time. So rather than a disorganized public melee or the empty rooms and new logins from yet another closed Slack, our communication could be part of a coherent scientific conversation.

Across from filtering what we receive, the same could be done to what we send by choosing where our posts are addressed and who can see them. The same multi-modality of "following" used to indicate the graph depth of the posts we see could let us indicate different kinds of relationships. We should be able to send global, undirected messages on a public feed, but we don't necessarily want to talk to our friends in the same way that we talk to strictly professional colleagues. We might want to organize privately with a few colleagues, or prevent trolls or hostile groups from accessing or making use of our work. Effectively, we should be able to direct our messages to different groups of peers to support the multiple registers of our communication.

The need for rapid and informal scientific communication being mediated by corporate social networks has the unfortunate byproduct of needing to carefully manage a "personal brand." To be seen as a "serious," we need to maintain some proximity to the stilted academic voice, forfeiting any approachability to science that might be gained from public communication. If we are to expand the scope of what we consider as the labor of scientific communication, we should also take seriously its many registers and contexts. Informal media like alt accounts, mailing lists, groupchats, zines, and whisper networks are also an integral part of science, particularly for marginalized and vulnerable scientists [295]. Parallel to organizing our communication in empirical professional communication, we might build systems that support our organization into federations to more effectively bargain over our working conditions and protect ourselves. The venues that organize our communication being limited to journals, and the accompanying regulation over the registers of communication that count as "real" science, is even more limiting than its profound effects on scientific literature proper. The absence of infrastructure to support the multiregister communication of science limits our ability to organize over the broader state of our work, form extended communities, and reduces what should be the collective project of making our work broadly understandable to the individualistic projects of "scicomm influencers." It shouldn't take a lot of additional critical analysis to say "shitposts are good, actually, for science."

There's a balance to be struck between a system of granular control over the messages we send and receive with the ease of a monolithic algorithmic feed. Mastodon sorts all posts purely chronologically, which translates into relatively steep limits on the size of communities as feeds become unintelligible washes of posts. Instead of forgoing algorithmic organization altogether, another means by which we could take advantage of the graph structure of our network is by being able to *choose* the sorting algorithms we use. We might want to prioritize posts from someone who we don't necessarily follow but is interacting with people that we do in contexts that we share, or be able to deprioritize posts that are "close" to us in our social graph in order to discover new things. This too could be a cumulative, community-driven project, where we might want to try out our friend's `@friends:sorting:NewAlgorithm`, tweak it a bit for our preferences, and republish a new version.

Generally, the impact of having a communication system that decouples hosting, identity, interface, and format on an underlying linked data graph gives us a broad space to build different views and tools to use the underlying data. Specifically, without predicting the infinite future of communication media, our system of linked, cell-like communication generalizes threadlike media like forums and feeds into a continuous system that can blend their features as needed. Durable, cumulative discussion about the state of our understanding should be able to live side-by-side with ephemeral, informal conversations. It should be possible for us to serendipitously discover people and information as well as for a newcomer to have a place to ask questions and build their understanding. It should be possible for us to form and dissolve communities fluidly without substantial technical start-up costs and the total loss of memory when they close. A system that supports the fullness of continuous communication would be an unfathomably richer way of building reliable, accessible, and multivalent understanding of our reality than the current system of a gladiatorial thumbs up/down indictment on years of your life that is journal-based peer review.

### Overlays & Adversarial Interoperability

We can't expect the entire practice of academic publishing to transition to cell-based text editors in a p2p linked data swarm overnight. In the same way that we discussed frameworks for integrating heterogeneous analytical and experimental tools, we need some means of **bridging** communication tools and **overlays** for interacting with existing communication formats. There are many examples of bridging communication protocols, eg. the many ways to use Matrix with Slack, email, Signal, etc. The overlays for websites, pdfs, and other more static media that we'll discuss are means to bring them into the system whether they support it or not: our interoperability should be willing to be adversarial if it needs to be [296, 297]. In representing the intrinsically interactive and social nature of reading (eg. see [298]), overlays as interfaces also supplement the "horizontal" connections between cells by injecting information into them or transcluding it elsewhere: creating a fuzzy boundary between writing *on* something *vs about* something.

We don't need to look far to find a well-trod interface for annotation overlays for document-like media: the humble highlighter. Hypothes.is, enabled on this page, lets readers highlight and annotate any webpage with a browser extension or javascript bookmarklet. This interface is a near match to the highlighting and review tools of Microsoft Word and Google Docs used for the same purpose. At its heart is a system for making anchors, references to specific places in a text, and the means of matching them even when the text changes or the reference is ambiguous [299]. For example, this anchor has three features, a `RangeSelector` that anchors it given the position within the paragraph, an absolute `TextPositionSelector`, and a contextual `TextQuoteSelector` that you can see with an API call. Anchors like these, along with references to the code that resolves them, could be the objects to which we could link from the rest of our communication system.

On its own, it serves to give a `Talk:` page to every website. With an integration into a system of linked data and identity, it also serves as a means of extending the notion of bidirectional transclusion described above to work that is not explicitly formatted for it. Most scientific work is represented as `.pdfs` rather than `.html` pages, and hypothes.is already supports annotating PDFs. With an integration into pdf reading software, for example Zotero's PDF reader, there would be a relatively low barrier to integrating collaborative annotation into existing workflows and practices.

Digital publishing makes imagining the social regulation of science as a much more broadly based and continuous process much easier, but the problem of moderation remains (as it has since at least the coiner of the terms "Gold" and "Green" open access lost faith in ahierarchical scientific communication after someone said poo-poo words at him on Internet while defending the use of they/them as gender-ambiguous pronouns [300, 301, 302]). Some movement has been made towards public peer review: eLife has integrated hypothes.is since 2016 [303], and bioRxiv had decided to integrate it as well in 2017 [304] before getting cold feet about the genuinely hard problem of moderation (among others [305]) and instead adopting the more publisher-friendly TRiP system of refereed peer-reviews [306].

Overlays raise basic questions about control over the representation of our work, about who is able to write what on it. As with potential incompatibility between interfaces, we should be able to control what comments appear *on* our work, but there is no way to control – even in our current communication systems – what someone says *about* it. Our system gives us some ability to identify bad actors and regulate the avenues of communication without overcorrecting into a system where criticism becomes impossible – even if we don't want to represent someone's comments on our work, it is possible to make them and for others to find them, but it's also possible to contextualize their context if they're made in bad faith.

Though a description of the norms and tools needed to maintain healthy public annotation is impossible here, our system *provides a space for having that conversation.* Authors could, for example, allow the display of annotations from a professional society like @sfn that has a code of conduct and moderation team, or annotations associated with comments on @pubpeer, or from a looser organization of colleagues and other @neurofriends. Conversely, being able to make annotations and comments from different federations gives us a rough proxy to different registers of communication and preserves the plurality of our expression. Social tools like these are in the hypothes.is team's development roadmap, but I intend it as a well-developed and mature example of a general type of technology[62] rather than a recommendation.

In addition to annotating other works, overlays can come in the form of bots or other tools for interacting with existing systems in a way that's compatible with a new one. One particularly impressive example of aggressive interoperability in this domain is Eduardo "flancian" Ivanec's agora [307, 308]. An agora is a wiki-like project with pages (or nodes) for each named concept, but it also allows for mul-tiple representations of a given node: so notes from multiple people across multiple mediums will be present on the same page. Accompanying the agora is the anagora bot (on Mastodon and Twitter), which makes links to, and backlinks from pages mentioned as `[[wikilinks]]` by accounts that follow them (for example: a post, the bot's reply, and one of the linked pages, `[[wikilinks everywhere]]`). This becomes natural quickly: it's common for people associated with the agora (or flancians) to speak with wikilinks, or index links and conversations that they come across for mutual discovery.

---

[62] cf. the genius.com overlay.

The agora makes linked annotation a basic part of using the web without requiring fundamental changes in communication practices. The agora is an exercise in radically permissive protocol-like thinking: rather than creating a new app or platform, theoretically any bot could be made to crawl different mediums for wikilinks and index them. It illustrates that interfaces can precede formal protocols and serve as a testing and development ground for them.

Another bridging overlay for more author-focused scientific communication would be to explicitly archive the threads that increasingly serve as companions to published work — or original works of scholarship on their own (eg. [309]). I have started experimenting with this with the `@threaddodo_bot`, a bot that converts a thread to a PDF[63] and uploads it to Zenodo when it is tagged beneath one. This bot is being programmed as a generalizable framework for bots that can accept parameterized commands. For example, someone can set their authorship information by tweeting "identify" at threaddodo, which accepts a series of key-value pairs to set your name, affiliation, and orcid. Future versions will support automatic reference generation for linked works, including previously archived threads, as well as setting prefixes for OWL schema for use in semantic `[[predicate::object]]` wikilinks.

[63] complete with markdown rendering!

When some recognizably different communication medium begins to coalesce, it should support bidirectional *crossposting* to and from existing mediums. Crossposting substantially eases transition — for example between Twitter and Mastodon — as patterns of usage that have been trained for years on hyperoptimized attention-capturing platforms are hard to break. Together with bridges, bots, and overlays for annotation, linking, and archiving, the dream of rewriting the norms of academic communication looks less like some "if you build it they will come" pipe dream and more like a transitional period of demonstrating what we can dream of together. Adversarial interoperability not only *works* [297], it's also a gift of the hacker mindset that teaches us how to make building a better world an act of unrepentant *joy*.

### Trackers, Clients, & Wikis

The final set of social interfaces are those for collective governance of the system. So far we have generalized documents "vertically" into recursive typed cells, "horizontally" into linked cells for communication, and then blurred their independence by and extended them into incompatible media with overlays. The remaining piece we need are multi-authored documents: **wikis**. We'll pick up the threads left hanging from our description of bittorrent trackers and knit them in with those from the wiki way to describe how systems for surfacing procedural and technical knowledge work can also serve as a basis of searching, indexing, and governing the rest of the system. Where the rest of our interfaces were means of creating particular kinds of structured links, we'll also describe wikis as a means of interacting directly with links to negotiate the relationships between the multiplicity of our folksonomic schema.

In the process we'll give some structure to the **clients** and **trackers** that serve and organize them.

Our notion of recursive cell-like documents is already a good basis for wiki pages. **Multi-author** documents should already be possible with a permission system that we have invoked previously to limit read access, and so the most radically open, publicly editable wikis would just have edit permissions open to anyone. The **version history** that makes the notion of SoftSecurity possible should also be a general property of links in our system. The other concept we'll borrow from traditional wikis is the model where **pages represent topics.** Practically, let's suppose this means that within documents beneath some namespace like `@jonny:wiki`, we can make wikilinks to [[New Pages]] that imply links to `@jonny:wiki:New_Pages` — though for the sake of simplicity in this section we will assume that our wiki starts at the root of the `@jonny` namespace.

We want to preserve two types of multiplicity: the multiplicity of *representations* (as in `Talk:` pages) and *instances* of a given topic, or the ability for multiple peers to have linked versions that potentially transclude content from other peers, but are ultimately independent. Both can use different components of a namespace: for multiplicity of representation we might follow the example of mediawiki and use parallel namespaces like `@jonny:talk`, and multiplicity of instances follows naturally from parallel peers by linking `@jonny:wiki:My_Page` to `@rumbly:wiki:My_Page`.

Wikis that represent multiple instances of a given page are already a subject of active experimentation. Flancian's Agora is one example, which is based on markdown files in git repositories, and markdown files with the same name in federated repositories are presented on the same page. A much older project[64], everything2 is built around multiple "writeups" for a given "node." Multiple instances of a page are also a defining feature of Ward Cunningham's federated wiki, which has a vertical "strip" based interface where clicking the colored squares at the bottom of a given page will open another strip to show another user's instance of the page. We'll borrow Ward's terminology and refer to this kind of wiki as a federated wiki.

Federated wikis already have some broader purchase as "personal knowledge graphs," [310] where people use tools like Notion or Obsidian to keep a set of linked, semistructured personal notes. Rather than thinking of a wiki as wikipedia, with pages that aspire to be uniformly named and written, personal knowledge graphs take whatever form is useful to the person maintaining them. This maps neatly onto our namespaces and recursive documents as a means of *organizing our system of links.*

Say we have a very simple project structure that consists of a dataset with two tables and a document with the date of the experiment and some short description of the data. In our pseudocode:

```
<#project>
   a @jonny:Project

   dataset
     @format:csv
       table1
       table2

   document
```

```
    a @jupyter:notebook

    @schema:Date dateCollected
    Description
      "This is the data that I collected"
```

This has a natural representation in our wiki as a set of nested cells: the `@jonny:project` page has two child cells, one for the dataset and one for the document, which have their own child cells that represent the tables, date, and description according to their types. Since the relationships between our cells can also typed, ie. have an associated predicate like `before`, `after`, or `inReplyTo`, we'll use two additional types to differentiate nested cells:

- `child` (and its inverse `parent`) cells correspond to a cell's position in our namespace, so we could find our data at `@jonny:project:dataset`.

- `transcludes` (and its inverse `transcluded`) indicates some other cell that we represent on a given wiki page, as we might want to do if we wanted to embed one of our plots in a post.

- And other cells linked with bare `[[wikilinks]]` are untyped.

This gives us a bidirectional representation of our link structure: and with it an interface for browsing and managing all the various types of objects that we have described so far.

Since schemas, or abstract representations of the links a type might have, are themselves made of links, these too can be managed with a wiki. Semantic mediawiki and its page schemas extension implement a system like this. For example, the Autopilot wiki has a form to submit build guides for experimental apparatuses. Build guides have a schema and an associated template that lays out the form input on the created page and makes the semantic wikilinks that declare its properties like `[[Is Version::2]]`.

This system is semantically rich while also being flexible, as everything reduces down to semantic wikilinks on a page, so free text can be used fluidly along with structured schemas, forms, and templates. The wide open structuring space of the wiki handles the messy iteration of technical knowledge work well while also having enough structure to be computer readable. A page for an amplifier makes the datasheet, serial protocol, and the GPIO pins it needs available via an API call while also carrying on a continuous effort to crudely defeat its low-pass output filter. A plugin page can credit the papers it was used in by DOI and the python packages needed to run it while also describing how to void the warranty of your oscilloscope to unlock additional functionality.

The page-centric model of semantic wikis poses a problem, though. The guide for building the Autopilot Behavior Box has semantic annotations describing the CAD schematics, materials, and tools it uses. This works fine for other assembled parts or schematics like 3d printed parts that have pages of their own, because their pages can contain the additional properties that describes them like the associated `.stl` files. Materials like screws are trickier. Each screw varies along about a dozen dimensions, and so that either requires making a separate page for each individual screw or use workarounds[65] that reduce the maximum depth of representation to

[65] like subobjects or record types

two layers and add other nasty complexities.

A recursive cellular system avoids these problems and provides a uniform interface to complex representations. We can create schema for experiments that allow for a build guide, which can contain assembled component descriptions, which can contain materials, etc. When using that schema to describe a new experiment, the researcher can be prompted for any of the possible available fields in the recursive model while also allowing for free space to write in the semi-structure of the building blocks. Extending an existing schema is just a matter of transcluding it and then modifying it as needed. With the ability for our interface to assign fixed IDs for these objects or generate unique hashes based on their contents, the tension of ephemeral object declaration with unique addresses disappears.

The tension of arbitrarily flexible personal knowledge graphs with multiscale organization with other peers remains, though. Approaching from the other side of discovery, rather than declaration of information leads back to considering the structure of our p2p client and tracker-like systems. The most immediate problem we face is the need to reconcile the differences between multiple instantiations of overlapping representations of concepts that change through time. That sounds a lot like version control system, and a VCS like git or mercurial should be a natural part of our client. Where IPFS is "a single bittorrent swarm, exchanging objects within one Git repository," [107] we make a mild modification and think of a single bittorrent swarm with a git repository per peer (also see IPLD [311]). Git stores files as content-addressed "blobs" of binary indexed by "trees" that represent the file hierarchy [312]. Our client can do something similar, except using the triplet link structure for trees rather than typical duplet links. Another peer querying our data would then resolve our identity to the top of the tree, our client would then either serve the parts of our tree that the peer has access to or else let them traverse some subsection of it, and they could then request any file "blobs" that the tree points to[66].

By itself this would have a lot of overhead as a large number of peers would need to be queried to find a particular subset of matching metadata. We can mediate that in a few ways. First, our clients could take advantage of the embedded social network to cache and rehost other peer's trees — either in their entirety or as shards distributed among other peers — depending on our relationship to them. Second, when making links, we could notify relevant and subscribed peers that we have made it (eg. see [315]). Combined with distributed caching, that would allow the peer responsible for the schema to direct queries to peers already known to have a particular kind of file: eg. the @nwb peer could track when @nwb datasets are declared.

We don't necessarily *want* to have an entirely autonomous protocol though, following the example of wikis and bittorrent trackers we want social systems for shared governance and maintenance of the system. Trackers first serve the technical need of indexing a particular community's data, eg. as @dandihub does with @nwb, in case peers go offline. We don't want to just track datasets, however, we want to track the many different kinds of metadata in our swarm. The second role of trackers is collective curation and negotiation over schema.

Say a group of my colleagues and I organize to set up a server as our tracker. As an interface, our tracker might allow us to browse schemas as a tree. For a given node, we might see "horizontally" across all the schemas that have modifications or extensions to that node, and "vertically" up and down their parent and children nodes. We notice that our colleague has made an extension to a schema that looks very similar to ours. We do a `diff` to see which nodes are similar and which are different between our schema. Both of us have some good ideas that the other doesn't have, so we open a conversation thread by creating a node that references both of our schemas as candidates for merging and send it to our colleague. We negotiate over a way to resolve their differences, similar to a pull request, and then `merge` them. Part of our merging process is indicating how to change either of our existing structures to become the third merged structure, so our clients are able to handle those changes for us and the update propagates through the network.

[66] That's sufficient detail for a sketch, but there is of course a great deal of subtlety that would need to be resolved in an implementation. For example, see [313, 314].

As our tracker grows and maybe even becomes the de-facto tracker for our subdiscipline, things start becoming a bit messier. Aside from the "tree" view for browsing metadata, we've built views that help it function as a forum for threaded conversations and a wiki for organization, tracking projects, and setting policies. The durable but plastic nature of wikis is exceptionally well suited for this. From Butler, Joyce, and Pike (emphasis mine):

> Providing tools and infrastructure mechanisms that support the development and management of policies is an important part of creating social computing systems that work. [...]
>
> When organizations invest in [collaborative] technologies, [...] their first step is often to put in place a collection of policies and guidelines regarding their use. **However, less attention is given to the policies and guidelines created by the groups that use these systems which are often left to "emerge" spontaneously.** The examples and concepts described in this paper highlight the complexity of rule formation and suggest that support should be provided to help collaborating groups create and maintain effective rulespaces.
>
> [...] **The true power of wikis lies in the fact that they are a platform that provides affordances which allow for a wide variety of rich, multi-faceted organizational structures.** Rather than assuming that rules, policies, and guidelines are operating in only one fashion, wikis allow for, and in fact facilitate, the creation of policies and procedures that serve a wide variety of functions
>
> *Don't Look Now, But We've Created a Bureaucracy: The Nature and Roles of Policies and Rules in Wikipedia* (2008) [316]

So we might have a set of policies that encourages a reporting system to notify other peers if their data is misformatted. Or we might reward contribution with a "peer of the week" award that highlights their work like What.cd's album of the week or Wikipedia's barnstars [317]. We might adopt a cooperative model where each peer pays their share of the server fees, or has to take shifts on moderation and cleanup duty for the week. Each tracker can adopt different policies to reflect their communities.

Trackers-as-wikis don't have to exist in isolation. Trackers for adjacent disciplines or purposes should be able to federate together to transclude pages: organizing multiple perspectives on the same topic, or supplementing each other into a broader base of knowledge.

What if consensus fails? Our system attempts to mitigate the potential damage of tyrannical moderators by making it extremely easy to *fork*. Since every link in the system "belong" to someone underneath a `@namespace`, links and the schemas they build are always a proposition: "something someone said that I don't necessarily have to agree with." If another peer doesn't like the `merge` that we did, they can fork the previous version and continue using it — for other peers the link to the merged version lets them translate between them. If we want to jump ship and go find a different tracker that better reflects our values, all our data, including relationships to the people that we liked there, guides we wrote on the wiki, etc. are still our own. The tracker just tracks, it isn't a platform.

Our joint tracker-wikis have many applications for scientific communication, and it's worth exploring a few.

### 3.4.3   Applications

Continuing the example of the Autopilot wiki, we could make an array of **technical knowledge wikis.** Wikis organized around individual projects could federate together to share information, and broader wikis could organize the state of our art which currently exists hollowed out in supplemental methods sections. The endless stream of posts asking around for whoever knows how to do some technique that should be basic knowledge for a given discipline illustrate the need. Across disciplines, we are drenched in widely-used instrumentation and techniques without coherent means of discussing how we use them. Organizing the technical knowledge that is mostly hard-won by early career researchers without robust training mechanisms would dramatically change their experience in science, whittling away at inequities in access to expertise. Their use only multiplies with tools that are capable of using the semantically organized information to design interface or simplify their operation as described in experimental frameworks.

Technical wikis could change the character of technical work. By giving a venue for technical workers to describe their work, they would be welcomed into and broaden the base of credit currently reserved only for paper authors. Even without active contribution, they would be a way of describing the unseen iceberg of labor that science rests on. Institutional affiliations are currently just badges of prestige, but they could also represent the dependence of scientific output on the workers of that institution. If I do animal research at a university, and someone has linked to the people responsible for maintaining the animal facility, then they should be linked to all of my work. Making technical knowledge broadly available might also be a means of inverting the patronizing approach to "crowdsourcing" "citizen science" by putting it directly in the hands of nonscientists, rather than at the whim of some gamified platform (see [318]).

Technical wikis blend smoothly into **methods wikis** for cataloguing best practices in experimental design and analysis. It is a damning indictment of our systems of training or review (or, more likely, both) that it is possible to publish a paper based on badly misused t-tests, yet the scientific literature is flooded with analytical and interpretive errors [319, 320, 321]. Analytical errors are not just a matter of lack of education, but also a complex network of incentives and disciplinary subcultures. Having the ability to discuss and contextualize different analytical methods elevates all the exasperated methods critiques and exhortations to "not use this technique that renders meaningless results" into something *structurally expressed in the practice of science.* See the `@methodswiki` page that summarizes this general category of techniques and the discussion surrounding their application in the relevant body of research. For implementation of analytical libraries, to move beyond fragile code reduplicated in every lab we need some means of reaching fluid consensus on a set of quasi-canonical implementations of fundamental analysis operations. Given a system where analysis chains are linked to the data they are used with, that consensus might come by negotiating over a semantically dense map of the analysis paths used in a research domain.

**Analysis wikis** would also be a natural means of organizing the previously mentioned Folding@Home-style distributed computing grids. Groups of researchers could organize computational resources and govern and document their use. For example, a tracker could implement a "compute ratio" where donated computing resources function as credit for "bounties." Analogously to private torrent trackers, where a bounty system might allow peers to trade their excess upload in exchange for someone uploading a rare album, linked tracker/wikis could translate that model to one where someone who has donated a lot of excess compute time could trade it for someone uploading or collecting a particular dataset. Since the kind of wikis we are describing combine free text with computer-readable data structures, policies for use could be directly implemented in the wiki in the same place they were discussed. This too is a means of collectivizing support for open-source initiatives that support basic infrastructure by donation and the mercy of cloud providers by integrating them in the basic social practices of science [286].

**Review wikis** could replace journals almost as an afterthought. Though an adequate infrastructure of scientific communication immediately antiquates traditional peer review, review wikis could facilitate it without recourse to an extractive information industry. In response to the almost unique profitability of publishing, some researchers have reacted, perhaps justifiably, by demanding payment for their reviews (eg. [322]). An alternative might be to organize review *ourselves*. Like the ratio requirements of private bittorrent trackers, we might establish a review ratio system, where for every review your work receives you need to review n other works. This would effectively function as a **reviewer co-op** that can make the implicit labor of reviewing explicit, and tie the reviews required for frequent publication with explicit norms around reciprocal reviewing.

**Library wikis** focused on curation, contextualization, and organization of information could be one modality of resisting the neoliberal drive to reduce librarians to stewards of subscriptions and surveillance data [19, 323]. Knowledge organization is hard practical and theoretical work, and reimagining the space of scientific communication as one that we actively *create* instead of one that we merely *suffer through* is a wide-open invitation for the comradeship and leadership of librarians. Linked data has been a mixed blessing for librarians, its promise obscured by intellectual property oligopolies and the complexity of linked data standards (see [187]). Given fresh tooling and a path away from structuring influence of for-profit publishers, the rest of us should be prepared to learn from those that have already been doing the work of curating our archives:

> [M]ake it easy to rely on linked data, easier than it is to rely on MARC, and the library world will shift, from the smallest and poorest libraries upward... and David will at last stone Goliath to death with his linked-data slingshot.
>
> *Stoning Goliath* (2022) The Library Loon [187]

Finally, **theory wikis** could "close the theoretical-experimental loop" to turn the buckshot of results into cumulative understanding of complex phenomena. In many (or maybe just the non-realist) scientific epistemologies, results do not directly re-flect some truth about reality, but instead are embedded in a system of meaning through a process of active interpretation (eg. [324, 325]). The model of grounding new research in existing understanding given by contemporary regimes of scientific communication is for each paper to synthesize and re-interpret the entire body of relevant prior research (formally, the "introduction"), which is bluntly impossible. We do the best we can alongside strong countervailing incentives to selectively engage with work in order to tell a publishable story in which we are the hero. Since the space of argumentation is built from scratch each time, cumulative progress on a shared set of theories is more of a myth for undergraduate introductions to the scientific method than a reality. Most fall far from the supposed ideal of hard refutation and can have long lives as "zombie theories." van Rooij and Baggio describe the "collecting seashells" approach of gathering many results and leaving the theory for later with an analogy:

> "In a sense, trying to build theories on collections of effects is much like trying to write novels by collecting sentences from randomly generated letter strings. Indeed, each novel ultimately consists of strings of letters, and theories should ultimately be compatible with effects. Still, the majority of the (infinitely possible) effects are irrelevant for the aims of theory building, just as the majority of (infinitely possible) sentences are irrelevant for writing a novel." [326]

They and others (eg. [327]) have argued for an iterative process of experiments informed by theory and modeling that confirm or constrain future models. Their articulation of the need for multiple registers of formality and rigidity is particularly resonant here. van Rooij and Baggio again, emphasis mine:

> **We should interpret any data in the context of our larger "web of beliefs,"** which may contain anything we know or believe about the world, including scientific or commonsense knowledge. One does not posit a function $f$ in a vacuum. [...] One can either cast the net wide to capture intuitive phenomena and refine and formalize the idea in a well-defined $f$ or, alternatively, make a first guess and then adjust it gradually on the basis of the constraints that one later imposes: The first sketch of an $f$ need not be the final one; what matters is how the initial $f$ is constrained and refined and how the rectification process can actually drive the theory forward. **Theory building is a creative process involving a dialectic of divergent and convergent thinking, informal and formal thinking.** [326]

Durable but plastic, referential and dialogic, structured and free mediums like our wiki-trackers could be a practical means of integrating theory in a loop with experimentation and interpretation. Many theories are formalizable, and our linked data system is a relatively arbitrary means of expressing complex constraints and inference logics. Others are not, and our mixed-format media also supports the dialectic of informal and formal, mathematized and non-mathematized theories.

In the most optimistic case, where we have a full provenance chain from interpretation of analytical results back through the viscera of their acquisition, we have a living means of formally evaluating the empirical contingencies that serve as the evidence for scientific theories. For a given theory, what kinds of evidence exist? As the state of the art in analytical tooling changes, how are the interpretations of prior results changed by different analyses? How do different experimental methodologies influence the form of our theories?

The points of conflicting evidence and unevaluated predictions of theory are then a means of distributed coordination of future experiments: guided by a distributed body of evidence and interpretation, rather than the number of papers individual researchers are able to hold in mind, what are the most informative experiments to do? This would be a fundamentally different way of approaching a new "unit" of scientific work that dissolves the scientific paper as such. Many calls for smaller units of scientific work amount to faster turnaround for shorter papers that preserve the unitary binding of an experiment, results, and interpretation. Instead new experiments could start *in medias res,* filling in some cracks in an ongoing experimental/interpretational network. A new node could be contributed already contextualized by the "introduction" of its position in a broader graph of understanding, its interpretation posed against a broader background of prior thought than the immediate data at hand. Given the means of directly applying accumulated technical knowledge, it would be possible for more than just the most resourced labs to be responsive to the nicks and burrs in the cutting edge.

The pessimistic case where we only have scientific papers in their current form to evaluate is not that much worse — it requires the normal reading and evaluation of experimental results of a review paper, but the process of annotating the paper to describe its experimental and analytical methods as a shared body of links makes that work cumulative. Even more pessimistic, where for some reason we aren't able to formulate theories even as rough schematics but just link experimental results to rough topic domains is still vastly better than the current state of proprietary disorganization in service of a surveillance-backed analytics industry.

A meta-organization of experimental results would change the way researchers and non-researchers alike interact with academic literature. It currently takes many years of implicit knowledge to understand any scientific subfield: finding canonical papers, knowing which researchers to follow, which keywords to search in table of contents alerts. Being able to locate a question in a continuous space of discussion, data, results, and theories — to say nothing of building a world without paywalls — would profoundly lower barriers to access to primary scientific knowledge for *everyone.* We might avoid the efforts to weaponize this gap into an ostensibly "helpful" algorithmic search platform that re-entrenches the very industries that make such a platform necessary by constraining the modes of our communication. We might instead arrive at a fluid, boisterous, collective project of explicitly organizing understanding. One sounds like science, the other sounds like industry capture.

### 3.4.4   Credit Assignment

I also think one of the big obstacles to freeing up scientific information remains the way in which we continue to pay allegiance to the idea that the most important work is published in so-called 'high-impact' journals [...]. These journals continue to thrive, despite a kind of anti-social policy, because **so many academic scientists evaluate each other's work and measure abilities and accomplishments based on where people have published.**

**The only way by which we'll eventually get out of the current situation is by changing the formula dramatically.** That means that we'll probably have to move to a world where the authors have full control – their work will be presented online together with expert reviews and perhaps accompanied by a new evaluation system in which members of the scientific community will pro-

vide qualitative and perhaps quantitative measures of the value of the paper. The current world of high- and low-impact journals will eventually dissolve, it's just taking a lot longer than I thought.

Harold Varmus, former director of the NIH (2019) *Of Oncogenes and Open Science* [328]

The reason we are (once again) having a fight about whether the producers of publicly available/published data should be authors on any work using said data is that we have a completely dysfunctional system for crediting the generation of useful data. [329] The same is true for people who generate useful reagents, resources and software. [330] And like everything, the real answer lies on how we assess candidates for jobs, grants, etc... **So long as people treat authorship as the most/only valuable currency, this debate will fester. But it's in our power to change it.** [331]

Michael Eisen, EIC eLife (2021)

The critical anchor for changes to the scientific infrastructure is the system of professional incentives that structure it. As long as the only way we operationalize scientific value is paper authorship and the prestige of the journals they are placed in, the system stays: Blog posts, software, analysis pipelines, wikis, forums, reviews, are nice, but they don't count as *science*.

Imagining different systems of credit assignment is easy: just make a new DOI-like identifier for my datasets that I can put on my CV. Integrating systems of credit assignment into commonly-held beliefs about what is valuable is harder. One way to frame solutions to the credit assignment problem is as a collective action problem: everyone/funding agencies/hiring committees just need to *decide* that publishing data, reviewing, criticism et al. is valuable without any serious changes to broader scientific infrastructure. As is hopefully obvious, the approach favored here is to *displace* the system of credit assignment by aligning the interests of the broad array of researchers, technicians, and students that it directly impacts to build an alternative that makes it *irrelevant*.

The sheer quantity of work that is currently uncredited in science is a structural advantage to any more expansive system of credit assignment. The strategic question is how to design a system that aligns the interests of enough people excluded by the current system. Belief, as always, is a tricky circular process: how would the people being evaluated come to believe in its value enough to contribute to it, and how would the people doing the evaluation believe in its value enough to ignore the analytics products by deeply embedded industries?

Everything that exists in this system is attributable to one or many equal peers. Rather than attempting to be an abstract body of knowledge, clean and tidy, that conceals its social underpinnings, we embrace its messy and pluralistic personality. We have *not* been focused on some techno-utopian dream of automatically computing over a system of universally linked data, but on representing and negotiating over a globally discontinuous body of work and ideas linked to people and groups. We have *not* been imagining new platforms and services to suit a limited set of needs, but on a set of tools and frameworks to let people work together to cumulatively build what they need. What is different about this set of ideas is that it is not a new metric, journal, or platform intended to be the new standard that replaces some small element of the system, leaving the rest unchanged. We are taking a broad view on the infrastructural deficits that define scientific work, learning from the broad histories of attempts to remedy them, and trying to chart a course to building systems that fill basic needs. The hope is to seed a critical mass of solidarity by organizing the work to fill the unmet needs that structure the current system of evaluation, in the process building a real alternative that makes the existing system look as ridiculous as it is.

Credit is woven through the heart of this system: the basic operations of interacting with someone else's work are tied to crediting it. While credit is currently meted out by proprietary scientometric tools like altmetric or Plum; downloading a dataset, using an analysis tool, and so on should be directly attributable to one or several digital identities that you control in the manner that you want.

The first-order effects for the usual suspects in need of credit are straightforward: counting the number of analyses and papers our datasets are cited in, seeing the type of experiments our software was used to perform. Control over the means of credit assignment also opens the possibility of surfacing the work that happens invisibly but is nonetheless essential for the normal operation of research. Why shouldn't the animal care technician receive credit for caring for the animals that were involved with a study, its results, and its impact on science more broadly?

A name prominently displayed on a wiki page and a permalink for a CV is ok, but clearly not enough. Foundational work like technical, communicative, and organizational work is useful in itself, but its impact is mostly felt *downstream* in the work it enables. Beyond first-order credit, a linked credit assignment system lets us evaluate *higher-order* effects of work that *more closely resemble* its impact. Say we find someone else's 3D Model, modify it for our use, and then use it to collect a dataset and publish a paper. Someone else sees it and links a colleague to it, and they too use it in their work. Over time someone else updates the design and puts it in some derivative component. Most of the linking is automatic, built into the interfaces of the relevant tools, and soon the network of links is dense and deep.

The incentive to "freeload" by making the use of the system without credit is changed by breaking apart the notion of unitary credit where one or a few people are responsible for "all" of a work. Our current obsession with utter novelty and closed credit removes incentives to extend someone else's work: why would I help patch their code? I won't be added as an author on their paper. For us, instead of just getting professional credit for our paper, we also get credit for extending someone else's work, for documenting it, and for the potentially large number of nth-order derivative uses. Our credit extends multimodally, including papers that cite papers that use our tool, and the "amount" of credit can be contextualized because the type of link between them is explicit – as opposed to the non-semantic links of citation. Our colleague that recommended our part gets credit as well, as they should since helpful communication is presumably something we want to reward. Rather than the scarcity mindset of authorship, a link-based system can push us towards abundance: "good" work is work that engages with and extends a broad array of techniques, technologies, and expertise.

From the perspective of the worker, their extended contribution graph will always be a superset of the things they would otherwise be credited for. The goal should make it be something we *prefer* to share because it's more reflective of our work. Unlike proprietary metrics that will be increasingly based on surveillance data, our system gives us control over which information we want to be part of our evaluative profile, and it's something that we own to do what we will with rather than the product of some platform.

It's easy to imagine extended credit scenarios for a broad array of workers: A grad student rotating in a lab might not get enough data to make a paper, but they might make some tangible improvement to lab infrastructure, which they can document and receive credit for. Open source software developers might get some credit from a code paper, but will be systematically undervalued from failure to cite it and undercounted in derivative packages. The many groups of workers whose work is formally excluded from scientific valuation are those with the most to gain by reimagining credit systems, and an infrastructural plan that actively involves them and elevates their work has a much broader base of labor, expertise, and potential for buy-in.

From the perspective of the evaluator, our contribution graph provides a much richer space of evaluation while also eroding the notion of a scalar-valued ranking. Some of my more communitarian colleagues might share my distaste for metricizing knowledge work — but hiring committees and granting agencies are going to use *some* metric, the question is whether it's a good reflection of our work and who controls it. Our problems with the h-index (eg. [332, 333]) are problems with paper citations being a bad basis for evaluating scientific "value", and their primacy is in turn a consequence of the monopoly over scientific communication and organization by publishers and aggregators. Their successors, black box algorithmic tools like SciVal with valuation criteria that are bad for science (but good for administrators) like 'trendiness' are here whether we like it or not. A transparent graph of scientific credit at least gives the *possibility* for reimagining the more fundamental questions of scientific valuation: assigning credit for communication, maintenance, mentorship, and so on. So some misguided reductions of the complexity of scientific labor to a single number are inevitable, but at least we'll be able to *see what they're based on* and *propose alternatives.* The presence of many simultaneous metrics on the same underlying graph would be itself a demonstration of the inability of any single metric to capture the value of our work. Conversely, spamming the graph to increase your "high score" with a large number of trivial contributions would be straightforward to detect because of the likely shallowness of the graph, so microcommodification of labor is less likely. The incentives are aligned to do work that is useful to others and positively affect the state of our understanding.

It's true that some of these extended metrics are already possible to compute. One could crawl package dependencies for code, or download the 100GB Crossref database [334] and manually crunch our statistics, but being *able* to compute some means of credit is very different than making it a *normal part* of doing and evaluating research. The multimodality of credit assignment that's possible with a linked data system is part of its power: our work *actually does* have impacts across modalities, and we should be able to represent that as part of our contribution to science.

Reaching a critical mass of linked tools and peers is not altogether necessary for them to be useful, but critical mass may trigger a positive feedback loop for the development of the system itself. Even in isolation, a semantic wiki is a better means of assigning credit than a handful of google docs, experimental tools that automatically annotate data are better than a pile of `.csv` files, etc. Bridging two tools to share credit is better than one tool in isolation, and more people using them are better than fewer for any given user of the system. Lessons learned from STS, Computer-Supported Cooperative Work (CSCW), pirates, wikis, forums, et al. make it clear that *the labor of maintaining and building the system can't be invisible.*

# 4

# *Conclusion*

To take stock:

To approach the deficits in the basic digital infrastructure of science, we divided them into three domains: systems for sharing **data, tools, and knowledge.** These map onto three rough patterns of infrastructure that define the current cloud orthodoxy era of the internet: **storage, computation, and communication.**

We traced the historical development of prior digital infrastructure projects to learn from their successes and failures, conditioned as they are by the contingency and combinatorics of the technologies that existed at the time. We started close at hand within science, and ranged more broadly into lessons from internet protocols, pirates, the semantic web and linked data, early wikis, and the fediverse/indieweb.

Our goal throughout was to sketch a **realistic plan** by which existing technologies could make an emergent interoperable system that was *expansive and evolving* beyond the isolated use of its quasi-independent parts. Our sketch was intended to be *specific* enough to be an actionable blueprint for dispersed groups to work in parallel, but *general* enough to allow refinement through inevitable complexity. We attempted to balance several constraints, primarily **technical capability** and **social compatibility,** but also simplicity and expressiveness, structure and permissiveness; systems that are personal and scalable, respect privacy and empower mutual organization. We are neither politically nor economically neutral, and see the infrastructural deficits of science as reflective of information's broader role as the currently dominant mode of capital accumulation. Accordingly we are searching for system design that can dismantle regimes of surveillance, extraction, and the commodification of information to **re-decentralize** our digital technologies for *people* not *profit*.

The system we arrived at is based on **p2p folksonomic linked data.** Using existing data formats as an initial onramp, and overlays to bridge to incompatible media, our p2p system blends ideas from bittorrent, IPFS, and the Linked Data Platform with metadata beneath a peer's **namespace** indicating content-addressed binary data. Our metadata uses **triplet links** as a means of specifying multimodal schema for data, tools, and social systems. We integrate our data in a complete provenance chain from collection to use with **metadata indicating code** in analytical frameworks and **code indicating metadata** in experimental frameworks. The **social reality** of infrastructure is designed into the core of our system, with peers forming overlapping **federations** with tracker-like overlays. A generalization of documents as systems of **recursive typed cells** serve as an interface to, and representation of the underlying data and metadata. From them we construct a fluid and continuous system of **documents, feedlike media,** and **wikis** for communication and governance of the system. With this system, we satisfy the design goal of a decentralized, protocol-driven infrastructure of linked data, tools, and knowledge.

So how do we build it?

## *4.1   Tactics & Strategy*

> Don't scab for the bosses / don't listen to their lies / us poor folks haven't got a chance / unless we organize
>
> Which side are you on?
>
> Florence Reece (1931) *Which Side Are You On?*

> **Oh but they will mock us and they will mistreat us til they can replace us all with an app or a kiosk,** [...]
>
> All of the energy that I end up expending, I will get back in spades when the systems that necessitate all of this work fall apart... **And we can work for ourselves for a change!**
>
> **So we gotta work!** Cuz none of our visions of a better tomorrow will come to fruition without **a whole lot of work!**
>
> RENT STRIKE (2021) Work! (Future Perfect) [335]

The primary ingredient needed to build decentralized infrastructure is **will.**  The incentive and professional systems of science are designed to make us build our own cage: play along, or lose your job.  We need to recognize that *the contemporary practice of science is unsustainable* without radical infrastructural, social, and economic reorganization.  As the logic of the digital enclosure movement transforms old enemies into new ones, publishers into surveillance conglomerates, the comfortable familiarity of science as we know it will evaporate into the cloud as we cede control over the direction of our work to for-profit companies with their gamified metrics and platforms that commodify every part of it.  The worst parts of scientific work are neither natural nor inevitable, but reflect the overwhelming structuring power of orbiting conglomerates.  We are *part of this world,* and the world is drowning in an algorithmic sea owned and operated by a rapidly consolidating cluster of information giants.  We need to see our place in a shared struggle, the relationship between our deinfrastructuring and the operation of science — and have the courage to do the work to counteract it.

The work doesn't need to be as dreary as its motivation: rebuilding our infrastructure will be *joyful.*  We have been starved for social and labor organization, for comradeship and compassion, isolated as we are on our workplace and disciplinary islands, crushed under the weight of cutthroat publish-or-perish schemes, secretive and distrustful from our culture of the heroic individual rushing through the gauntlet of credit before our enemies do.  What we might lose in prestige we will regain in collaboration with new and unexpected colleagues building tools to make our work *more fun.*  We can trade artificial scarcity for abundance.

### *4.1.1   Starting Points*

Much of the tactical and strategic vision for our new infrastructure is embedded in its design.  We have taken pains to articulate its components as elaborations of existing and widely-used systems, keeping them separable so each can be independently useful before a fuller system is realized, exemplifying them with the real problems that they can remedy.  Still, some more scaffolding for how to get there from here is useful.

The core of our strategy should be to organize alongside each other in a series of independent groups working in parallel. We don't need a new leadership council to become a single point of failure. We should try and organize the many existing groups working in different related areas to pull in the same direction towards interoperability. We should avoid the pitfalls of designing our infrastructures "in theory," building crystal palaces removed from the reality of their use. We should seek to embed in existing projects, using their existing mass to lessen the need to prospect for abstract "early adopters."

We should look outside our usual circles for collaborators, and there we might find unexpected energy and expertise. Though the miserable academic fleeing to the greener pasture of "industry" is now a well-trod trope, there is plenty of disaffection on the other side. We shouldn't underestimate the number of extremely talented engineers that would do *anything* to not have to build tools so that Facebook can mine your thoughts to target ads [336] or maximize the time people spend watching YouTube by recommending them increasingly toxic videos [337]. Academic science is relatively unique in that it can marshal funding and labor for projects not bound to profit. Resources and applications are two potent missing ingredients in developing technologies that are intended to be anti-profitable, and we should work to provide them. We should trawl the places where the decentralized messaging, former semantic web, indieweb, and small tech people are already working on building better infrastructure and invite them to work with us.

The three broad domains of our infrastructure could, but don't necessarily need to, correspond to a division of development labor. The serial order of this piece is primarily a byproduct of the constraint of the medium, and there is no reason we can't proceed in parallel. I want to avoid being too prescriptive here in order to invite input from the many people that might potentially be involved — the purpose of this document as a pre-development plan is to provide direction and a high-level design so that the details can be sorted out as we work. For the sake of illustration, though I'll drop down from the level of strategy to tactics to flesh out some of the more proximal possibilities for development, but the remainder of this section should be considered non-normative.

A promising context to develop a p2p linked data client is existing collaborations or tools that have a base of users that handle overlapping but variable data. For example, the users or developers of a tool like OpenEphys [338] or Miniscope [339] that has data acquisition software that outputs semi-structured data might be interested in making it possible for everyone who uses the tool to share data with one another from the time of acquisition. The situation is similar for other types of tools like analysis tools, or for collaborations where people are sharing data frequently. Since the output data is relatively simple (eg. videos and timestamps) with some variation (eg. configuration and notes), it would be a smaller climb to prototype generating a metadata model linked to the data. Since the group of people that would be sharing data might initially be relatively small with room to grow, the several components of the p2p client could be worked out separately: eg. manually index repositories of metadata from some frontend while figuring out how to strap a git server to a p2p client like hypercore, etc. Being able to be plugged into a group of people sharing data by using a tool might be a reasonably attractive idea to get people to adopt the tool, so it would be worthwhile to the developers while being a useful feature for the users. Being able to do very tight loops of development and field testing might make the tool more robust than if it were developed strictly in-house, and would be a good small-scale demonstration of the utility of p2p.

At the same time, work could happen in the other direction from data standards towards p2p. Datalad [207] would be an excellent candidate to add linked data and p2p support to, as it already supports JSON-LD with a metadata extension and has a generalizable data storage backend. In neuroscience, DANDI hosts data formatted in NWB, and interoperability with IPFS is on its development timeline. Working from multiple directions towards aligned projects could encourage a small set of modular tools that can accommodate the variation in each of the approaches, and the process would be useful for navigating the fine-scale constraints to the system without putting all of the development eggs in one basket, so to speak.

Aside from p2p, a toolset that's desperately needed across disciplines is an generalizable, approachable means of modeling and ingesting data. The work of building an interface that lets researchers create a JSON-LD metadata model and a declarative description of where that metadata can be located in whatever lab-idiosyncratic format already exists would supplement all other parts of the system. There is no reason for each format to develop a separate schema language and storage mechanism, and this might be one way to spark collaboration between format maintainers.

One of the major reasons for bootstrapping the system with existing formats is to be able encourage analytical and experimental tool interoperability before the means of creating and negotiating over arbitrary schema are developed. This is already starting to happen to a degree in neuroscience with datajoint elements [216] and NWB (see pynapple), but since the conversion tooling for NWB at the moment is still relatively opaque there isn't strong incentive for analysis libraries to support it for seamless input. A wrapper framework to be able to specify an analysis pipeline from metadata that combines a few of these tools might be useful to kick off the positive feedback loop of analysis toolbuilders building towards interoperability, incentivizing format conversion, etc.

The other major starting point for development I see is generalizing cellular documents with JSON-LD and mixing them with ActivityPub. With some relatively minor extensions to the jupyter document format we could add the ability to create new cell types with elaborated linked metadata. From there, we could build an ActivityPub client that allows researchers to post their notebooks and invite comment on them in a document/threaded communication medium. The support of an existing organization would be useful here too: they could apply to be a crossref member and make use of the very general specification [72] such that each post can be given a hierarchical DOI like `doi:10.<registrar>/user/post/version`. Along with the ability to automatically submit to legacy journals with the conversations attached as supplemental material, this might attract a reasonable critical mass towards a model that would make the move towards a p2p document/communication a much smaller step. Neuromatch [340, 341, 342] has expressed interest in work in this area, though at the time of writing their plans are still in development.

I'll leave the remainder of the organization project to the work of the future.

### 4.1.2    *To Whom It May Concern...*

This project should benefit everyone, but we all have different roles to play. Without enumerating every possible category, a few love letters:

**PIs:** Infrastructure is everyone's responsibility! Diverting time towards organizing the development of basic infrastructure seems expensive and risky, but the truly expensive thing is to do nothing. The quantity of time spent rebuilding everything from scratch, debugging local code, contending with the journal system, resurrecting old data, etc. for all but the most efficient labs is truly staggering. The absence of collective organization makes PIs a sitting duck for profiteering: seeing the difficult resistance posed by library consortia, the open access model shifted towards payments from individual PIs because they have little choice but to pay them on their own. It is in your best interest to commit time to organize with others in your discipline to build generalizable tools that you can share with other labs. We will need you to help shake down funding to pay for development — it will be worth it.

It is also in your best interest to start closing ranks and collectively disavowing the for-profit journals. The rationalization that you need prestige publications for the sake of your trainees is plainly self-fulfilling: what it actually accomplishes is guaranteeing they have to endure the same grim circumstances you do. The best way to support your trainees is to fight to fix our broken infrastructure! Individually you may have little power, but if you organized your colleagues, starting in your department and working out, to agree to never publish for-profit, the problem starts looking very different. In tenure and hiring decisions, having no Nature papers looks very different when you have been loudly organizing with your colleagues for the health of science. Except for those at the extreme heights of the prestige economy, you have perhaps the most to gain by getting off the treadmill.

**Early Career Researchers:** We don't have much, but we can have each other! We don't have the leverage to make huge changes quickly, but but since we're the ones doing most of the work of building the tools for our research anyway, we should also start organizing with our colleagues to share techniques and methods. As we build infrastructure, coalescing into institutional collaboratives makes it that much easier to organize across institutions. We shouldn't fall into the trap laid for us working to the bone for a prestige publication — if we want to make academic science something that we would actually want to work in, we can help shake the researchers trained in prior generations out of complacency. A better world is out there!

**Scientific Open Source Developers:** We've got work to do! First, we need to start making alliances with people we're not necessarily used to, but that's the fun part! For those of us working outside the few major projects, the best thing we can do is to start organizing our tools as broader infrastructure instead of one-off, single-purpose tools. We should focus on designing our tools in such a way that they can be integratable: as a small sample, that means spending time on good packaging rather than throwing everything in a docker container, making APIs that are clear in what they expect and return, and well-contained configuration and parameterization. If our tools aren't already part of a broader framework, we should work on that first! We should avoid cloud dependency when possible: if it is necessary, make sure that it can also be deployed locally just as easily. We should also emphasize multi-scale interfaces: instead of just exposing a set of top-level functions, it should also be possible for someone else to understand its internal operations, otherwise interoperability becomes a distant dream. At the risk of being preachy as a younger developer, the most important thing we can do is organize and be organizable.

**Funding Agencies:** You are being swindled! Partnership with the cloud industry is a recipe for burning ever-larger portions of your funding allocations on systems that only become harder to walk away from with time. Open source is your friend.

Rather than funding projects piecemeal, or funding massive top-down projects with little user engagement, it needs to be possible to receive funding for projects that fill fundamental cross-disciplinary infrastructural gaps. We need to figure out some way of breaking the catch-22 of scientific software development where only projects with demonstrated uptake can receive funding, but it is difficult to start projects intended to address large problems without funding. Scientific funders already do fund a large amount of open source development, and I am not proposing an alternative model here, except to say that the energy and expertise is there to build open-source infrastructure that avoids creating another triple-dip industry.

**University Administrators:** You're also being swindled! The disorganized smattering of SaaS that serves as the infrastructure of many universities [343] is a short-run bandaid that makes operations more fragile in the long-term! Putting your resources behind organizing institutional and regional infrastructure is a better PR story and far more attractive when recruiting than how large of an AWS subscription you have [104]. University libraries shoulder a huge burden of the cost of the for-profit publishing system, and so you should have every incentive to cut ties — instead of open access mandates, we need you lobbing on behalf of all of us to end the for-profit system.

## 4.2   Limitations

To get a few big and obvious limitations out of the way first: - Everyone could ignore this piece entirely and it is dead on arrival. - This project would be a direct threat to some of the most powerful entities in science, and they will likely actively work against it. - Despite my best efforts, I could be completely misinformed and missing something fundamental about the underlying technologies. - The social tensions between the relevant development communities could be too great to overcome.

Beyond those, there are several open questions that deserve further consideration, particularly those things concerning cryptography as it is squarely outside my domain of expertise:

**Identity:** Identity is extremely challenging for any decentralized system. An identity needs to be unique, difficult to counterfeit, easy to verify, easy to manage or recover, and also recognizable if not memorable — and several of these requirements are clearly in conflict. A satisfying resolution of identity will require guidance from cryptographers, but the design of our system has some features that make identity a less-than-intractable problem. The actual raw identifier itself will likely need to be a cryptographic hash of a public key (as in IPFS) for uniqueness and verifiability, but they are very far from memorable. One approach might be to have each peer provide a signed identification object that can be publicly queried with a shorter handle or username, which can then be stored by the peers that follow or befriend them. When peer A refers to peer B's namespace, then, it would be in reference to peer A's follow/friends list. Another approach is to use an RDF-like prefixing idea: in a given context, a short name for a peer's hash is always explicitly declared first before using it. Neither of these are entirely satisfying, and will require a bit of experimentation to get right.

The problem of managing keys and recoverability is also tricky: there's no "forgot password" link if you lose your private key. Since our system is designed to be intrinsically social, we can relax some of the more stringent requirements of zero-trust, totally-anonymous networks like IPFS and lean more on a "web of trust." We might share additional private keys with other peers that we trust to verify or recover our identity, which might be particularly useful in the case of a more stable federation of peers. We want to avoid peers operating like identity systems, as that lends itself to centralization of power and returning to a more activitypub-like style of identity, so it would be a tricky balance. Another strategy might be to use an out-of-band mechanism, like storing a URL in the signed identity object that can be used to update the public key associated with a particular identity – if you lose yours, you can generate a new keypair and update the public key stored at the URL, which another peer could check to verify that you are who you say you are. These too are not very satisfying, and so more work will be needed to draft a satisfying identity system, the practicality and usability of which will be critical for its success.

**Privacy:** Closely related to identity, in a p2p system any message that isn't intended to be public will need to be encrypted so that secondary peers can't just reshare something we intended to be only for them. In our system, it's not desirable to be able for some data-greedy entity to scrape all the data, we want peers to be able to make friction as-needed. To some degree this is not a solvable problem, as it's always possible to take a screenshot of the most secure end-to-end encrypted chat. It's possible even in analog social systems for secrets to slip, or for people to lie about something that another person said, so arguably the question is how to protect the things that can be verified to be from a person. Another practical problem is communicating which peers are allowed to see something so that a secondary peer knows whether or not they can help seed something: we don't want to have to transmit a list of a thousand peers along with every message, and if they have to ask the primary peer every time then the redundancy of the system is lost. Capability-based security, where permissions for a given object are conferred by having a hard to guess reference to it rather than by checking an easy to guess reference against a permissions list, seems like a good approach (see [344]). This would look a bit like generating (revocable) sharing links for different groups. Here too we might lean a bit on the social nature of our system, where peers that routinely violate the privacy requirements of other peers can be labeled untrustworthy.

**Security:** Most parts of this system are relatively low-risk as they are based on metadata that only relies on defined actions programmed into the receiving client — you don't get viruses merely by torrenting something. Several of the more interesting applications, though, involve a message or link being able to self-identify some code used to run or display it, and whenever you introduce running arbitrary code you introduce significant security risks. This is largely mitigated by our emphasis on non-automaticity: the default for any of these cases should be to *not* do the action. That's cold comfort, though, given the high clickthrough rates for phishing emails. More mitigation can be had by executing code in containers or virtual machines, but that too is not total. We manage to get by extraordinarily well with a very informal reputation system in open source computing. For the most part, people don't think twice about running some Python package without reading the full source code. Our system is one very conducive to soft security [345], which is based more on accountability and resilience than strict guarantees of security. Where typically an untrustworthy platform will do whatever it can to avoid people being able to leave comments or talk about it, in our linked data system it's always possible to issue a link saying that something is not to be trustworthy. The ability to mirror shards of our data makes any particular attack more likely to be recoverable, but special care will be needed to ensure the whole network is not subject to rolling waves of ran-somware. Like cryptographers, we'll need consultation and input from the infosec crowd to make it safe, but there's nothing intrinsically more dangerous than, say, pip being able to run arbitrary code inside a `setup.py` file.

**RDF Standards:** RDF is highly polarizing, and many people have written it off as a lost cause because it is too complex. Much of the computing world runs off of table and relational databases rather than graphs. Though we tried to be careful to avoid endorsing any particular technology in favor of thinking about triplet links as such, the question of the literal implementation of the standards is an inevitable one. JSON-LD is, thankfully, a relatively humane standard that should be the first point of exploration. We should consider interconvertibility and interoperability with existing standards a top priority of the system in general, so we will need to make interfaces to make it trivial to interact with commonly used formats, even if it is just a wrapper that indicates the format rather than one that can convert it to JSON-LD. Interface design is one of the major missing pieces in the linked data story, and that too should be a top priority so that as little of the system as possible needs to rely on directly interfacing with the underlying data model.

**Performance:** We have more or less explicitly cast performance aside as the wrong thing to optimize for: we want to have *autonomy* more than be able to blaze through the network in microseconds. Still, it's possible for technologies to be so inefficient to be nonfunctional. In a world where we have been conditioned to expect to be able to speak with a manager when our apps are not immediately responsive, or to be able to just buy whatever server performance we want, it will take some collective unlearning to rethink the internet along the lines of cooperatively managed resources. Unlearning performance will take time and has no boardroom-friendly KPIs to measure. There's no reason to believe the system will be slow before it exists, and we ultimately don't imagine this system running from residential connections and personal computers, but being a mixture of institutional and personal resources. Decentralization is a continuum, rather than a binary: we don't have to *ban* large servers from the network, but instead want to make sure that there is a healthy mix so that the system doesn't *depend* on them. This is another place where it is useful to seed this from academia: internet service providers have historically leaned on their oligopolistic control over the underlying hardware of the internet to crush threatening technologies [346], and we should expect no different this time. We will need to have access to commercial connections, and will likely need to convince our institutions to lobby on our behalf.

**Manipulation:** What if people lie? What if people purposefully rig the system by seeding it with a bunch of fake data and bad papers? People already lie! People already game the system! What we are hoping to change is to make a system where manipulation isn't built into the system as a self-reinforcing partnership between its proprieters and beneficiaries. The real dangerous thing is a system that *presents* itself as being infallible or neutral through its automaticity and glittering PR campaign. This is why we have baked the social contingency of the system so thoroughly into its design (and should investigate making triplet links into quadruple links with each having an explicit author to make it even more concrete). This is why, I believe, it is so difficult for some people to imagine a world without pre-publication peer review vouched for by a journal: the social contingency of information is scary! It is, however, preferable to the economic contingency of factuality-as-a-service.

The last set of concerns are diffuse rumblings about uptake and whether or not it is even still *possible* to challenge entrenched economic powers in science. It is true that people are cynical, and busy, and some benefit immensely from the present system, and so on. It is also true that we are likely to be met with stiff resistance if we start posing a credible threat to their dominance — the danger of opposing a set of companies who are the primary data brokers to federal law enforcement agencies, credit rating, and insurance agencies is not lost on me. I don't have any good answers to these sets of questions except that the work from here is about organizing people, adapting and responding to their needs, and making something that is useful enough for even the most complacent to adopt. I don't present this blueprint for infrastructure as infallible, and intend it to be mutated and merged with other ideas as we progress. The only thing that *isn't* an option is doing *nothing*.

## 4.3    In Closing

Infrastructure isn't just a technical, or even social project: it's also ethical. We started by outlining the harms of our infrastructural deficits for science, many of which are widely seen as normal, or otherwise inevitable. Some harms are only possible to recognize when it's possible to imagine an alternative to the system that causes them. This project was an attempt to help us imagine what science can be like as a guide and inspiration for us to organize to make it real. I didn't get everything right, and I probably raised more problems than I addressed. My goal more than to be right was to give a fistful of threads to pull for those that are eager to, and to make it impossible to say that a better future for science is impossible. If all we can imagine science to be is a system where we scrape by, forcing a chain of papers through a rapacious machine that turns curiosity into a treadmill of funding and prestige, playing out the clock as our working conditions deteriorate to the point where publicly funded science is nothing more than a training program for pharmaceutical and advertising companies — what are we even doing here?

Infrastructure isn't a distraction from science or something to put off as the work of a diffuse *someone else.* It's not even an ill-defined alternative cynics use to grandstand about how much they know about how bad everything is. Collectively built infrastructure is the best way for us to make science continue to be possible. We often focus on the problems of science in isolation: what do we do about the *journals,* how do we make *scientific software* more sustainable, why is it so hard to share *data.* My central argument is that the only way we will address any of these problems is by considering the system as a whole. Rather than being a utopian vision of ripping it out from the root and starting anew in one fell swoop, considering the whole system is how we turn nibbles around the edges into coordinated mass movement. It's less about this vision matching exactly what we end up building, but making it possible for the many diverse and dispersed people who care about different facets of the problem to see it as a shared project.

Beyond any technology, my hope is that by organizing together to build something that helps us organize better, that we can re-commit to working for, instead of against each other. Some of our deeper problems like the neoliberalization of universities will only be possible to approach with a renewed sense of ourselves as often privileged, but nonetheless exploited labor. I am the first to admit my naïveté, but I think a nontrivial part of the lack of labor consciousness in science is the way our systems of work, communication, and evaluation feed back into an individualist celebration of the hero of knowledge. Maybe by rebuilding those systems to support the abundance of cooperation and make collective organization a central part of our work can help us both do better science and make science better (also see [347]).

Science for science's sake also misses the point. The dominant stories we tell of how science can give back to society are also shot through with market individualism: become a scicomm influencer, or found a start-up. Instead of giving back to a society that we are somehow separate from, we can take our part in shared struggle seriously, like the graduate workers at Columbia and Harvard who did us all proud by fighting like hell through the strike wave this year and last [348, 349]. Even for the most basic research-oriented, the problem of informational dominance in the 21st century tolls for thee. Systems like those described here could serve as a basis for a new kind of digital infrastructure that challenges the basic platform model of the internet more broadly [350]. What are the three to five remaining websites but data storage, computation, and communication systems? By organizing to make our own work better, we might also seed the systems that help reclaim digital infrastructure as something that empowers everyone, rather than uses our urge to connect with each other to control us.

It was scientists[1] looking for a better way to communicate that created the internet in the first place, radically rewriting the course of history [351] — and we can do it again.

[1] With funding from the military

## 4.4   Contrasting Visions of Science

Through this text I have tried to sketch in parallel a potentially liberatory infrastructural future with the many offramps and alternatives that could lead us astray, but to make two of those futures clearer, it's worth imagining them outright.

### 4.4.1   What if we do nothing?

You're a researcher with dead-center median funding at an institute with dead-center median prestige, and you have a new idea.

The publishing industry has built its surveillance systems into much of the practice of science: their SeamlessAccess login system and browser fingerprinting harvest your reading patterns across the web [352, 353, 354, 355, 356], Mendeley watches what you highlight and how you organize papers, and with a data sharing agreement with Google crossreference and deanonymize your drafts in progress [5]. Managing constant surveillance is a normal part of doing science now, so when reading papers you are careful to always use a VPN, stay off the WiFi whenever possible, randomly scroll around the page to appear productive while the PDF is printing to read offline. The publishers have finally managed to kill sci-hub with a combination of litigation and lobbying universities to implement mandatory multifactor authentication, cutting off their ability to scrape new papers. The few papers you're able to find, and fewer that you're able to access, after several weeks of carefully covering your tracks while hopping citation trees make you think your hunch might be right — you're on to something.

This is a perfect project for a collaboration with an old colleague from back in grad school. Their SciVal Ranking is a little low, so you're taking a risk by working with them, but friendship has to be worth something right? "Don't tell me I never did nothing for you." You haven't spoken in many years though, so you have to be careful on your approach. The repackaged products of all their surveillance are sold back to the few top-tier labs able to afford the hype-prediction products that steer all of their research programs [357, 358]. The publishers sell tips on what's hot, and since they sell the same products to granting agencies and control the publishing process, every prediction can be self-fulfilling — the product is plainly prestige, and the product is good. If you approach your colleague carelessly, they could turn around and plug the idea into the algorithm to check its score, tipping off the larger labs that can turn their armies of postdocs on a dime to pounce. There is no keeping up with the elites anymore.

Even if you do manage to keep it a secret, it'll be a hard road to pull off the experiment at all. There are a few scattered open source tools left, but the rest have been salami sliced into a few dozen mutually incompatible platforms (compatibility only available with the HyperGold Editions). The larger labs are able to afford all the engineers they need to build tools, but have little reason to share any of the technical knowledge with the rest of us — why should they spoil the chance to spin it off into a startup? There aren't any jobs left in academia anyway.

Industry capture has crept into ever more of the little grant funding you have, all the subscriptions and fees add up, so you can only afford to mentor one grad student at a time while keeping plausibly up to date with new instrument technology. You can't choose who they are anymore really. The candidate ranking algorithms have thoroughly baked the exclusionary biases of the history of science into the pool of applicants [5, 13], so the only ones left are those who have been playing to the algorithm since they were in middle school. Advocates for any sort of diversity in academia are long gone. We've never been able to confirm it, but everyone knows that the publishers tip the scales of the algorithm to downrank anyone who starts organizing against them.

Your colleague and you manage to coordinate. they're the same as they've always been, trustworthy. You really need someone from a different field at least in consultation, but there isn't really a good way to find who would be a good fit. Somehow Twitter is still the best way to communicate at large, but you've never really gotten how it works and the discourse has gotten *dark* so you don't have enough followers to reach outside your small bubble of friends. You decide to go it your own, and find the best papers you can from what you think is the right literature base, but there's no good way of knowing you're following the right track. Maybe that part of the paper is for the supplement.

Data is expensive, if you can find it. Who can pay the egress costs for several TB anymore? You forego some modeling that would help with designing the experiment because you don't have the right subscription to access the data you need. You'll have to wait until there is a promotional event to to get some from a Science Influencer.

You experiment in public silence until you've collected your data. Phew, probably safe from getting scooped. You start the long slog of batch analysis with the scraps of Cloud Compute time you can afford.

Papers are largely unchanged, still the same old PDFs. They're a source of grim nostalgia, at least we'll always have PDF. What has changed is citation: since it's the major component of the ranking algorithm, nobody cites to reference ideas anymore, just to try and keep their colleagues afloat. The researchers who still care about the state of science publish a parallel list of citations for those who still care to read them, but most just ignore them — the past is irrelevant anyway, the only way to stay afloat is hunting hype. You know this is distorting the literature base, feeding the algorithm junk data that will steer the research recommendations off course, but you don't want to see your colleague down the hall fired [13]. Their rankings have been sinking lately.

Uploading preprints is expensive now too, and they charge by the version, so you make sure you've checked every letter before sending it off. It's a really compelling bit of science, some of that old style science, fundamental mechanisms, basic research kind of stuff. You check your social media metrics to perfectly time your posts about it, click send, and wait. Your friends reply with their congratulations, glad you managed to pull it off, but there's not really a lot that can be made a meme of, and it's not inflammatory enough to bait a sea of hot takes. You watch your Altmetric idle and sigh. You won't get a rankings boost, but at least it looks like you're safe from sinking for awhile.

You're going to take a few weeks off before starting the multi-year process of publication. Few researchers are willing to review for free anymore, everyone is sick of publisher profiteering, but we didn't manage to build an alternative in time, and now it's too dangerous to try. Triage at the top of the journal prestige hierarchy is ruthless. Most submissions not pre-coordinated with the editor are pre-desk rejected after failing any one of the dozen or so benchmarks for "quality" and trendiness crunched by their black box algorithms. Instead we ping-pong papers down the hierarchy, paying submission fees all along the way. Don't worry, there's always some journal that will take any work — they want the publication fees in any case. If you're cynically playing the metrics game, you can rely on the class of blatantly sacrificial junk journals that can be hastily folded up when some unpaid PubPeer blogger manages to summon enough outrage on social media. We haven't managed to fix the problems with peer review that favor in-crowd, clickbait-friendly, though not necessarily reproducible, research. It turned out to have been a feature, not a bug for their profit model all along.

You're not sure if you've made a contribution to the field, there isn't any sense of cumulative consensus on basic problems. People study things that are similar to you, lots of them, and you talk. You forget what they've been doing sometimes, though, and you catch what you can. You like your work, and even find value in it. You can forget about the rest when you do it. And you like your lab. The system isn't perfect but everyone knows that. Some good science still gets done, you see it all the time from the people you respect. It's a lot of work to keep track of, at least without the subscription. But you managed to make it through another round. That feels ok for now. And it's not your job, your job is to do science.

The attention span of your discipline has gotten shorter and shorter, twisting in concentric hype cycles, the new *rota fortuna*. It's good business, keeping research programs moving helps the other end of the recommendation system. It started with advertising that looked like research [171], but the ability to sell influence over the course of basic science turned out to be particularly lucrative. Just little nudges here and there, you know, just supply responding to demand. They turn a blind eye to the botnets hired to manipulate trending research topics by simulating waves of clicks and scrolls. More clicks, more ads, the market speaks, everybody wins.

The publishers are just one piece of the interlocking swarm of the information economy. The publishers sell their data to all the others, and buy whatever they need to complete their profiles. They move in lockstep: profit together, lobby together. The US Supreme Court is expected to legalize copyrighting facts soon, opening up new markets for renting licenses to research by topic area. No one really notices intellectual property expansions anymore. There are more papers than ever, but the science is all "fake news." Nobody reads it anyway.

### 4.4.2   What we could build

You're a researcher with dead-center median funding at an institute with dead-center median prestige, and you have a new idea.

You are federated with a few organizations in your subdiscipline that have agreed to share their full namespaces, as well as a broader, public multidisciplinary indexing federation that organizes metadata more coarsely. You navigate to a few nodes in the public index that track work from some related research questions. You're able to find a number of forum conversations, blog posts, and notebooks in the intersection between the question nodes, but none that are exactly what you're thinking about. There's no such thing as paywalls anymore, but some of the researchers have requested to be credited on view, so you accept the prompts that make a `read` link between you and their work. You can tell relatively quickly that there is affirmatively a gap in understanding here, rather than needing to spend weeks reading to rule it out by process of elimination — you're on to something.

You request access to some of the private sections of federations that claim to have data related to the question nodes. They have some writing, data, and code public, but the data you're after is very raw and was never written up — just left with a reference to a topic in case someone else wanted to use it later. Most accept you since they can see your affiliation in good standing with people and federations they know and trust. Others are a little more cagey, asking that you request again when you have a more developed project rather than just looking around so they can direct their permissions more finely, or else not responding at all. The price of privacy, autonomy, and consent: we might grumble about it sometimes, but all things considered are glad to pay it.

Your home federations have a few different names for things than those you've joined, so you spend a few hours making some new mappings between your communities, and send them along with some terms they don't have but you think might be useful for them and post them to their link proposals inbox. They each have their own governance process to approve the links and associate them with their namespace, but in the meantime they exist on yours so you use them to start gathering and linking data from a few different disciplines to answer some preliminary questions you have. In the course of feeling out a project, you've made some new connections between communities, concepts, and formats, and made incremental improvements on knowledge organization in multiple fields. You're rehosting some of their data as a gesture of good faith, because you're using it and it's become part of your project, (and because a few of the federations have ratio requirements).

You do some preliminary analysis to refine your hypotheses and direct the experimental design. You are able to find some analysis code from your new colleagues in a notebook linked to the data of theirs that you're using. It doesn't do *exactly* what you want, but you're able to extend it to do a variation on the analysis and link it from their code in case anyone else wants to do something similar.

You post a notebook of some preliminary results from your secondary analysis and a brief description of your idea and experimental plan in a thread that is transcluded between the forums of the now several federations involved in your project. There's little reason to fear being scooped: since you're in public conversation with a lot of the people in the relevant research areas, and have been linking your work to the concepts and groups that any competitor also would have to, it doesn't really make sense to try and rush out a result faster than you to take credit for your ideas. All the provenance of your conversations and analyses is already public, and so if someone did try and take credit for your idea, you would be able to link to their work with some "uncredited derivation" link.

In the thread, several people from another discipline point out that they have already done some of what you planned to do, so you link to their post to give them credit for pointing you in the right direction and transclude the relevant work in your project. Others spitball some ideas for refinements to the experiment, and try out alternate analysis strategies on your preliminary results. It's interesting and useful, you hadn't thought about it that way. They give you access to some of their nonpublic datasets that they never had a chance to write up. It'll be useful in combination with your experimental results, and in the process you'll be helping them analyze and interpret their unused data.

You're ready to start your experiment. They say an hour in the library is worth a thousand at the bench, and your preliminary work has let you skip about a third of what you had initially planned to do. The project gives credit and attribution to the many people whose work you are building on and who have helped you so far, and has been made richer from the discussion and half dozen alternative analyses proposed and linked from your thread.

Some of the techniques and instruments are new to you, but you're able to piece together how they work by surfing between the quasi-continuous wikis shared between federations. Hardware still costs money, but since most people able to make do with less specialized scientific instruments because of the wealth of DIY instrument documentation, and scientists are able to maintain grant funded nonprofit instrument fabrication organizations because their work is appropriately credited by the work that uses them, it's a lot less expensive. You try out some parameter sets and experiment scripts in your experimental software linked by some technical developers in the other fields. You get to skip a lot of the fine tuning by making use of the contextual knowledge: less dead ends on the wrong equipment, not having to rediscover the subtleties of how the parameters interact, knowing that the animals do the experiment better if the second phase is delayed by a second or two more than you'd usually think. Your experimental software lets you automatically return the favor, linking your new parameters and experimental scripts as extensions of the prior work.

While you were planning and discussing your experiment you had been contributing your lab's computing hardware to a computational co-op so other people could deploy analyses on it while it was idle. Now you have some credit stored up and distribute the chunks of your analysis across the network. It takes a little bit of tweaking to get some of the more resource-intensive analysis steps to work on the available machines. You don't have time to organize a full pull request to the main analysis code, but if someone wants to do something similar they'll be able to find your version since it's linked to the main library as well as the rest of your project.

You combine the various intermediary results you have posted and been discussing in the forums into a more formal piece of writing. You need to engage with the legacy scientific literature for context, so you highlight the segments you need and make direct reference to and transclude the arguments that they are making in your piece. While you're writing you annotate inline how your work `[[extends::@oldWork]]` because it `[[hasPerspective::@newDiscipline]]`. Some of your results `[[contradict::@oldWork:a-claim]]` and so the people who have published work affirming it are notified and invited to comment.

There isn't any need for explicit peer review to confirm your work as "real science" or not. The social production of science is very visible already, and the smaller pieces you have been discussing publicly are densely contextualized by affirmative and skeptical voices from the several disciplines you were engaging with. You have `@public` annotations enabled on my writing, so anyone reading my work is able to see the inbound links from others highlighting and commenting on it. Submitting in smaller pieces with continual feedback has let you steer your work in more useful directions than your initial experimental plan, so you've already been in contact with many of the people who would have otherwise have been your biggest skeptics and partially addressed their concerns. People are used to assessing the social context of a work: the interfaces make it visually obvious that work that has few annotations, a narrow link tree, or has a really restricted circle of people able to annotate it has relatively less support. When a previously well-supported set of ideas is called into question by new methods or measurements, it's straightforward to explore how its contextual understanding has changed over time.

It's rare for people to submit massive singular works with little public engagement beforehand. There isn't a lot of reward for minimal authorship because the notion of "authorship" has been dissolved in favor of fluid and continuous credit assignment — engaging with little prior work and making few contributions to the data and tooling where it would have been obvious to do so is generally seen as antisocial. They are in the unenviable position of having sunk several years of work into a flawed experimental design that many people in the community could have warned about and helped with, but now since the criticisms are annotated on their work they likely will have to do yet more work if they can't be adequately addressed or dismissed. We don't miss the old system of peer review.

It's clear that you have made a contribution to not only your field, but several that you collaborated with. Your project is a lot more than a single PDF: you can see (and be credited for) the links between data formats, communities, forum posts, notebooks, analytical tools, theories, etc. that you created. It's clear how your work relates to and extends prior work because you were engaging with the structure of scientific research throughout. Your work implies further open questions in the open spaces in the concept graphs of several different research communities, and can organize future experiments without the need for explicit coordination.

There are a dozen or so metrics that are used to evaluate research and researchers. None of them are exactly neutral, and there is ongoing debate about the meaning and use of each since there are so many modalities of credit in a given person's graph. There isn't such a thing as a *proprietary* metric though, because no company has a monopoly on proprietary information that they could say makes it unique, and why would you trust a random number given by a company when there are plenty of ways to measure the public credit graphs? It's relatively hard to game the system, there aren't any proprietary algorithms to fool, and trust is a social process based on mutual affiliation instead of a filter bubble.

The public inspectability of scientific results, the lowered barriers to scientific communication, and ability to find research and researchers without specialized training has dramatically changed between science and the public at large. It's straightforward to find a community of scientists for a given topic and ask questions in the public community forums. Scientific communication resembles the modes of communication most people are familiar with, and have shed some of the stilted formality that made it impenetrable. There isn't such a firm boundary between 'scientist' and 'nonscientist' because anyone can make use of public data and community clusters to make arguments on the same forums and feeds that the scientists do with the same mechanism of credit assignment.

Scientists, building new systems of communication and tooling and then seeding them with their communities has provided alternatives to some of the platforms that dominated the earlier web. The scientists were able to use some of their labor and funding to overcome the development problems of prior alternatives, so they are just as easy to use as (and much more fun than) platforms like Twitter and Facebook. Their well-documented and easily deployed experimental hardware and software has empowered a new generation of DIY enthusiasts, making it possible for many people to build low-cost networked electronics to avoid the surveillance of the ad-based "Internet of Things," air quality sensors, medical devices, wireless meshnets, and so on. The scientists helped make controlling and using personal data much more accessible and fluid. We now control our own medical data and selectively share it as-needed with healthcare providers. Mass genetics databases collected by companies like 23andme and abused by law enforcement slowly fall out of date because we can do anything the geneticists can do.

By taking seriously the obligation conferred by their stewardship of the human knowledge project, the scientists rebuilt their infrastructure to serve the public good instead of the companies that parasitize it. In the process they did their part ending some of the worst harms of the era of global information oligopoly.

Most things aren't completely automatic or infinite, but you don't want them to be. It's nice to negotiate with your federations and communities, it makes you feel like a person instead of a product. Being in a silent room where algorithms shimmer data as a dark wind friction-free through the clouds sounds lonely. Now we are the winds and clouds and the birds that gossip between them, and all the chatter reminds us that we forgot what we were taught to want. You take the hiccups and errors and dead links as the work of the world we built together.

Everything is a little rough, a little gestural, and all very human.

# 5